\documentclass[aps,prb,twocolumn,showpacs,amsmath,amssymb,superscriptaddress,floatfix]{revtex4-2}

\usepackage{graphicx}
\usepackage{amsmath,amsfonts}
\usepackage{xcolor}
\usepackage{color}
\usepackage{mathtools}
\usepackage{eufrak}
\usepackage{pgfplots}
\usepackage{multirow}
\usepackage{textcomp}
\usepackage{array}

\usepackage{comment}

\usepackage{placeins}

\usepackage[normalem]{ulem} 
\usepackage{times}

\definecolor{darkgreen}{rgb}{0,0.6,0}

\def\presuper#1#2%
  {\mathop{}%
   \mathopen{\vphantom{#2}}^{#1}%
   \kern-0.5\scriptspace%
   #2}

\usepackage{hyperref}
\hypersetup{colorlinks=true,breaklinks,linkcolor=blue,urlcolor=blue,citecolor=blue}
\begin{document}
%\pgfplotsset{compat=1.8}
%\pgfmathdeclarefunction{gauss}{3}{%
%    \pgfmathparse{1/(#3*sqrt(2*pi))*exp(-((#1-#2)^2)/(2*#3^2))}%
%    }

    \pgfmathdeclarefunction{gauss}{2}{%
          \pgfmathparse{1/(#2*sqrt(2*pi))*exp(-((x-#1)^2)/(2*#2^2))}%
          }
    \pgfmathdeclarefunction{mgauss}{2}{%
          \pgfmathparse{-1/(#2*sqrt(2*pi))*exp(-((x-#1)^2)/(2*#2^2))}%
          }
    \pgfmathdeclarefunction{lorentzian}{2}{%
        \pgfmathparse{1/(#2*pi)*((#2)^2)/((x-#1)^2+(#2)^2)}%
          }
    \pgfmathdeclarefunction{mlorentzian}{2}{%
        \pgfmathparse{-1/(#2*pi)*((#2)^2)/((x-#1)^2+(#2)^2)}%
          }

\newcommand{\TUVienna}{\affiliation{Institute of Solid State Physics, TU Wien, 1040 Vienna, Austria}}
\newcommand{\UniTueb}{\affiliation{Institut f\"ur Theoretische Physik and Center for Quantum Science, Universit\"at T\"ubingen, Auf der Morgenstelle 14, 72076 T\"ubingen, Germany}}
\newcommand{\LMU}{\affiliation{Arnold Sommerfeld Center for Theoretical Physics, 
Center for NanoScience,\looseness=-1\, and Munich Center for \\ Quantum Science and Technology,\looseness=-2\, Ludwig-Maximilians-Universit\"at M\"unchen, 80333 Munich, Germany}}
\newcommand{\Rutgers}{\affiliation{Department of Physics and Astronomy, Rutgers University, Piscataway, New Jersey 08854, USA}}

\title{Fulfillment of sum rules and Ward identities in the multiloop functional renormalization group solution of the Anderson impurity model}

\author{Patrick Chalupa-Gantner}       \TUVienna
\author{Fabian B. Kugler}      \Rutgers
\author{Cornelia Hille}        \UniTueb
\author{Jan von Delft}         \LMU
\author{Sabine Andergassen}    \UniTueb
\author{Alessandro Toschi}     \TUVienna

\begin{abstract}
We investigate several fundamental characteristics 
of the multiloop functional renormalization group (mfRG) flow by hands of its application to a prototypical many-electron system: the Anderson impurity model (AIM).
We first analyze the convergence of the algorithm in the different parameter regions of the AIM.
As no additional approximation is made, the multiloop series for the local self-energy and response functions converge perfectly to the corresponding results of the parquet approximation (PA) in the weak- to intermediate-coupling regime. 
Small oscillations of the mfRG solution as a function of the loop order gradually increase with the interaction, hindering a full convergence to the PA in the strong-coupling regime, where perturbative resummation schemes are no longer reliable.
By exploiting the converged results, we inspect the fulfillment of (i) sum rules associated to the Pauli principle and (ii) Ward identities related to conservation laws. For the Pauli principle, 
we observe a systematic improvement by increasing the loop order and including the multiloop corrections to the self-energy. This is consistent with the preservation of crossing symmetries and two-particle self-consistency in the PA. 
For the Ward identities, 
we numerically confirm 
a visible improvement by means of the Katanin substitution.
At weak coupling, violations of the Ward identity are further reduced by increasing the loop order 
in mfRG.
In this regime, we also determine the precise scaling of the deviations of the Ward identity as a function of the electronic interaction.
For larger interaction values, the overall behavior
becomes more complex, and 
the benefits of the higher-loop terms are mostly 
present in the 
contributions at large frequencies.
\end{abstract}

\maketitle

\section{Introduction}

The many-electron problem poses a formidable challenge to modern solid-state physics, 
involving a large number of degrees of freedom at different energy scales. 
In general, although the exact solution cannot be computed, some of its fundamental properties are known \textit{a priori}. Specifically, the exact solution is guaranteed to obey the Pauli principle, which manifests itself in sum rules and the crossing symmetry of four-point correlators. At the same time, it also fulfills Ward identities (WIs) related to thermodynamic and quantum-mechanical principles. This knowledge usually provides an important ``compass" for constructing suitable approximation schemes.
For a given approximation, 
however, the preservation of \textit{all} 
fundamental features of the exact solution cannot be  guaranteed \cite{Bickersbook2004}. For instance, approximate schemes constructed from the Luttinger--Ward functional, so-called conserving approximations, maintain WIs---an important aspect when comparing with spectroscopic experiments---but violate the crossing symmetries. 
On the other hand, it is known that approximate approaches specifically designed to guarantee the crossing symmetries, such as the parquet approximation (PA)  \cite{Bickersbook2004, Yang2009, Tam2013, Valli2015, Li2016, Wentzell2020}, violate the WIs to a certain degree \cite{Smith1992,Janis2017,Kugler2018b}.
Hence, investigating how this trade-off actually manifests itself in advanced quantum many-body methods will provide significant theoretical insight. 

In this work, we analyze these issues within the functional renormalization group (fRG) for interacting Fermi systems
\cite{Metzner2012,Salmhofer1999,Berges2002,Kopietz2010,Dupuis2021}, which can be used as a framework for introducing  powerful new approximation schemes.
Specifically, we consider the recent multiloop extension (mfRG) of fermionic fRG in the vertex expansion \cite{Kugler2018,Kugler2018a,Kugler2018b} 
and apply it to the Anderson impurity model (AIM), a paradigmatic model of many-body physics. Reasons for focusing on this particular model are given below.

Computation schemes based on the fermionic fRG 
can be designed to treat the characteristic scale-dependent behavior of correlated electrons in a flexible and unbiased way. 
The most commonly used implementations employ the one-loop ($1\ell$) truncation of the exact hierarchy of flow equations. 
There, one neglects three-particle and higher vertices, which can be justified, e.g., from a perturbative perspective.
Several studies have discussed the noncon\-ser\-ving nature of $1\ell$ fRG-based schemes,
and possible routes for mitigating the violation of the associated WIs
have been proposed \cite{Katanin2004,EnssThesis,Metzner2012,Veschgini2013,Caltapanides2021}. 
An important example is the widely used Katanin substitution \cite{Katanin2004}. 
In its most common form, it incorporates some contributions
from the three-particle vertex as two-loop contributions to
the flow of the two-particle vertex via (one-particle)
self-energy corrections. The Katanin substitution was
designed to better fulfill WIs---%
we here present first numerical results 
to quantitatively assess 
this aspect. Conversely, WIs have also been used to propose new truncation schemes \cite{Schuetz2005,Bartosch2009,Streib2013}.

The multiloop extension of the fRG   
approach, mfRG, includes all
contributions of the three-particle vertex to the 
flow of the two-particle vertex and self-energy that can be computed
with numerical costs proportional to the $1\ell$ flow. In doing so, it sums up all parquet diagrams, formally reconstructing the PA if loop convergence is achieved \cite{Kugler2018,Kugler2018a,Kugler2018b}.
This ensures self-consistency at the one- and two-particle level, in that the PA is a solution of the self-consistent parquet equations \cite{Bickersbook2004}. It also ensures the validity of one-particle conservation laws, but not of two-particle ones \cite{Kugler2018b}.

Whether or not mfRG yields quantitative improvements over the $1\ell$ truncation depends on the context.
For a zero-dimensional model with a logarithmically divergent perturbation theory, 
it was recently shown \cite{Diekmann2020} that the leading logarithms can be obtained in the 1$\ell$ truncation, 
 in which case the higher-loop contributions incorporated via mfRG thus are subleading.  Similarly, $1\ell$ fRG treatments of the interacting resonant level model \cite{Karrasch2010a,Karrasch2010b,Kennes2013,Kennes2013a} as well as inhomogeneous 
Tomonaga--Luttinger liquids \cite{Meden2002,Meden2003,Andergassen2004,Meden2005,Enss2005,Andergassen2006a,Meden2008} 
should yield a proper summation of the leading logs governing the infrared behavior of these systems. 
By contrast, a quantitatively precise description of the weakly interacting two-dimensional Hubbard model could only be achieved with a full multiloop computation \cite{Tagliavini2019,Hille2020,Schaefer2021}.
It is thus of interest to analyze the multiloop series for a model whose perturbation series lacks a leading-log classification, but which is less complex than the Hubbard model.

This criterion is satisfied by the AIM.
We study it here at finite temperature in the imaginary-frequency Matsubara formalism. 
A Matsubara treatment of the AIM suits our purpose for two further reasons. 
First, a numerically exact solution is available as a benchmark via Quantum Monte Carlo (QMC) methods \cite{Gull2011a}. 
Second, recent algorithmic and methodological advances \cite{Wentzell2020,Li2016}
make it possible to track the full frequency dependence
of the 
two-particle vertex functions of the AIM \cite{Rohringer2012,Tagliavini2018,Rohringer2018},
 including their non-trivial asymptotic structure \cite{Wentzell2020}.
The numerical (m)fRG equations can be then solved  to great accuracy and without any further approximations. 
This sets our study apart from 
recent mfRG applications \cite{Tagliavini2019,Hille2020} to more complex systems (where additional approximations for the momentum dependence \cite{Lichtenstein2017,Eckhardt2020} of the problem were necessary)
and builds upon
previous frequency-dependent fRG 
studies of the AIM 
\cite{Hedden2004,Karrasch2008,Jakobs2010,Isidori2010,Rentrop2016,Yirga2021},
paving the way for a systematic inspection of sum rules and WIs in mfRG and parquet approaches.

From a more general perspective, we note that 
flows of the truncated fermionic fRG or mfRG
can \textit{a priori} be expected to be reliable for weak to intermediate interaction strengths only. 
However, nonperturbative \cite{Chalupa2021} parameter regimes of, e.g., the Hubbard model
can be accessed \cite{Vilardi2019,Bonetti2022} 
via fRG  by proceeding as follows: 
first evoke dynamical mean-field theory (DMFT) \cite{Georges1996}
to solve a self-consistent AIM (by non-fRG means, e.g.\ QMC or the numerical renormalization group \cite{Bulla2008}); then use fRG to systematically include nonlocal correlations missed by DMFT \cite{Rohringer2018}.
This procedure defines the so-called DMF$^2$RG scheme \cite{Taranto2014,Wentzell2015}.
So far, it 
has been implemented in the $1\ell$ truncation, but multiloop extensions are conceivable, too.
Our careful investigation of the 
mfRG solution of the AIM may also provide valuable methodological information for future multiloop DMF$^2$RG developments. For example, analogous vertex frequency parametrizations are needed 
for a mfRG treatment of the AIM and for the mfRG part of DMF$^2$RG computations. Furthermore, the study of the mfRG convergence properties as well as of crossing symmetries and WIs for different coupling strengths will represent an important guidance for DMF$^2$RG calculations relying on multiloop resummations.

The structure of our paper reflects
the main scientific questions raised above.
After introducing the required formalism in Sec.~\ref{sec:formal}, we present a detailed analysis of the mfRG solution of the AIM in Sec.~\ref{sec:mfRG=PA}. We illustrate how the convergence to the corresponding results of the PA is perfectly achieved in the weak- to intermediate-coupling regime and also discuss the appearance of increasing multiloop oscillations in the strong-coupling regime. 
Having defined the parameter region of convergence for mfRG applied to the AIM, we analyze in Sec.~\ref{sec:PPandWI} the fulfillment of the sum rules associated to the Pauli principle as well as of the WIs related to conservation laws. 
We discuss the systematic effects observed as a function of loop order, and separately consider the low- and high-frequency parts of the WIs.
Throughout, we also include results obtained via the
Katanin substitution, allowing its merits to 
be compared to those of $1\ell$ or higher-loop schemes.
We summarize our conclusions and 
perspectives for future developments 
in Sec.~\ref{sec:concl}, and discuss additional technical aspects relevant to a more specialized readership 
in the Appendices.

%%%%%%%%%%%%%%%%%%%%%%%%%%%%%%%%%%%%%%%%%%%%%%%%%%%%%%%%%%%%%%%%%%%%%%%%%%%%%%%%%%%%%%%%%%

\section{Formalism}
\label{sec:formal}

In this section, we concisely introduce the methods and concepts underlying the calculations presented in the following sections.
For brevity, we
reduce the formal derivations to a minimum, 
referring to prior works 
for more explicit discussions.
In Sec.~\ref{sec:PPandWI}, we extend the formalism where needed for the 
analysis of the Pauli principle and WIs.

\subsection{Anderson impurity model}

Throughout this paper, we consider the AIM \cite{Anderson1961} 
close to the wide-band limit \cite{Hewson1993}. 
The Hamiltonian is given by
\begin{align}
\label{equ:ham_AIM}
\hat{\mathcal{H}} 
& = 
\sum_{\sigma} \epsilon_d^{\phantom \dagger} 
\hat{d}^{\dagger}_{\sigma} \hat{d}^{\phantom \dagger}_{\sigma} + \sum_{\mathbf{k},\sigma} \epsilon^{\phantom \dagger}_{\mathbf{k}} 
\hat{c}^{\dagger}_{\mathbf{k},\sigma} 
\hat{c}^{\phantom \dagger}_{\mathbf{k},\sigma}
\\ 
& \ + 
U \hat{n}_{\uparrow} 
\hat{n}_{\downarrow}
+ \sum_{\mathbf{k},\sigma} \big( V^{\phantom \dagger}_{\mathbf{k}} \hat{d}^{\dagger}_{\sigma} 
\hat{c}^{\phantom \dagger}_{\mathbf{k},\sigma} + \mathrm{H.c.} \big) 
, \nonumber 
\end{align}
where $\hat{d}_{\sigma}^{\dagger}$ ($\hat{d}^{\phantom \dagger}_{\sigma}$) is the creation (annihilation) operator of electrons localized on the impurity site, and 
$\hat{c}_{\mathbf{k}, \sigma}^{\dagger}$, $\hat{c}^{\phantom \dagger}_{\mathbf{k}, \sigma}$ are the corresponding operators for the bath electrons. The energy on the impurity site is denoted by $\epsilon_d$ and the dispersion relation in the bath by $\epsilon_{\mathbf{k}}$.
The first term in the second line represents the local interaction, where $U$ is the interaction strength and $\hat{n}_{\sigma} = \hat{d}_{\sigma}^{\dagger}\hat{d}^{\phantom \dagger}_{\sigma}$ 
with $\sigma \in \{ \uparrow, \downarrow \}$. The second term accounts for the hopping onto/off the impurity site, where we consider a $\mathbf{k}$-independent hybridization strength 
$V_\mathbf{k} = V$. We set $V=2$, thereby measuring energy in units of $V/2$.
We use a box-shaped DOS for the bath electrons, $\rho(\epsilon) = 1/(2D) \Theta(D-|\epsilon|)$, with half-bandwidth $D=10$. 
Further, we consider half filling, where $\epsilon_d = -U/2$ is exactly canceled by the Hartree self-energy. 
Thus, the (bare) propagator is purely imaginary:
\begin{align}
G_{0,\nu} & = \frac{1}{i\nu - \Delta_\nu},
\quad
\Delta_\nu = - i \frac{V^2}{D} \arctan \frac{D}{\nu}
.
\label{eq:AIM-G0}
\end{align}
For $|\nu| \!\ll\! D$, we find $\Delta_\nu \!\approx\! - i \mathrm{sgn}(\nu) \Delta_0$
with the characteristic hybridization strength $\Delta_0 \!=\! \pi V^2/(2D) \!=\! \pi / 5 \!\simeq\! 0.63$. 
For prior works using this specific AIM, see Ref.~\cite{Chalupa2018} and especially Ref.~\cite{Chalupa2021}, where the physical regimes relevant for this paper are also discussed. For a more general introduction of the physics of the AIM, we refer to Refs.~\cite{Hewson1993,Coleman2015}. The values $U= 1$, $1.5$, $2$, $3$ and $4$ studied below correspond to $U/\Delta_0 \simeq 1.59$, 2.39, 3.18, 4.77 and 6.37, respectively.
Throughout we fix the inverse temperature to $\beta=10$.

\subsection{Numerical approaches}

\textit{fRG flows and PA}---%
We briefly discuss here the structure of the flow equations,
both on the one- and multiloop level, 
as well as the PA, 
for the
one-particle self-energy $\Sigma$, the two-particle vertex $F$,
and the susceptibilities $\chi$ of the AIM. 

The fRG flow describes the evolution of $\Sigma$, $F$, $\chi$
upon tuning the scale or flow parameter $\Lambda$
from an initial value $\Lambda_i$ to a final value $\Lambda_f$.
The flow parameter $\Lambda$ is
introduced in the quadratic part of the action, i.e., the bare (one-particle) propagator [Eq.~\eqref{eq:AIM-G0}].
We consider two cutoff functions: 
the frequency flow (or 
$\Omega$-flow for short), 
\begin{equation}
G^\Lambda_{0,\nu}  = 
\frac{\nu^2}{\nu^2+\Lambda^2} G_{0,\nu}
 \quad
\textrm{with}\; \Lambda_i = \infty, \ 
\Lambda_f = 0
,
\label{eq:G0-Oflow}
\end{equation}
and the interaction flow (or $U$-flow) \cite{Honerkamp2004},
\begin{equation}
G^\Lambda_{0,\nu} = 
\Lambda G_{0,\nu}
 \quad
\textrm{with}\; \Lambda_i = 0, \ 
\Lambda_f = 1
.
\label{eq:G0-Uflow}
\end{equation}
With $G_0^{\Lambda_i} \!=\! 0$, the initial values of $\Sigma$ and $F$ are 
$\Sigma^{\Lambda_i} \!=\! 0$, 
where the Hartree term is absorbed in $G_0$, 
and $F^{\Lambda_i} \!=\! F_0$, the bare vertex of magnitude $U$ (in our convention $F_0^{\sigma\sigma'} \!=\! -U\delta_{\bar{\sigma}\sigma'}$,
where $\bar{\uparrow}=\downarrow$ and vice versa). 
The fRG flow of $\Sigma$ is determined by the two-particle vertex $F$ contracted with the single-scale propagator $S \!=\! -G (\partial_\Lambda G_0^{-1}) G$, which is related to the differentiated propagator $\dot{G} \!\equiv\! \partial_\Lambda G$ by $\dot{G} \!=\! S + G \dot{\Sigma} G$. For simplicity, we omit the superscript $\Lambda$ here and henceforth. 
The flow equation for $F$ further involves the
three-particle vertex $\Gamma^{(6)}$. If $\Gamma^{(6)}$ was known at all scales, the flow of $\Sigma$ and $F$ would be exact. This would imply, in particular, that every specific $\Lambda$ dependence or cutoff choice, as in Eqs.~\eqref{eq:G0-Oflow} or \eqref{eq:G0-Uflow}, yields the same result at the end of the flow. In practice, however, $\Gamma^{(6)}$ can hardly be treated numerically and its effect on the flow of $\Sigma$ and $F$ can only be accounted for approximately. As a consequence, the results of such truncated fRG flows will generically depend on the choice of the cutoff.

The most widely used fRG implementations 
neglect $\Gamma^{(6)}$, 
yielding approximate 
$1\ell$ flow equations for $\Sigma$ and $F$.
The contributions of $\Gamma^{(6)}$ that amount to 
self-energy derivatives can be added to the vertex flow by substituting $S \!\rightarrow\! \dot{G}$. This ``Katanin substitution'' \cite{Katanin2004}, labeled by $1\ell_K$ in the following, was argued to yield a better fulfillment of WIs.
A further refinement, which effectively incorporates the three-particle vertex to third order in the renormalized interaction, is obtained by the two-loop ($2\ell$) vertex corrections \cite{Katanin2004,Eberlein2014,Rueck2018}. 

Subsequently, the mfRG extension \cite{Kugler2018,Kugler2018a,Kugler2018b} was introduced to 
incorporate all those contributions of $\Gamma^{(6)}$ to the flow of $\Sigma$ and $F$ ensuring that their right-hand sides are \textit{total} scale derivatives---which
is not the case for the $1\ell$ flow---thus guaranteeing by construction that the final results are independent of the choice of cutoff.
In fact, the corresponding higher-loop terms of the mfRG represent the \textit{minimal}
additions to the conventional 
1$\ell$ flow required to obtain cutoff-independent results.
They also provide the \textit{maximal} 
amount of diagrammatic contributions that can be added in a numerically feasible manner.
Indeed, due to the iterative structure based on successive $1\ell$
computations, these higher-loop contributions can be computed very efficiently \cite{Tagliavini2019}.
Besides the two-dimensional Hubbard model \cite{Hille2020}, recent applications also include spin models \cite{Thoenniss2020,Kiese2020} within the pseudofermion representation.

The mfRG was shown to formally reproduce the diagrammatic resummation of the PA.
We use the vertex decomposition
\begin{equation}
\label{eq:decomp}
F = R_{\mathrm{2PI}} +\sum_r \gamma_r, 
\end{equation}
where $\gamma_r$ are the two-particle reducible vertices in channel
$r \!\in\! \{a, p, t\}$ and $R_{\mathrm{2PI}}$ the fully two-particle irreducible (2PI) vertex 
(notation as in Ref.~\cite{Kugler2018b} \footnote{The translation of the notation used in this work to the notation used in many other works, among them Ref.~\cite{Rohringer2018}, is the following: The diagrammatic channels relate to one another as $a=\overline{ph}$, $p=pp$, $t=ph$; the vertex two-particle reducible in channel $r$ is referred to as $\gamma_r = \Phi_r$, the vertex irreducible in channel $r$ as $I_r = \Gamma_r$ and the fully two-particle irreducible vertex as $R_{\mathrm{2PI}} = \Lambda_{\mathrm{irr}}$.}). 
The PA then corresponds to the approximation $R_{\mathrm{2PI}}=F_0$.
An analogous approximation is performed in truncated fRG flows:
Neglecting $\Gamma^{(6)}$, the vertex flow equation is of the form
$\dot{F} \!=\! \sum_r \dot{\gamma}_r$.
Thus, only the reducible parts are renormalized, while the 
fully irreducible part does not flow and remains at its initial value
$R_{\mathrm{2PI}} \!=\! F_0$.

In the following, we 
recall the 
flow equations for the self-energy, two-particle vertex, and susceptibilities, as well as the parquet equations.
We will 
use the compact symbolic notation introduced in Ref.~\cite{Kugler2018b}; the explicit dependence on spin and frequencies will be given where needed.

\begin{figure}[t!]
\centering
\includegraphics[width=0.483\textwidth]{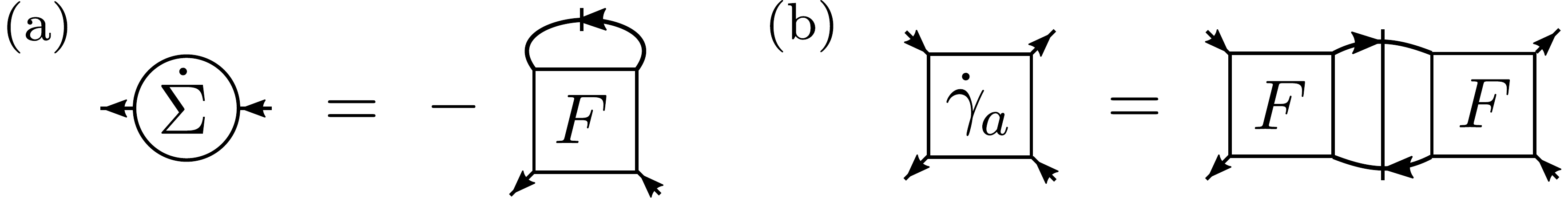}
\caption{Diagrammatic representation of the one-loop flow of (a) the self-energy and (b) the vertex in the $a$ channel. The slashed line denotes $S$, the slashed pair of lines $\dot{\Pi}_{a,S}$.}
\label{fig:1l-flow}
\end{figure}

\textit{One-loop flow}---%
The `standard' fRG self-energy flow \cite{Metzner2012}
is 
\begin{align}
\dot{\Sigma}_{\mathrm{std}}
= - F \cdot S
,
\label{eq:SE-flow_std}
\end{align}
as illustrated in Fig.~\ref{fig:1l-flow}(a).
Figure~\ref{fig:1l-flow}(b) shows an exemplary depiction of the $1\ell$ flow of the vertices, given by
\begin{align}
\dot{\gamma}^{(1)}_r = F \circ \dot{\Pi}_{r,S} \circ F
,
\label{eq:vertex-flow_1l}
\end{align}
and $\dot{F} \!=\! \sum_r \dot{\gamma}_r$.
$\dot{\Pi}_{r,S}$ corresponds to the differentiated two-particle propagator 
in channel $r$, 
with $S$ used instead of $\dot{G}$.

The flow equation for the susceptibilities $\chi_r$ can be derived from the corresponding reducible vertex in the limit of large fermionic frequencies, i.e., the so-called $K_1$ contribution \cite{Wentzell2020},
\begin{align}
\dot{\chi}_r^{(1)} = - \lambda_r \circ \dot{\Pi}_{r,S} \circ \lambda_r
,
\label{eq:chi-flow_1l}
\end{align}
where $\lambda_r$ are the 
renormalized three-point vertices 
(for further details and the $1\ell$ flow equation of the latter, see Ref.~\cite{Metzner2012}).

The $1\ell_K$ flow with Katanin substitution is
obtained by replacing 
$S \!\rightarrow\! \dot{G}$, i.e., $\dot{\Pi}_{r,S} \!\rightarrow\! \dot{\Pi}_{r}$, in Eqs.~\eqref{eq:vertex-flow_1l} and \eqref{eq:chi-flow_1l}.
Since it includes self-energy (and not vertex) corrections from $\Gamma^{(6)}$,
we will 
display the $1\ell_K$ results
between those for $\ell \!=\! 1$ and $\ell \!=\! 2$.

\textit{Multiloop flow}---%
The multiloop flow further 
includes the contributions 
from $\Gamma^{(6)}$
which are generated by vertex corrections. These can be 
ordered by loops, leading to the expansion
$\dot{\gamma}_r \!=\! \sum_{\ell \geq 1} \dot{\gamma}_r^{(\ell)}$
\cite{Kugler2018,Kugler2018a}.
Here, $\dot{\gamma}^{(1)}$ already includes the Katanin substitution to account for the self-energy corrections as above.
The higher-loop terms, $\ell>1$, are determined by
\begin{subequations}
\begin{alignat}{2}
\dot{\gamma}_r^{(\ell)}
& =
\dot{\gamma}_{\bar{r}}^{(\ell-1)} \circ \Pi_r \circ F
+
F \circ \Pi_r \circ \dot{\gamma}_{\bar{r}}^{(\ell-1)}
\quad && (\ell \geq 2)
\label{eq:ml-flow-LR}
\\
& \ +
F \circ \Pi_r \circ \dot{\gamma}_{\bar{r}}^{(\ell-2)} \circ \Pi_r \circ F
\quad && (\ell \geq 3)
,
\label{eq:ml-flow-C}
\end{alignat}
\end{subequations}
where
$\gamma_{\bar{r}} = \sum_{r' \neq r} \gamma_{r'}$.
Equation~\eqref{eq:ml-flow-LR} 
with $\ell=2$
corresponds to the $2\ell$ flow, 
while the so-called center part $\dot{\gamma}^{(\ell)}_{r,\mathrm{C}}=F \circ \Pi_r \circ \dot{\gamma}_{\bar{r}}^{(\ell-2)} \circ \Pi_r \circ F$ of Eq.~\eqref{eq:ml-flow-C} contributes only for $\ell \!\geq\! 3$.

In order to fully generate all parquet diagrams, the self-energy flow also acquires a multiloop correction \cite{Kugler2018},
\begin{align}
\dot{\Sigma} 
=
\dot{\Sigma}_{\mathrm{std}} + 
(1 + F \circ \Pi_t ) \circ \dot{\gamma}_{\bar{t},\mathrm{C}} \cdot G
,
\label{eq:multiloop_self-energy_flow}
\end{align}
where 
$\dot{\gamma}_{\bar{t},\mathrm{C}} \!=\! \sum_{\ell\geq 3} \dot{\gamma}^{(\ell)}_{\bar{t},\mathrm{C}}$ (see above).
While not relevant for our AIM study, we note that additional approximations, such as the low-order expansion in form factors for the momentum-dependence of the vertex functions, useful for reducing the numerical effort in treating lattice problems, require  extra adaptations of the flow equations for the mfRG solution to converge to the PA \cite{Hille2020,Hille2020a}.

\begin{figure}[t!]
\centering
\includegraphics[width=0.483\textwidth]{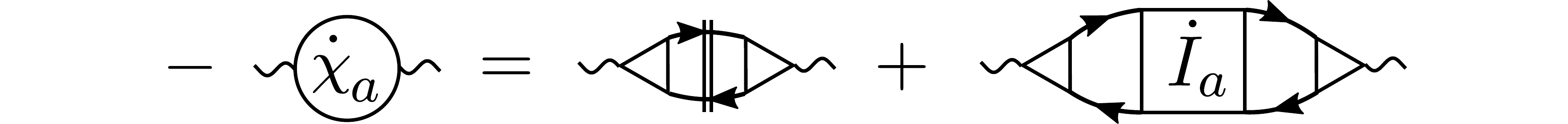}
\caption{Diagrammatic representation of the multiloop flow of the susceptibility in the $a$ channel. The doubly slashed lines denote $\dot{\Pi}_a$.}
\label{fig:chi-ml-flow}
\end{figure}

\begin{figure*}[t]
\centering
{{\resizebox{17.8cm}{!}{\includegraphics {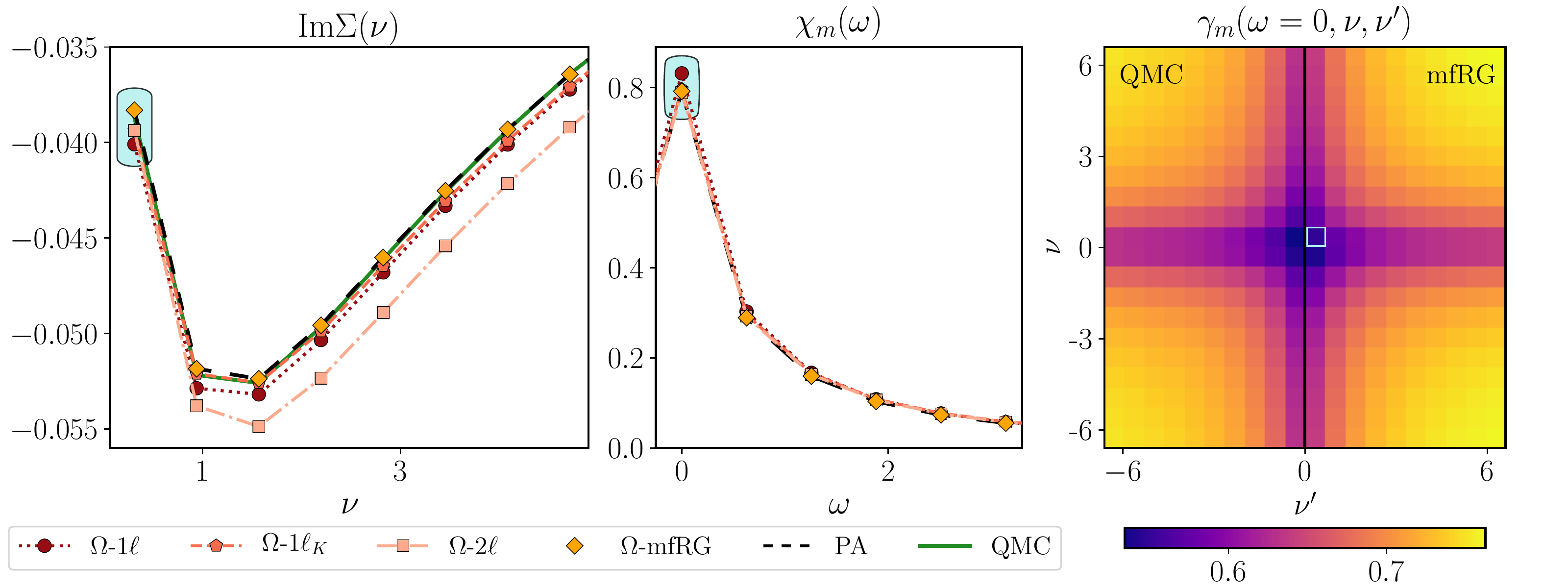}}} 
\caption{Self-energy Im$\Sigma(\nu)$ (left), magnetic susceptibility $\chi_m(\omega)$ (center), and reducible vertex $\gamma_m = K_{1m} + K_{2m} + K_{2'm} + K_{3m}$ (right) as obtained by different approaches, for $U=1$. We consider $\beta=10$ and half filling throughout. 
The fRG results shown here are computed with the $\Omega$-flow [Eq.~\eqref{eq:G0-Oflow}]. The shaded areas in the first two panels and the frame in the right one mark the frequencies used to study the loop convergence in Secs.~\ref{subsec:mfrg-conv} and \ref{subsec:mfrg-params}. 
}
\label{fig:Schaufenster-DATA}  
}
\end{figure*} 
The multiloop flow equation for the susceptibilities 
reads
\begin{align}
\dot{\chi}_r = - \lambda_r \circ (\dot{\Pi}_{r} + \Pi_{r} \circ \dot{I}_r \circ \Pi_{r} ) \circ \lambda_r
,
\label{eq:chi-flow_ml}
\end{align}
with the scale derivative of the two-particle irreducible vertex 
$\dot{I}_r \!=\! \sum_{\ell \geq 1} \dot{\gamma}_{\bar{r}}^{(\ell)}$, 
see Fig.~\ref{fig:chi-ml-flow} for an exemplary diagrammatic representation.
For more details 
and the equations for
$\lambda_r$, we refer to Refs.~\cite{Kugler2018b,Tagliavini2019}.

\textit{PA}---%
In parquet approaches \cite{Bickersbook2004}, a set of self-consistent
equations for the self-energy and vertex is solved by iteration.
First, $\Sigma$ is related to $F$ by the Schwinger--Dyson equation (SDE)
\begin{align}
\Sigma 
& = 
- F_0 \cdot G 
- \tfrac{1}{2}
(F_0 \circ \Pi_a \circ F) \cdot G
.
\label{eq:SDE-general}
\end{align}
Second, in the decomposition~\eqref{eq:decomp}, 
the two-particle reducible vertices $\gamma_r$ are related to
two-particle irreducible vertices $I_r$ 
by the Bethe--Salpeter equations (BSEs)
\begin{align}
\gamma_r 
& = 
I_r \circ \Pi_r \circ F
, \qquad 
I_r = F - \gamma_r
= R_{\mathrm{2PI}} + \gamma_{\bar{r}}
.
\label{eq:BSE-general}
\end{align}
In the PA, $R_{\mathrm{2PI}} \!=\! F_0$. 
Finally, the susceptibilities $\chi_r$ can be directly deduced
from $F$ (and $\Sigma$ via the propagators) by
\begin{align}
\chi_r = - \lambda_{r,0} \circ ( \Pi_r + \Pi_r \circ F \circ \Pi_r ) \circ \lambda_{r,0}
.
\label{eq:chi-vertex-general}
\end{align}
Here, $\lambda_{r,0}$ are the bare three-point vertices
encoding the relation of the composite bosonic degrees of freedom
of $\chi_r$ to the original fermionic ones.

In the parquet context, Eqs.~\eqref{eq:SDE-general}--\eqref{eq:chi-vertex-general} 
do not involve a scale parameter $\Lambda$.
However, as they hold for any underlying bare propagator, 
they can also be applied when the bare propagator is $G_0^\Lambda$. 
These relations can then be used to derive the multiloop flow equations \cite{Kugler2018b},
and thus Eqs.~\eqref{eq:SDE-general}--\eqref{eq:chi-vertex-general} are fulfilled exactly in mfRG \cite{Kugler2018a,Tagliavini2019}.
In other truncated schemes, they can be exploited 
as additional
\textit{post-processing} (PP) relations for computing
(i)
the self-energy from 
the SDE~\eqref{eq:SDE-general},
(ii)
the
reducible vertices from the BSEs~\eqref{eq:BSE-general},
and (iii)
the susceptibilities 
using Eq.~\eqref{eq:chi-vertex-general}, instead of using the corresponding results of the flow.
We recall that, for a generic truncated fRG scheme (including the standard $1\ell$ truncation), the PP values of $\Sigma$, $\gamma_r$, and $\chi_r$ differ from their counterparts obtained directly from the flow.
In fact, the equivalence between the flowing and PP results for $\Sigma$, $\gamma_r$, and $\chi_r$  (upon convergence) represents, besides the independence from the choice of the cutoff function, a hallmark of the mfRG \cite{Tagliavini2019}.
For this reason, we will also compute the PP results for $\Sigma$ and $\chi_r$, and analyze their evolution with loop order.

\textit{QMC}---%
Next to the fRG 
and 
PA described 
above, we use a state-of-the-art Quantum Monte Carlo \cite{Gull2011a} (QMC) solver 
to obtain numerically exact benchmark results of the AIM.
We employ continuous-time QMC in the hybridization expansion (CT-HYB)\cite{Gull2011a} provided by the open-access \textsc{w2dynamics} \cite{w2dynamics} package. 
Further details on the calculations are provided 
in Appendix~\ref{sec:APP-tech-QMC}.

%%%%%%%%%%%%%%%%%%%%%%%%%%%%%%%%%%%%%%%%%%%%%%%%%%%%%%%%%%%%%%%%%%%%%%%%%%%%%%%%%%%%%%%%%

\section{$\text{mfRG}$ solution of the AIM}
\label{sec:mfRG=PA}
We now apply the mfRG, briefly summarized in Sec.~\ref{sec:formal}, to the half-filled AIM at the inverse temperature $\beta = 10$ and discuss the results. 
For details on the implementation, 
we refer to Refs.~\cite{Tagliavini2019,Wentzell2020}. 
We just note here that, for the reducible vertices, we adopt the parametrization $\gamma_r \!=\! K_{1r}\! +\! K_{2r}\! +\! K_{2'r} \!+\! K_{3r}$ proposed in Ref.~\cite{Wentzell2020}. The $K_{1r}$ and $K_{2^{(\prime)}r}$ functions with one and two frequency arguments, respectively, describe the high-frequency asymptotics, while the remaining full dependence at low frequencies is contained in $K_{3r}$.
This reduces the numerical cost, allowing for the calculation of 
the vertices on a larger Matsubara frequency range 
(see Appendix~\ref{subsec:APP-tech-frequencies} for 
computational details). 
The (flowing) susceptibilities are conveniently extracted through $\chi_r = - K_{1r}/U^2$.

We start the presentation of our numerical results by showcasing the central quantities of our study of the AIM, i.e., the self-energy
$\Sigma$, 
the magnetic susceptibility
$\chi_m$
($=- \chi_a^{\uparrow\downarrow}$),
and the reducible vertex $\gamma_m$ of the impurity site in the magnetic channel, computed in the weak-coupling regime ($U=1$) by means of all the approaches mentioned in Sec.~\ref{sec:formal}.
Figure~\ref{fig:Schaufenster-DATA} displays our results for 
$\Sigma$, $\chi_m$ and  $\gamma_m$
as a function of fermionic (bosonic) Matsubara frequencies.
The corresponding numerical data would also allow one to estimate important physical quantities (e.g., the quasiparticle mass renormalization and life time) relevant for the description of the Fermi-liquid state of the impurity problem \cite{Hewson02,coleman2006heavy} as well as to quantify the temporal fluctuations of the local magnetic moment on the impurity site \cite{Watzenboeck2020,Tomczak2021,Watzenboeck2021}.

Consistent with the small $U$ value of these illustrative calculations, 
all approaches yield qualitatively
the same behavior and deviations to numerically exact QMC data are hardly visible. 
In particular, we note that the converged mfRG solution (orange squares), 
perfectly 
matches
the PA 
(dashed black line) 
for all quantities,
$\text{Im}\Sigma$, $\chi_m$, and $\gamma_m$ (not shown).
The results at the highlighted Matsubara frequencies are then used in the following Sec.~\ref{subsec:mfrg-conv} for a quantitative study of
the mfRG convergence as a function of 
loop order $\ell$.
There, we also showcase two hallmark qualities of the converged mfRG solution: 
(i) It is cutoff-independent, reflecting the fact that it reproduces the PA solution, which, as a self-consistent diagrammatic resummation, by construction is defined without reference to any cutoff.
(ii) For quantities that can be computed either via their own RG flow equations or via PP relations, 
the results agree. 
(If the susceptibility flow is computed separately, and not via that of the $K_{1r}$ part of the vertex, this requires to further use multiloop flow equations for the susceptibilities and the three-point vertices \cite{Tagliavini2019,Kugler2018b}.)
In Sec.~\ref{subsec:mfrg-params}, we extend this analysis to larger values of $U$.

%%%%

\subsection{Multiloop convergence to PA}
\label{subsec:mfrg-conv}

\begin{figure*}[t]
\centering
{{\resizebox{17.8cm}{!}{\includegraphics {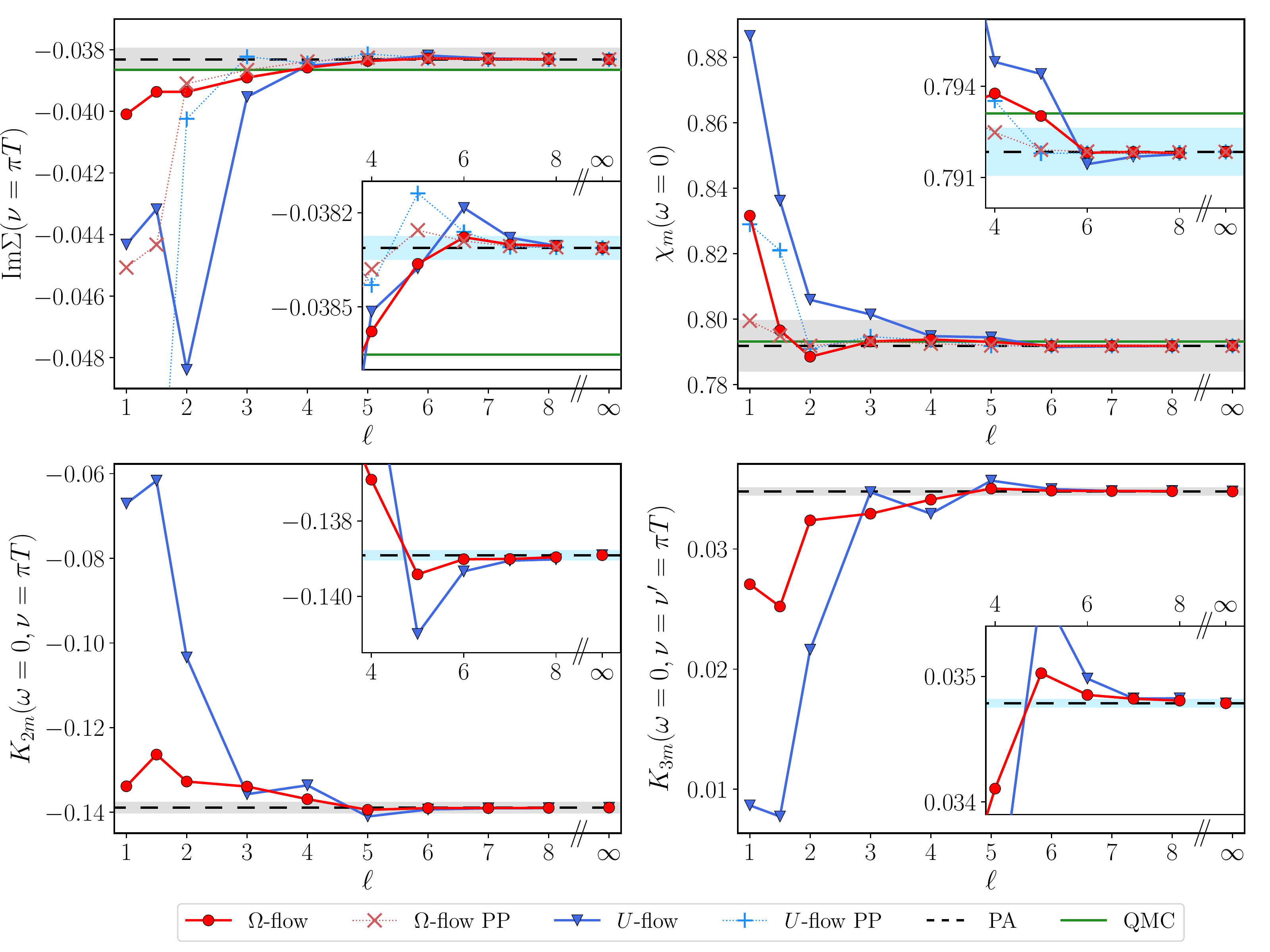}}} 
\caption{$\Omega$-flow (red) and $U$-flow (blue) mfRG results as a function of loop order $\ell$ in comparison with the PA (black, dashed) and the numerically exact QMC data (green), for 
$U=1$. Upper panels: 
$\text{Im}\Sigma(\nu=\pi T)$ and $\chi_m(\omega=0)$, showing perfect agreement between post-processed (PP) and flowing results of both cutoffs and the PA result. 
Lower panels: 
asymptotic vertex functions $K_{2m}$ and $K_{3m}$ for the lowest Matsubara frequencies.
Insets show a zoom for $\ell \ge 4$.
The gray areas mark $1\%$ deviation from the PA result, the blue ones in the insets 
$0.1\%$. 
The label `$\infty$' represents the fully converged mfRG result. In this and similar figures below, the data points plotted between those at $\ell=1$ and $\ell=2$ represent the $1\ell_K$ results (Katanin substitution). 
}
\label{fig:mloop-convergence}  
}
\end{figure*} 

%%%

In Fig.~\ref{fig:mloop-convergence}, we analyze in detail the loop convergence of the mfRG flow for $U\!=\!1$. The four panels display both the flowing and PP results for $\text{Im}\Sigma(\nu\!=\!\pi T)$ and $\chi_m(\omega\!=\!0)$ as well as the flowing results for $K_{2m}(\omega=0,\nu\!=\!\pi T)$ and $K_{3m}(\omega=0,\nu\!=\!\nu'\!=\!\pi T)$, as a function of loop order $\ell$ obtained with the two cutoffs, i.e., the $\Omega$-flow (red circles) and the $U$-flow (blue triangles). 
For comparison, the PA (black dashed line) and QMC (green solid line) solutions are also reported. 
One readily notices that the mfRG solution for \textit{both} cutoffs converges to the PA for all considered quantities. Throughout the paper, the label `$\infty$' refers to the infinite loop-order mfRG solution (see Appendix~\ref{subsec:APP-mfRGcalc-loop} for its numerical definition).
The high quality of the mfRG convergence can be appreciated by looking at the corresponding insets, showing the data restricted to higher loop orders. While the gray area in the main panels marks $1\%$ deviation with respect to the PA, the blue area in the insets corresponds to $0.1\%$. 

It is worth stressing that for some quantities and specific values of $\ell$, the mfRG and PA solution may be accidentally close, e.g.\ the $3\ell$ $\Omega$-flow result for $\chi_m(\omega=0)$ or the $3\ell$ $U$-flow result for $K_{3m}(\omega=0,\nu=\nu'=\pi T)$. 
Of course, this does not mean that the mfRG procedure has already converged at $3\ell$: Full convergence implies the 
equivalence of mfRG and PA for {\sl all} quantities and {\sl both} cutoffs up to differences smaller than a given $\epsilon$, e.g., here $0.1\%$. For the $U=1$ calculations, this is clearly achieved for $\ell \ge 8$. 
Looking at the insets, 
the $\Omega$-flow appears to converge systematically faster 
than the $U$-flow. 
We note that all $U$-flow results shown in the paper are obtained via a frequency extrapolation (see Appendix~\ref{subsec:APP-tech-frequencies}), which is required to achieve the highly precise convergence to PA demonstrated in the inset. 

Another important property of the converged mfRG solution is the equivalence of the flowing and PP results, shown both for Im$\Sigma(\nu=\pi T)$ and $\chi_m(\omega=0)$ in the upper panels of Fig.~\ref{fig:mloop-convergence}. 
Except for the $1\ell$ and $1\ell_K$ results for the self-energy, the PP data (dotted lines with `$\times$' or `$+$' symbols) are always found to be closer to the PA than the flowing data 
(for the susceptibility, this trend was previously reported in Ref.~\cite{Hille2020}).
For both cutoffs, flowing and PP results agree with the PA for $\ell\ge 8$, highlighting the perfect convergence of the mfRG scheme in this parameter regime.
The loop convergence can also be seen from calculations with a single cutoff, as there are no more changes larger than a small $\epsilon$ in all quantities when going from $\ell$ to $\ell+1$, and flowing and PP results agree with one another.
Finally, let us note that adopting the PP procedure has also important implications for the fulfillment of sum rules, which are studied in Sec.~\ref{subsec:sumrules}.

%%%%%%%%%%%%%%%%%%%%%%%%%%%%%%%%%%%%%%%%%%%%%%%%%%

\subsection{Towards strong coupling}
\label{subsec:mfrg-params}

\begin{figure*}[t]
\centering
{{\resizebox{17.8cm}{!}{\includegraphics {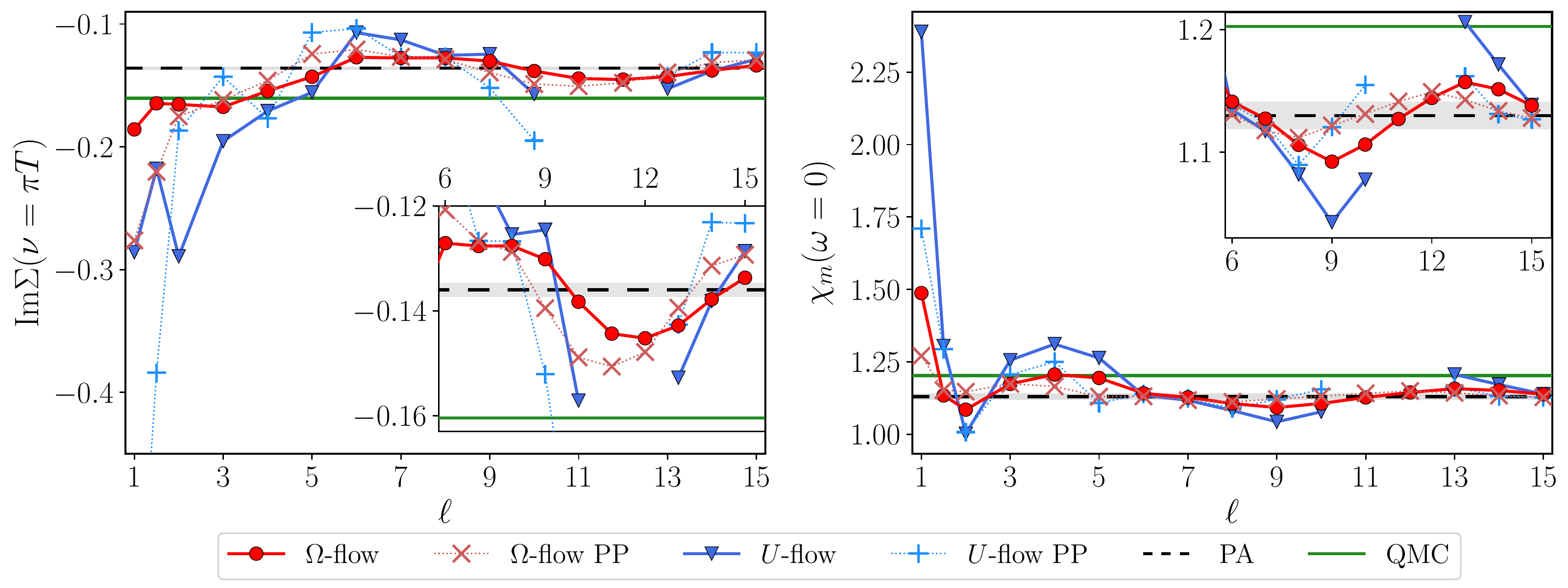}}} 
\caption{$\text{Im}\Sigma(\nu = \pi T)$ and $\chi_m(\omega = 0)$ as in Fig.\,\ref{fig:mloop-convergence} but for $U=2$. 
Insets show a zoom for $\ell\ge6$.
The gray area 
indicates $1\%$ deviation from the PA. For $\ell\!=\!11,12$ we were unable to converge the $U$-flow calculations.  
}
\label{fig:mloop-convergenceU20}  
}
\end{figure*} 

\begin{figure*}[t]
\centering
{{\resizebox{17.8cm}{!}{\includegraphics {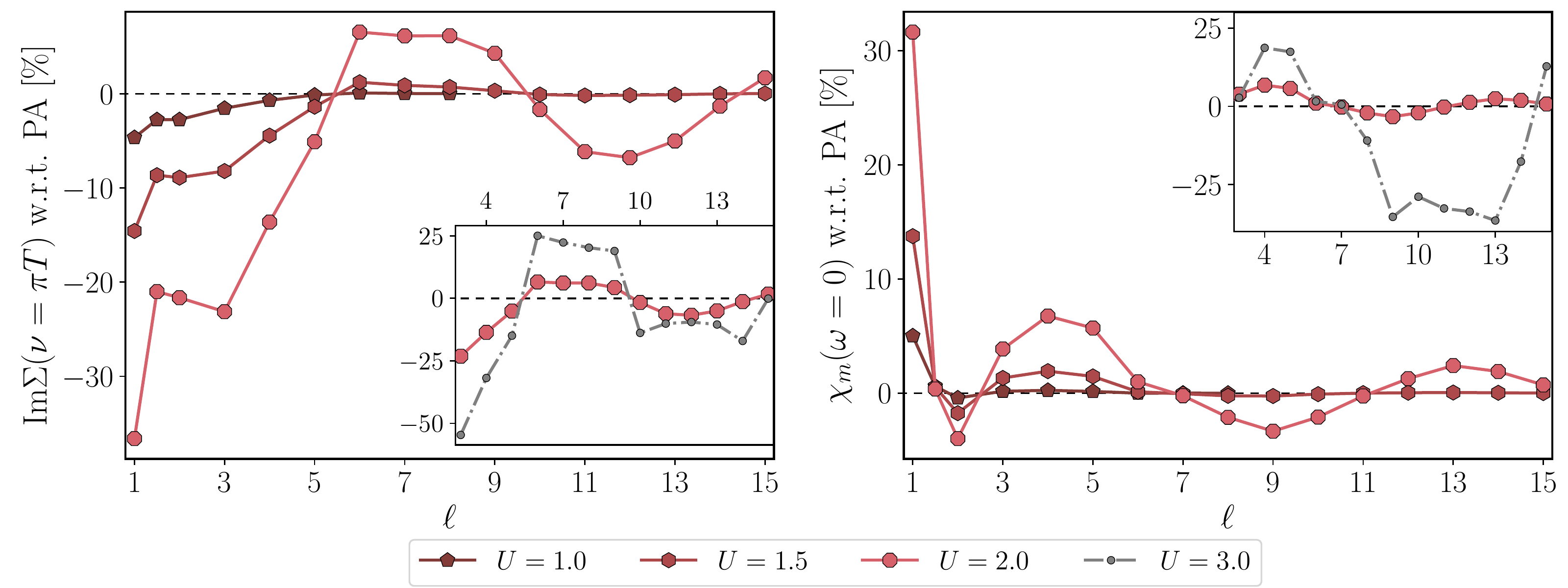}}} 
\caption{Relative difference between $\Omega$-flow mfRG calculations (flowing) and the corresponding PA solutions for Im$\Sigma(\nu=\pi T)$ (left) and $\chi_m(\omega=0)$ (right), as a function of loop order $\ell$ and different values of the interaction $U$. 
Main panels show $U \leq 2$, insets $U \geq 2$.
}
\label{fig:mloop-Ucomparison}  
}
\end{figure*} 

\begin{figure}[tb]
\centering
{{\resizebox{8.0cm}{!}{\includegraphics {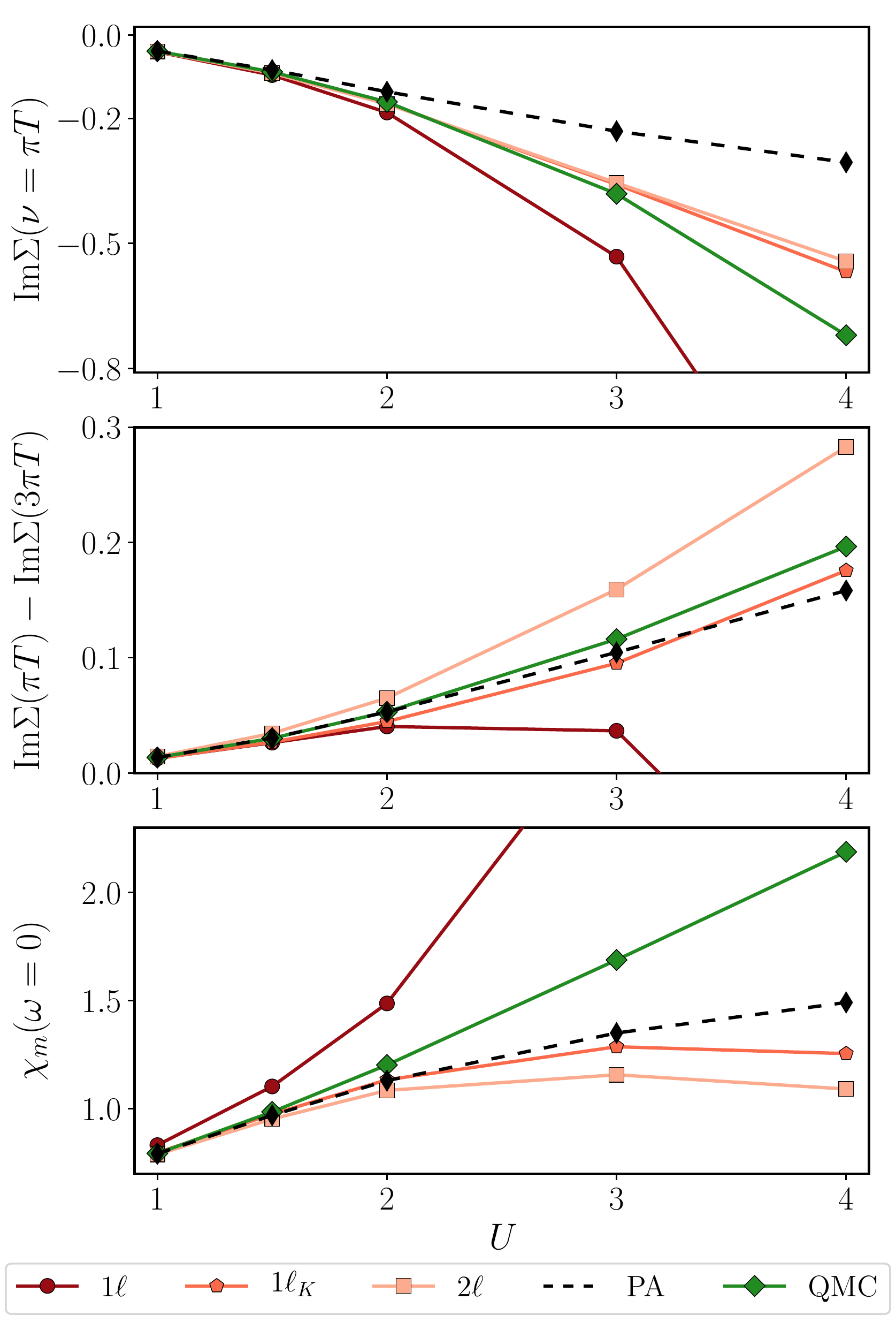}}} 
\caption{$\text{Im}\Sigma(\pi T)$ (top), $\text{Im}\Sigma(\pi T) - \text{Im}\Sigma(3\pi T)$ (center) and $\chi_m(\omega = 0)$ (bottom) as a function of $U$, obtained from $\Omega$-flows at low loop order, the PA, and QMC. 
}
\label{fig:phys-Ularge}  
}
\end{figure}

We now analyze 
how the convergence of the mfRG flow is affected by increasing the interaction $U$.
In Figs.~\ref{fig:mloop-convergenceU20}--\ref{fig:phys-Ularge}, we focus on the results for the physical quantities Im$\Sigma(\nu \!=\! \pi T)$ and $\chi_m(\omega \!=\! 0)$, but we also checked for convergence of $K_{2m}(\omega \!=\! 0,\nu \!=\! \pi T)$ and $K_{3m}(\omega \!=\! 0,\nu \!=\! \nu' \!=\! \pi T)$.

For values of $U$ slightly larger than $U=1$, the convergence behavior is qualitatively the same 
(see Fig.~\ref{fig:mloop-convergenceU15} in Appendix~\ref{sec:APP-add} for $U=1.5$),
albeit with increasing interaction, as expected, more 
loop orders are required to reach 
convergence. 

For $U=2$, the dependence on loop order is shown in Fig.~\ref{fig:mloop-convergenceU20}. While the mfRG solution quickly approaches the PA 
for low $\ell$, 
the path towards full convergence for higher $\ell$ becomes visibly slower as the curves describing the loop dependence of the mfRG calculations keep oscillating around the PA solution. 
The $\Omega$-flow results are generally found to be more accurate than the $U$-flow data
(note that for the $U$-flow at $\ell=11,12$, no solution could be obtained; see also Appendix~\ref{subsec:APP-mfRGcalc-loop}).
Yet, even with the $\Omega$-flow, 
we did not reach \textit{perfect} convergence 
up to $\ell=40$. 
Different from the situation at $U\!=\!1$ and $U\!=\!1.5$, 
the results obtained by PP do not show a clear improvement. 
Instead, they seem to follow a 
slightly different oscillation pattern, somewhat shifted from the flowing data (see insets of Fig.~\ref{fig:mloop-convergenceU20}). 
Further insight on the oscillations characterizing the mfRG convergence with increasing interaction can be gained from Fig.~\ref{fig:mloop-Ucomparison}. Here, we show the relative difference between the mfRG results and the corresponding PA solutions for different values of $U$. 
By comparing the (flowing) results of the $\Omega$-flow for different interaction strengths $U\!=\!1,1.5$ and $2$, one notices 
the presence of
``nodes" in the multiloop oscillations, i.e., of loop orders at which mfRG and PA yield numerically very similar results for the quantity under consideration. The location of these nodes, however, depends on the observable.
[While, e.g., $\ell=7$ for $\chi_m(\omega=0)$ is close to the PA for all values of $U$, for Im$\Sigma(\nu=\pi T)$ this is not the case.] 
For larger interactions, the 
oscillations 
become stronger. 
Already for $U\!=\!2$, the amplitude of the self-energy oscillations hardly decreases  
with increasing loop order, making a full convergence numerically challenging as discussed above.
(The $U$-flow shows similar behavior, see Fig.~\ref{fig:APP_UFlow_UCOMP} in Appendix~\ref{sec:APP-add}.)
This effect 
gets even more pronounced for $U=3$ displayed in the insets, together with $U=2$ for comparison.
There, 
higher loop orders, especially for $\chi_m(\omega=0)$, 
yield a progressively enhanced deviation 
from the PA for increasing loop order. 
Therefore, we conclude that, within our current implementation 
and the 
given settings of the AIM, 
the mfRG loop resummation 
ceases to converge for $U=3$.
Such a lack of loop convergence serves as a built-in red-flag indicator that a parameter regime lies outside the zone of safe applicability of the approach.
This outcome, however, is not entirely unexpected since, for the specific AIM considered, the interaction strength $U\!=\! 3$ already corresponds to the strong-coupling regime, where nonperturbative \cite{Kozik2015,Gunnarsson2017,Reitner2020,Chalupa2021} 
divergences of two-particle irreducible vertices \cite{Schaefer2013,Janis2014, Ribic2016, Schaefer2016c,Vucicevic2018,Thunstroem2018,Springer2019,Kotliar2020}, 
which are---per construction---beyond the PA, were detected by means of QMC calculations
\cite{Chalupa2018,Chalupa2021}.
More speculatively, one might then suppose a relation between 
the breakdown of the mfRG convergence 
and the entrance into the nonperturbative parameter regime, where the PA itself yields results significantly different from the exact solution \cite{Chalupa2021}.
In this respect, the oscillations of increasing size could be seen as a precursor of the breakdown of perturbative resummation schemes.

%%%%%%%%%%%%%%%%%%%%%%%%%%%%%%%%%%%%%%%%%%%%%%%

We finally compare the results of low loop orders to the PA and the exact solution, as a function of $U$.
For very low values of $U$, the deviations of mfRG and PA schemes from QMC can be qualitatively understood from general perturbation-theory considerations.
Already for $U>1$, however, the interpretation becomes more complicated, and the accuracy of the different schemes depends on the observable considered.
Among the $\Omega$-flow results up to $U=4$ in Fig.~\ref{fig:phys-Ularge}, the plain 1$\ell$ flow performs worst for all quantities.
Comparing 1$\ell_K$ and the PA to the exact QMC for large $U \geq 2$,
we find the best results for Im$\Sigma(\nu=\pi T)$ with 1$\ell_K$,
similar deviations for $\mathrm{Im}\,\Sigma(\nu=\pi T)-\mathrm{Im}\,\Sigma(\nu=3\pi T)$
with 1$\ell_K$ and the PA,
and the best results for $\chi_m(\omega=0)$ with the PA.
For the physical interpretation of the strong-coupling regime, 
we refer to Ref.~\cite{Chalupa2021} and the corresponding supplemental material. There it was shown that both the PA and fRG schemes yield a qualitatively correct description of the magnetic channel; 
in particular, the proper 
behavior of $\chi_m(\omega=0)$ as a function of $T$ is found, reflecting the formation of a local magnetic moment and its screening. However, both methods fail in describing the associated suppressed fluctuations in the charge sector, which are heavily affected by the emergence of the local magnetic moment. 
Hence, at strong-coupling, the truncated fRG, mfRG or PA resummations of diagrams describe the formation of a local moment without the intrinsic physical implications onto the charge channel. This can be regarded \cite{Chalupa2021,AdlerSBE} as an insufficient transfer of information between the magnetic and the charge sector, formally corresponding to the impossibility of generating the irreducible vertex divergences in these approximate methods.

On a more general perspective, we note that the loop convergence of the mfRG procedure is mostly controlled by the ratio between the local interaction $U$ and other relevant energy scales of the system under consideration (e.g., in the case of the AIM: $\pi \Delta$ or the temperature $T$) rather than by the ratio between the temperature and the Kondo temperature \cite{Chalupa2021}. In future dedicated studies, it may be interesting to verify to what extent the grade of the loop convergence itself might be regarded as an additional independent marker of central physical aspects of the underlying exact solution of the problem.

%%%%%%%%%%%%%%%%%%%%%%%%%%%%%%%%%%%%%%%%%%%%%%%%%%%%%%%%%%%%%%%%%%%%%%%%%%%%%%%%%%%%%%%%%%%%%%%%%%%%%%%%%

\section{Pauli principle and Ward identity}
\label{sec:PPandWI}
Both the Pauli principle and the WIs are fundamental features of the many-electron physics.
They are deeply rooted in quantum mechanics and pose important constraints on 
many-body correlation functions.
An exact solution must evidently obey all such constraints.
In approximate treatments, however, their fulfillment is not guaranteed {\sl a priori}.
As mentioned in the Introduction, it is commonly reckoned \cite{Bickersbook2004} that approximate many-body approaches 
either obey sum rules imposed by the Pauli principle \textit{or} satisfy WIs.
Hence, fulfilling both the Pauli principle and the WIs would represent a specific hallmark of the exact solution.
On a more formal level, a pertinent example of such a trade-off in the context of parquet-based approximations can be obtained by exploiting
explicit relations between the self-energy and four-point vertices
\cite{Smith1992,Kugler2018b, Janis2017} in the parquet formalism.

In the following, we utilize our converged numerical results for the AIM to analyze, on a quantitative level, to what extent the Pauli principle and WIs are fulfilled for the important class of approximate many-body approaches ranging from the conventional fRG to the mfRG and PA. 

\subsection{Pauli principle}
\label{subsec:sumrules}

\textit{Sum rule of $\chi^{\sigma\sigma}$: Formal aspects}---%
The Pauli exclusion principle states that two electrons cannot occupy the same quantum state. 
On the operator level, this 
corresponds to the fact that a fermionic occupation-number operator can only have eigenvalues zero and one.
On the diagrammatic level, such a constraint 
affects the many-body correlation functions in several ways, e.g., 
through 
sum rules they must 
obey.

In this context, a relevant correlation function for the physics of the AIM is the equal-spin density-density susceptibility,
\begin{align}
\chi^{\sigma\sigma}(\tau) 
= 
\langle \mathit{T}_\tau \hat{n}_\sigma(\tau) \hat{n}_\sigma \rangle
-
n_\sigma^2
.
\label{eq:chi}
\end{align}
Here, 
$n_\sigma = \langle \hat{n}_\sigma \rangle$,
and $\mathit{T}_\tau$ denotes (imaginary) time ordering 
(for brevity,  we omit here the particle-hole channel label).
This susceptibility is directly affected by the Pauli principle through the operator identity 
$\hat{n}_\sigma^2 = \hat{n}_\sigma$. 
Indeed, an evaluation at $\tau=0$ yields
\begin{align}
\chi^{\sigma\sigma}(\tau=0) 
=
\langle \hat{n}_\sigma^2 \rangle
-
n_\sigma^2
=
n_\sigma (1 - n_\sigma)
,
\end{align}
a value, which is fully determined by the single-particle expectation value $n_\sigma$.
Furthermore, as the equal-time correlator $\chi^{\sigma\sigma}(\tau=0)$ is identical to the sum over all its Fourier components $\chi^{\sigma\sigma}_\omega$, the following sum rule \cite{Vilk1997} must hold:
\begin{align}
\frac{1}{\beta} \sum_\omega \chi^{\sigma\sigma}_\omega
=
\chi^{\sigma\sigma}(\tau=0) 
=
n_\sigma (1 - n_\sigma)
.
\label{eq:chi-sum}
\end{align}
At SU(2) spin symmetry and half filling, the result is $1/4$.

For the purposes of the subsequent discussions, it is useful to elaborate on the quantum-field-theoretical relations
which underlie Eq.~(\ref{eq:chi-sum}).
To this end, we recall that the Pauli principle can be translated from an operator identity
($\{ \hat{d}_\sigma, \hat{d}_{\sigma'} \} = 0$,
$\{ \hat{d}_\sigma, \hat{d}_{\sigma'}^\dag \} = \delta_{\sigma\sigma'}$)
to the crossing symmetry of four-point correlators. 
For illustration, let us briefly use a compact notation where all arguments of an electronic operator are summarized in a single index. 
Then, for $G^{(4)}_{1,2;1',2'} \propto \langle \mathit{T}_\tau d_1 d_2 d^\dag_{1'} d^\dag_{2'} \rangle$, 
the crossing symmetry implies $G^{(4)}_{1,2;1',2'} = -G^{(4)}_{2,1;1',2'} = -G^{(4)}_{1,2;2',1'}$.

\begin{figure}[t!]
\centering
\includegraphics[width=0.483\textwidth]{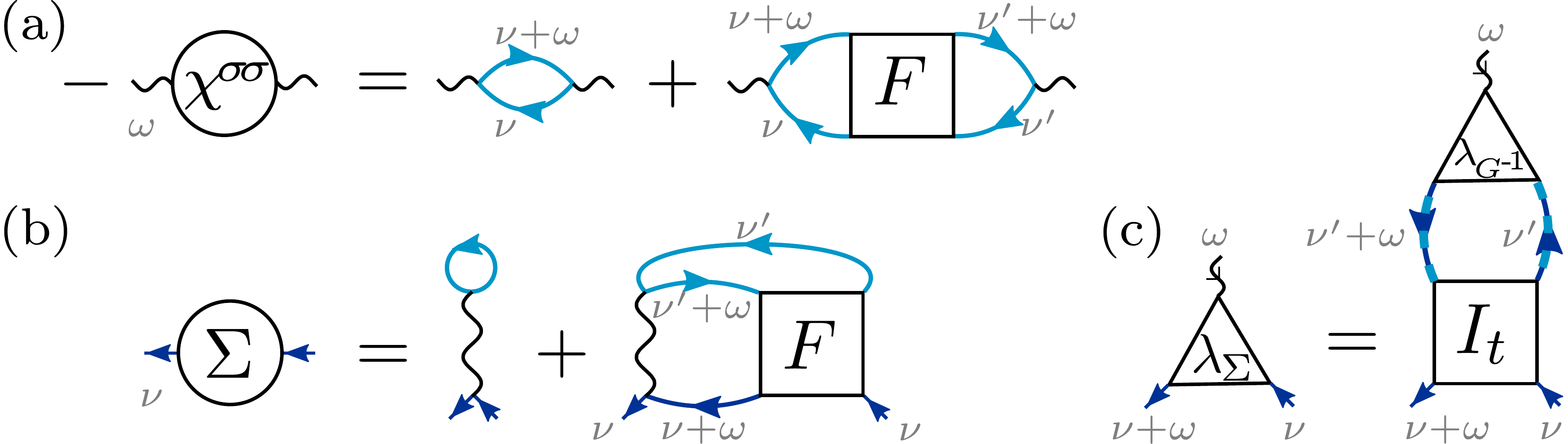}
\caption{Identities between many-body correlation functions.
Dark and light colors on electron propagators distinguish the two spins.
Frequency labels are given for clarity.
(a) The susceptibility $\chi^{\sigma\sigma}$ consists of a bubble term and corrections in terms of the full vertex $F$, see Eq.~\eqref{eq:chi-GG-vtx}.
(b) 
SDE~\eqref{eq:SDE} for the self-energy $\Sigma$,
consisting of the static Hartree part and additions containing 
$F$. 
(c) The WI 
of Eq.~\eqref{eq:WI2} relating $\Sigma$ to $I_t$,
the vertex irreducible in the transverse (vertical) particle-hole channel.
Dashed dark and light colors indicate a sum over spin. Triangles represent objects defined in Eqs.~\eqref{eq:WI} and \eqref{eq:WI2}.} 
\label{fig:chi-vtx_SDE_WI}
\end{figure}

Furthermore, the susceptibility can be represented through (full) propagators $G$ and the (full) two-particle vertex $F$ by 
\begin{align}
\label{eq:chi-GG-vtx}
\chi^{\sigma\sigma}_\omega
& =
-
\frac{1}{\beta} \sum_\nu G^\sigma_{\nu+\omega} G^\sigma_\nu\nonumber
\\
& \quad
-
\frac{1}{\beta^2}
\sum_{\nu\nu'} G^\sigma_{\nu+\omega} G^\sigma_{\nu} G^\sigma_{\nu'+\omega} G^\sigma_{\nu'}
F^{\sigma\sigma}_{\nu,\nu'+\omega;\nu';\nu+\omega}
,
\end{align}
as illustrated in Fig.~\ref{fig:chi-vtx_SDE_WI}(a).
The first term of Eq.~\eqref{eq:chi-GG-vtx} summed over all frequencies $\omega$, 
i.e., taken at $\tau=0$, gives
\begin{align}
\chi^{\sigma\sigma}_{GG}(\tau=0) 
& =
- 
G^\sigma(\tau=0^-)
G^\sigma(\tau=0^+)
.
\label{eq:chi-GGsum1}
\end{align}
Upon inserting
$G^\sigma(\tau) = - \langle \mathit{T}_\tau \hat{d}(\tau) \hat{d}^\dag \rangle$,
one finds
\begin{align}
\chi^{\sigma\sigma}_{GG}(\tau=0) 
& =
\langle d_\sigma^\dag d_\sigma \rangle
\langle d_\sigma d_\sigma^\dag \rangle
=
n_\sigma (1 - n_\sigma)
,
\label{eq:chi-GGsum2}
\end{align}
which yields already  the \textit{entire} sum rule [Eq.~\eqref{eq:chi-sum}].
Consequently, the vertex contributions must vanish when summed over all frequencies $\omega$. 
This is indeed guaranteed by the crossing symmetry, as we show below.

Consider the summed vertex contribution of Eq.~\eqref{eq:chi-GG-vtx},
\begin{align}
\frac{1}{\beta}
\sum_\omega \chi^{\sigma\sigma}_{\mathrm{vtx};\omega}
& \!=\!
-
\frac{1}{\beta^3}
\sum_{\omega\nu\nu'} G^\sigma_{\nu+\omega} G^\sigma_{\nu} G^\sigma_{\nu'+\omega} G^\sigma_{\nu'} 
F^{\sigma\sigma}_{\nu,\nu'+\omega;\nu';\nu+\omega}
.
\label{eq:chi-vtx-derivation1}
\end{align}
For $F^{\sigma\sigma}$, the vertex with equal spins on all legs, the crossing symmetry simply gives
$F^{\sigma\sigma}_{\nu_1',\nu_2';\nu_1,\nu_2} =
- F^{\sigma\sigma}_{\nu_1',\nu_2';\nu_2,\nu_1}$.
After inserting this into Eq.~\eqref{eq:chi-vtx-derivation1},
we relabel the summation indices
according to 
$\tilde{\omega} = \nu' - \nu$,
$\tilde{\nu} = \nu+\omega$:
\begin{align}
\frac{1}{\beta}
\sum_\omega \chi^{\sigma\sigma}_{\mathrm{vtx};\omega}
 =& \;
\frac{1}{\beta^3}
\sum_{\omega\nu\nu'} G^\sigma_{\nu+\omega} G^\sigma_{\nu} G^\sigma_{\nu'} G^\sigma_{\nu'+\omega}
F^{\sigma\sigma}_{\nu,\nu'+\omega;\nu+\omega,\nu'}
\nonumber
\\
 =& \;
\frac{1}{\beta^3}
\sum_{\tilde{\omega}\tilde{\nu}\nu} G^\sigma_{\tilde{\nu}} G^\sigma_{\nu} G^\sigma_{\tilde{\omega} + \nu} G^\sigma_{\tilde{\omega}+\tilde{\nu}}
F^{\sigma\sigma}_{\nu,\tilde{\omega}+\tilde{\nu};\tilde{\nu},\tilde{\omega}+\nu}
.
\label{eq:chi-vtx-derivation2}
\end{align}
This reproduces the original expression for the summed vertex correction
$\frac{1}{\beta} \sum_\omega \chi^{\sigma\sigma}_{\mathrm{vtx};\omega}$ [Eq.~\eqref{eq:chi-vtx-derivation1}]
with opposite sign, so that
\begin{flalign}
&
\textstyle
\frac{1}{\beta}
\sum_\omega \chi^{\sigma\sigma}_{\mathrm{vtx};\omega}
=
-
\frac{1}{\beta}
\sum_\omega \chi^{\sigma\sigma}_{\mathrm{vtx};\omega}
\ \Rightarrow \
\frac{1}{\beta}
\sum_\omega \chi^{\sigma\sigma}_{\mathrm{vtx};\omega} = 0
.
\hspace{-1cm}
&
\label{eq:chi-vtx-derivation3}
\end{flalign}

\begin{figure}[t!]
\centering
\includegraphics[width=0.483\textwidth]{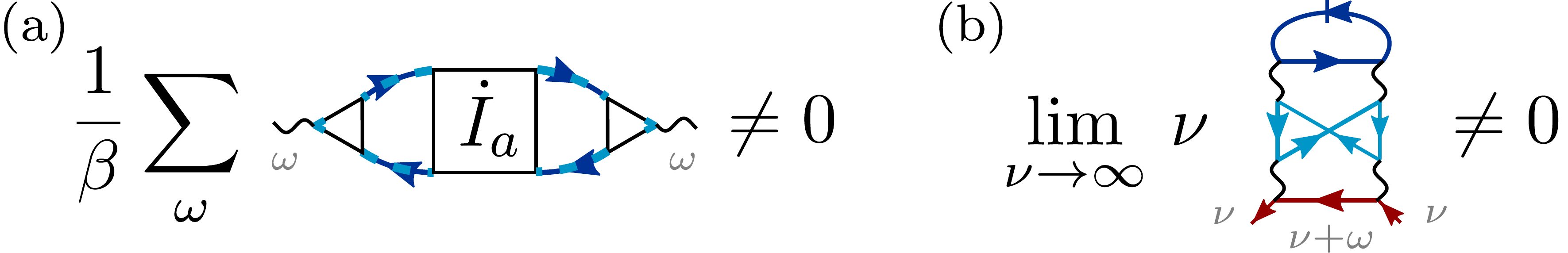}
\caption{(a) The multiloop corrections to the flow of $\chi^{\sigma\sigma}$ do not vanish when summed over $\omega$,
since $I_a$ itself is not crossing symmetric.
(b) Totally irreducible `envelope' vertex diagrams inserted into the standard self-energy flow
contribute to the $1/\nu$ asymptote of $\Sigma$.
Red colors indicate propagators that carry the large frequency $\nu$.}
\label{fig:nonzero_contr}
\end{figure}

\textit{Sum rule of $\chi^{\sigma\sigma}$: Numerical results}---%
As mentioned in Sec.~\ref{sec:formal}, there are two ways \cite{Metzner2012,Tagliavini2019} of computing susceptibilities in fRG:
(i) one can use Eq.~\eqref{eq:chi-GG-vtx}
to obtain $\chi$ from $\Sigma$ and $F$ in a PP fashion,
or (ii) one can deduce $\chi$ from its own flow equation.
In the former approach the sum rule of $\chi^{\sigma\sigma}$ is fulfilled per construction,
as long as the vertex used in the computation obeys the crossing symmetry, see Eqs.~\eqref{eq:chi-GGsum2} and \eqref{eq:chi-vtx-derivation3}, 
while, in the latter scheme, this property is not guaranteed.

Not surprisingly, strategies (i) and (ii) then yield different results within $1\ell$ fRG
(see Figs.~\ref{fig:mloop-convergence} and \ref{fig:mloop-convergenceU20}), 
suggesting that the susceptibility computed from a $1\ell$ flow does not fulfill the sum rule.
Indeed, one can easily convince oneself that the multiloop vertex corrections to the flow
of $\chi^{\sigma\sigma}$ do not vanish when summing over all frequencies,
cf.\ Fig.~\ref{fig:nonzero_contr}(a).
On the other hand, we already noted that, for a converged  mfRG calculation, both schemes of computing susceptibilities become equivalent \cite{Tagliavini2019,Kugler2018b}.
Therefore, the sum rule of $\chi^{\sigma\sigma}$ will be
consistently fulfilled, no matter the strategy employed.

\begin{figure}[t!]
\centering
{{\resizebox{8.3cm}{!}{\includegraphics {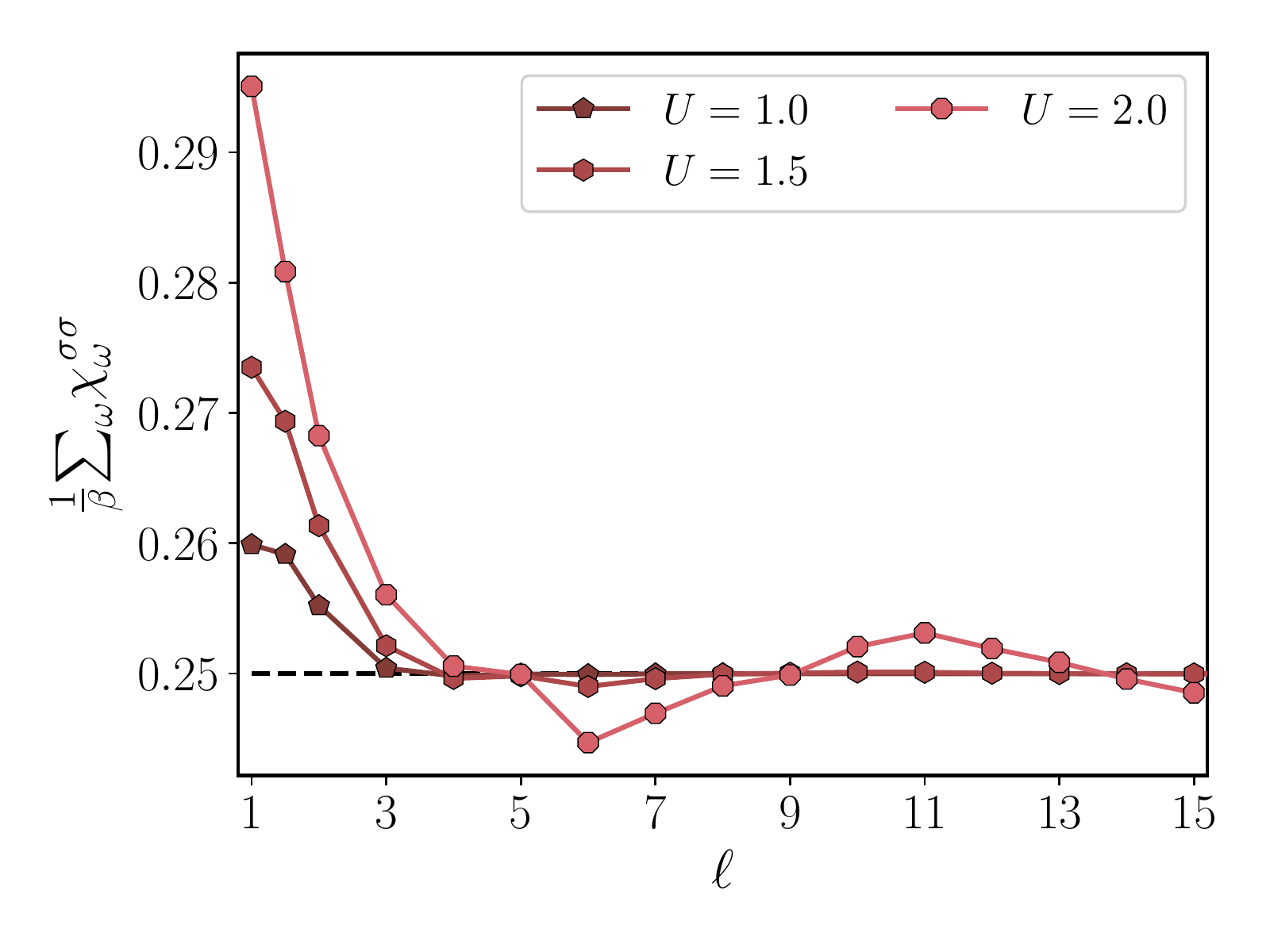}}} 
\caption{
The frequency sum of $\chi^{\sigma\sigma}_\omega$
obtained for different values of $U$ and loop order $\ell$ (in the $\Omega$-flow).
The multiloop corrections systematically improve the 
fulfillment of the sum rule [Eq.~\eqref{eq:chi-sum}]. 
Upon multiloop convergence, 
the sum rule is exactly fulfilled, as in the PA (dashed black line).
}
\label{fig:sum-rule}  
}
\end{figure}

On the basis of these considerations, we now turn to our numerical mfRG data.
In Fig.~\ref{fig:sum-rule}, we show the loop dependence of $\frac{1}{\beta}\sum_\omega\chi^{\sigma\sigma}_\omega$
for the flowing susceptibility (obtained in the $\Omega$-flow) 
for different values of $U$. With increasing loop order, the fulfillment of the sum rule [Eq.~\eqref{eq:chi-sum}], indicated 
by a dashed black line, is approached. Altogether, we observe a similar behavior as in Sec.~\ref{sec:mfRG=PA}:
While, 
at low interaction values, the exact value is quickly reached,
multiloop oscillations characterize the behavior at larger interaction ($U\!=\!2$).
Nevertheless, 
even for large $U$, 
the results at large $\ell$ are much closer to the fulfillment of the sum rule than the ones at low loop order. 
As for the PP susceptibility (not shown), we confirmed numerically that it fulfills the sum rule for all $\ell$, consistent with the above explanations.

\begin{figure*}
\centering
{{\resizebox{16.8cm}{!}{\includegraphics {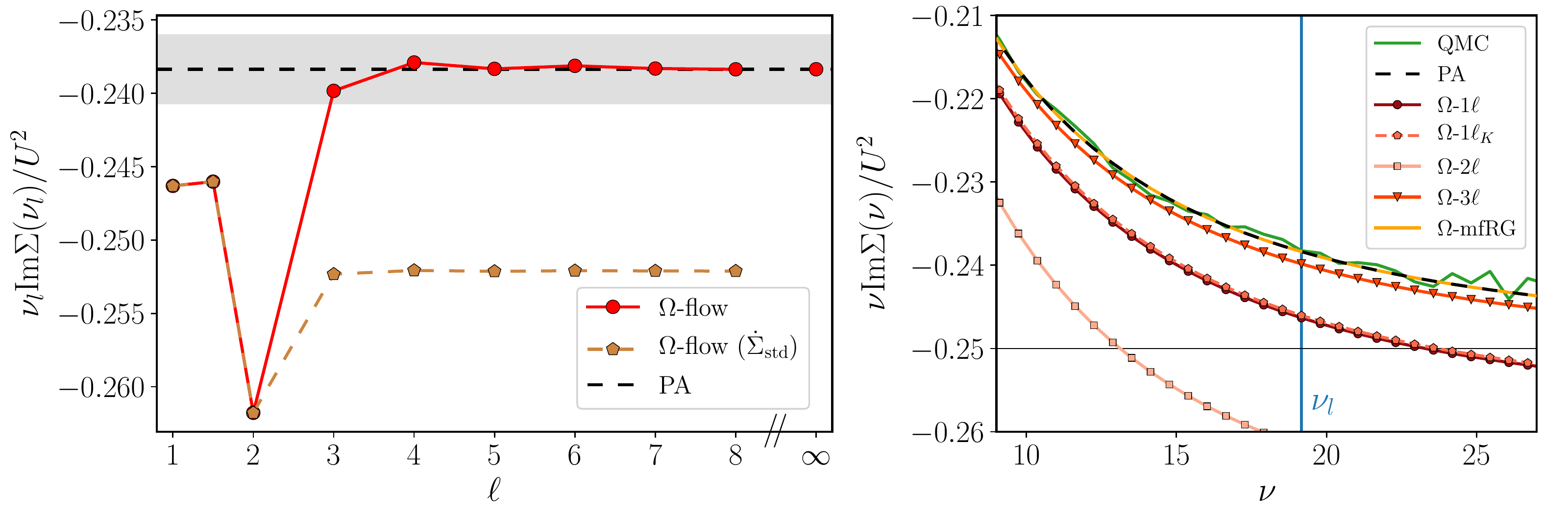}}}
\caption{%
High-frequency asymptote of the self-energy, $\nu_l \mathrm{Im}\Sigma(\nu_l)/U^2$,
for a large but finite frequency $\nu_l \!\approx\! 19.16$ 
($\nu_l \mathrm{Im}\Sigma(\nu_l)/U^2 \to -\frac{1}{4}$ for $\nu_l \to \infty$), for $U\!=\!1$.
Left: Flowing result as a function of loop order, for an $\Omega$-flow with (red circles) and without (gold pentagons) multiloop corrections to the self-energy flow, compared to the PA (black dashed). The gray area represents $1\%$ deviation from PA.  
Right: Frequency dependence around $\nu_l$ for different methods, 
with QMC, PA, and mfRG following the exact asymptote.}
\label{fig:asymptote}
}
\end{figure*}

\textit{High-frequency asymptote of $\Sigma$: Formal aspects}---%
Beside its natural link to the density susceptibility, the Pauli principle also affects the self-energy, albeit more indirectly.
From the moments of the single-particle spectral function,
known through expectation values of operators,
one can determine the high-frequency expansion of the propagator $G$, and thereby of the self-energy $\Sigma$ \cite{Vilk1997}. 
One finds
\begin{align}
\Sigma^\sigma_\nu
=
U n_{\bar{\sigma}} + \frac{U^2 n_{\bar{\sigma}} (1 - n_{\bar{\sigma}})}{i\nu}
+ 
\mathit{O} \Big( \frac{1}{\nu^{2}} \Big)
.
\label{eq:SE-asymp}
\end{align}

Next to the constant Hartree shift $U n_{\bar{\sigma}}$, the $1/\nu$ coefficient coincides with the r.h.s.~of the
sum rule for $\chi^{\bar{\sigma}\bar{\sigma}}$ [Eq.~\eqref{eq:chi-sum}]. Indeed, Eq.~(\ref{eq:SE-asymp}) can be equivalently rewritten \cite{Rohringer2016} as
\begin{align}
\Sigma^\sigma_\nu
=
U n_{\bar{\sigma}} +
\frac{U^2}{i\nu} 
\frac{1}{\beta}
\sum_\omega \chi^{\bar{\sigma}\bar{\sigma}}_\omega
+ 
\mathit{O} \Big( \frac{1}{\nu^{2}} \Big)
.
\label{eq:SE-asymp_chi}
\end{align}

More insight about the quantum-field-theoretical relations underlying the asymptotic behavior of $\Sigma$ can be gained from the SDE,
\begin{align}
\Sigma^\sigma_\nu
=
U n_{\bar{\sigma}}
+ 
\frac{U}{\beta^2} \sum_{\omega\nu'}
G^\sigma_{\nu+\omega}
G^{\bar{\sigma}}_{\nu'+\omega} G^{\bar{\sigma}}_{\nu'}
F^{\sigma\bar{\sigma}}_{\nu'+\omega,\nu;\nu+\omega,\nu'}
,
\label{eq:SDE}
\end{align}
see Fig.~\ref{fig:chi-vtx_SDE_WI}(b).
To this end, let us replace the vertex by its bare contribution,
$F_0^{\sigma\bar{\sigma}} = -U$, and use the first propagator in Eq.~\eqref{eq:SDE},
$G^\sigma_{\nu+\omega}$, to factor out the
dominant contribution for large $\nu \gg \omega$, 
$G^\sigma_{\nu+\omega} \sim 1/(i\nu)$.
The remainder is a $GG$ bubble summed over both frequencies
$\omega$ and $\nu'$.
Hence, we find that the second-order contribution,
\begin{align}
\Sigma^\sigma_\nu
\overset{2^{\mathrm{nd}}}{\sim}
\frac{-U^2}{i\nu} 
G^{\bar{\sigma}}(\tau \!=\! 0^-)
G^{\bar{\sigma}}(\tau \!=\! 0^+)
\!=\!
\frac{U^2 n_{\bar{\sigma}} (1 \!-\! n_{\bar{\sigma}})}{i\nu}
,
\label{eq:SE_asymptote_2ndOrder}
\end{align}
already provides the correct asymptotic behavior~\eqref{eq:SE-asymp}.
This is similar to the sum rule of $\chi^{\sigma\sigma}$,
where Eqs.~\eqref{eq:chi-GGsum1}--\eqref{eq:chi-GGsum2} give the entire result,
while the summed vertex corrections vanish [Eq.~\eqref{eq:chi-vtx-derivation3}].
Via Eq.~\eqref{eq:SE-asymp_chi}, the same cancellation of vertex corrections occurs for the self-energy asymptote, as we explicitly show in Appendix~\ref{sec:chi-sum_Sigma-asymptote}.

Within an fRG treatment, the standard flow equation for the self-energy 
$\dot{\Sigma}_{\mathrm{std}}$
in terms of the vertex $F$
is \textit{in principle} 
exact, as long as the exact vertex $F$ is available.
As this is almost never the case, the flow
$\dot{\Sigma}_{\mathrm{std}}$ must be considered approximate.
In mfRG, the multiloop corrections to the self-energy flow 
[cf.\ Eq.~\eqref{eq:multiloop_self-energy_flow}]
effectively generate
contributions to $\dot{\Sigma}_{\mathrm{std}}$ which \textit{would} require---%
when using the term $\dot{\Sigma}_{\mathrm{std}}$ only---%
vertex diagrams beyond the PA (and thus beyond $1\ell$ fRG).
Indeed, one can generally show that vertex diagrams beyond the PA 
(and thus beyond $1\ell$ fRG),
such as the envelope diagram, do contribute to $\dot{\Sigma}_{\mathrm{std}}$
to order $1/\nu$ in the large-frequency limit [cf.\ Fig.~\ref{fig:nonzero_contr}(b)].
Therefore, the $\Sigma$ asymptote [Eq.~\eqref{eq:SE-asymp}]
is violated when using a $1\ell$ or multiloop vertex flow
while keeping the standard self-energy flow.
This problem is circumvented by including the multiloop corrections to the self-energy flow \cite{Kugler2018a}, which guarantee a perfect equivalence to the SDE and, thereby, that the correct asymptote will be restored.

\textit{High-frequency asymptote of $\Sigma$: Numerical results}---% 
In Fig.~\ref{fig:asymptote}, we show (flowing) results for the asymptotic behavior of $\Sigma$ as obtained from $\Omega$-flow calculations
for $U\!=\!1$. 
The left panel displays $\nu \mathrm{Im} \Sigma_\nu/U$ as a function of $\ell$ for a fixed, large value of $\nu_l \!\approx\! 19.16$.
At this frequency, $\nu \mathrm{Im} \Sigma_\nu$ is expected to be slightly lower (in absolute value) than the corresponding asymptotic value of $- 1/4$ for $\nu \!\to\! \infty$.
The correct asymptotic description of the  mfRG results (red circles) for large $\ell$ is demonstrated by their perfect match with the corresponding PA results, as the latter yield the correct high-frequency asymptotic by construction.
As explained above, this would have not been the case without multiloop corrections to the self-energy flow.
In fact, the gold pentagon line shows results which are obtained by
$\dot{\Sigma}_{\mathrm{std}}$ without multiloop additions to the self-energy flow (these start at $\ell \!=\! 3$) and notably deviate from the correct value.

The right panel shows  the frequency dependence of $\nu \mathrm{Im} \Sigma_\nu$  
in a frequency window around $\nu_l$ ($\nu_l$ is represented by the vertical blue line). 
For fRG results at lower loop order, the high-frequency asymptote is incorrect, reflecting the fact that the SDE relation is not fulfilled.  
For the same reason, all approaches satisfying the SDE lie on top of each other, i.e., the PA (black dashed line), mfRG (orange solid line), and QMC (green line) \footnote{The QMC result was obtained using w2dynamics \cite{w2dynamics} with Worm sampling \cite{Gunacker15,Gunacker2016} and symmetric improved estimators \cite{Kaufmann2019}, designed to reduce the high-frequency noise, see further Appendix \ref{sec:APP-tech-QMC}. However, the noise cannot be suppressed completely, and thus the QMC result fluctuates around the PA and mfRG solution.} yield the correct high-frequency behavior of $\Sigma$. 
While the improvement of the high-frequency results is 
not monotonous for the lowest loop orders, we observe that rather accurate results are obtained already at the $3\ell$ level, where the first 
multiloop corrections to the self-energy flow appear. 
In this respect, it is also interesting to note that the standard self-energy flow 
$\dot{\Sigma}_{\mathrm{std}}$
provides a large-frequency asymptote in agreement with Eq.~\eqref{eq:SE-asymp_chi},
but with $\chi^{\sigma\sigma}$ obtained by a one-loop flow and thus violating the sum rule [Eq.~\eqref{eq:chi-sum}].
We derive this result in Appendix~\ref{sec:chi-sum_Sigma-asymptote} and show explicitly which multiloop additions to $\dot{\Sigma}_{\mathrm{std}}$ contribute to the asymptote.

\begin{figure*}
\centering
{{\resizebox{16.8cm}{!}{\includegraphics {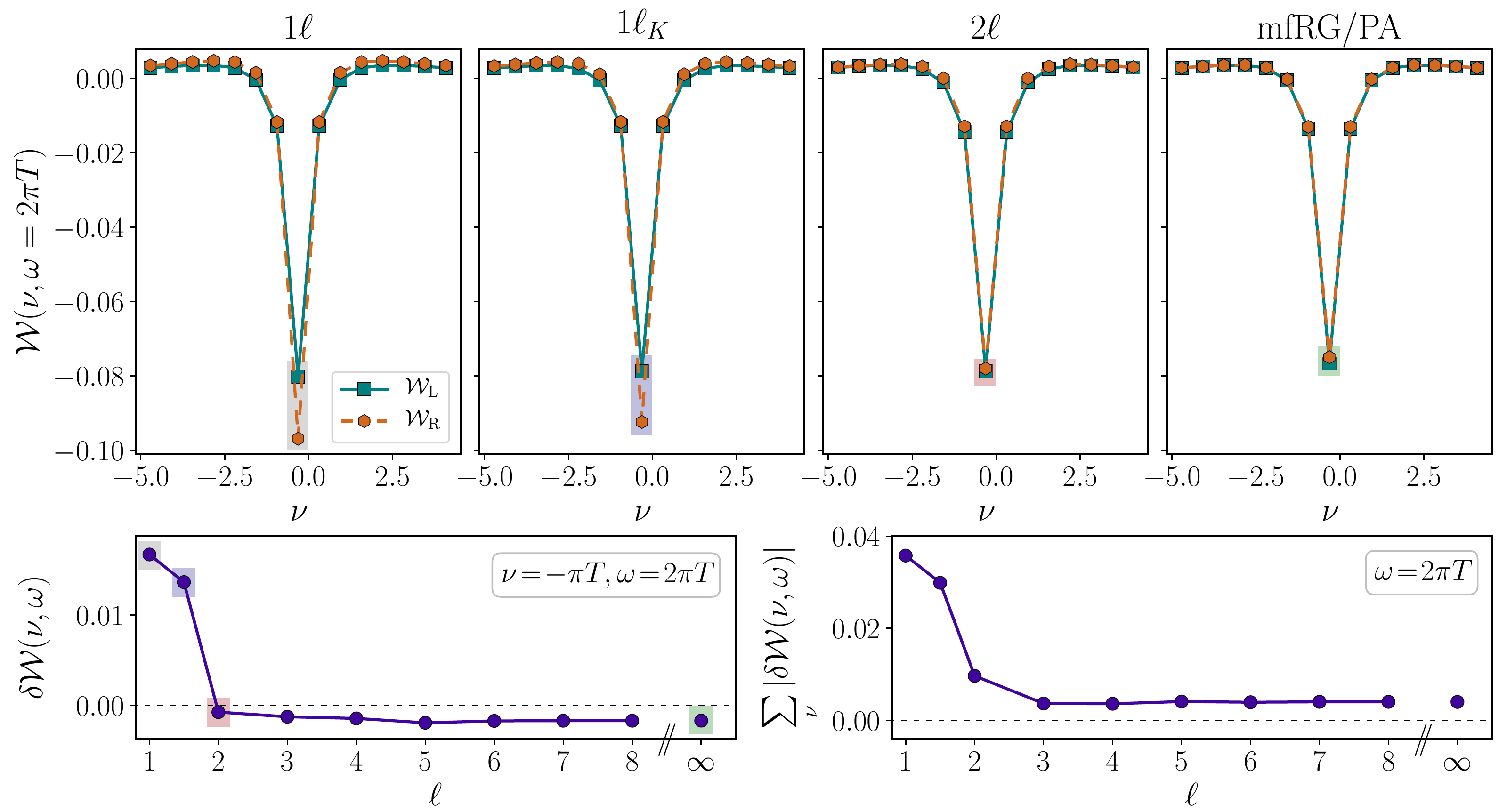}}}  
\caption{Top: Comparison of the left side  $\mathcal{W}_{\mathrm{L}}$ (teal, squares) and the right side $\mathcal{W}_{\mathrm{R}}$ (brown, hexagons) of the WI~\eqref{eq:WI} for $\Omega$-flow mfRG calculations, as a function of $\nu$ for $\omega=2\pi T$ and $U=1$ ($\beta=10$ throughout).
Bottom: Fulfillment of the WI estimated by $\delta\mathcal{W}=\mathcal{W}_{\mathrm{L}}-\mathcal{W}_{\mathrm{R}}$ as a function of loop order, for $\nu=-\pi T$ (left) and for $\nu$ summed over a finite box (right, see text).
Colored areas in the upper and lower left panel mark equivalent data points.
}
\label{fig:Ward-1}  
}
\end{figure*} 

\subsection{Ward identities}

\textit{Formal aspects}---%
The WIs play an essential role in the many-electron theory 
as they define how the information encoded in the continuity equations at a microscopical level is reflected onto response functions and macroscopic quantities.
More specifically, a continuity equation is an operator relation of the form
$\partial_\tau \hat{\rho} = - [ \hat{\rho}, \hat{H} ]$.
If $\hat{\rho}$ is a symmetry of the Hamiltonian, 
$[ \hat{\rho}, \hat{H} ] = 0$,
then $\hat{\rho}$ is a conserved quantity, 
$\partial_\tau \hat{\rho} = 0$.
In this case, the continuity equation describes a conservation law.
However, even if this is not the case,
continuity relations can be used 
for deriving relevant WIs,
in particular when $[ \hat{\rho}, \hat{H} ]$---albeit nonzero---yields a simple expression.

In practice, WIs can be derived for $n$-point correlation functions of arbitrary $n$.
If $\hat{\rho}$ and $[ \hat{\rho}, \hat{H} ]$
involve $n_1$ and $n_2 = n_1 + \delta n$ fermionic operators, respectively,
then
\begin{align}
\langle \mathit{T}_\tau \hat{c}_1 \cdots \hat{c}^\dag_{n-n_1}
\partial_\tau \hat{\rho}
\rangle
= 
-
\langle \mathit{T}_\tau \hat{c}_1 \cdots \hat{c}^\dag_{n-n_1}
[ \hat{\rho}, \hat{H} ] 
\rangle
\end{align}
relates an $n$ to an $(n+\delta n)$-point function.
Typically, one mostly considers the WI connecting
two- and four-point functions (i.e., the WIs ensuring the physical consistency between the one- and the two-particle description)
and restricts oneself to the (local or global)
charge or spin operators, substituting them for $\hat{\rho}$.
A recent derivation, applicable to lattice and impurity systems,
as well as references to prior work
can be found in Refs.~\cite{Krien2017,KrienThesis}.
Here, we consider explicitly the (local) charge, 
$\hat{\rho} = \sum_\sigma \hat{n}_\sigma$,
as done in several preceding works \cite{Katanin2004,Hafermann2014a}.
The resulting WI for the AIM, formulated 
in a way that allows for an optional SU(2) spin symmetry breaking (e.g.\ by a Zeeman field), 
reads
\begin{flalign}
\underbracket[0.5pt]{\vphantom{\sum_{\sigma'}}
\Sigma^\sigma_{\nu+\omega}
-
\Sigma^\sigma_\nu}_{\mathcal{W}_{\rm L}}
& =
\underbracket[0.5pt]{
-
\frac{1}{\beta}
\sum_{\sigma'\nu'}
I^{\sigma\sigma'}_{t;\nu+\omega,\nu';\nu,\nu'+\omega}
( G^{\sigma'}_{\nu'+\omega} - G^{\sigma'}_{\nu'} )
}_{\mathcal{W}_{\rm R}}
.
\hspace{-1cm}
&
\label{eq:WI}
\end{flalign}
We introduce the short-hand $\mathcal{W_{\rm L}(\nu,\omega)} $ for the left and $\mathcal{W_{\rm R}(\nu,\omega)} $ for the right side of the above equation, 
which is illustrated diagrammatically in Fig.~\ref{fig:chi-vtx_SDE_WI}(c).
There, we use
$\lambda^\sigma_{\Sigma;\omega,\nu} \!=\! 
\Sigma^\sigma_{\nu+\omega} \!-\! \Sigma^\sigma_\nu$
and
$\lambda^{\sigma'}_{G^{-1};\omega,\nu'} \!=\! 
(G^{\sigma'}_{\nu'+\omega})^{-1} \!-\! (G^{\sigma'}_{\nu'})^{-1}$,
such that Eq.~\eqref{eq:WI} becomes
\begin{flalign}
\lambda^\sigma_{\Sigma;\omega,\nu}
& \!=\!
\frac{1}{\beta}
\sum_{\sigma'\nu'}
I^{\sigma\sigma'}_{t;\nu+\omega,\nu';\nu,\nu'+\omega}
G^{\sigma'}_{\nu'+\omega} G^{\sigma'}_{\nu'}
\lambda^{\sigma'}_{G^{-1};\omega,\nu'}
.
\hspace{-1cm}
&
\label{eq:WI2}
\end{flalign}
For our numerical results we exploit the SU(2) spin symmetry,
which---together with the crossing symmetry---entails
\begin{align}
I^{\uparrow\uparrow}_{t;\nu'_1,\nu'_2;\nu_1,\nu_2}
& =
I^{\uparrow\downarrow}_{t;\nu'_1,\nu'_2;\nu_1,\nu_2}
-
I^{\uparrow\downarrow}_{a;\nu'_1,\nu'_2;\nu_2,\nu_1}
.
\end{align}

Eventually, we briefly recall that one often refers to \textit{functional} WIs, such as  
$\frac{\delta \Sigma}{\delta G} = - I_t$. These are a cornerstone of $\Phi$-derivable approaches \cite{Baym1962},
where $\frac{\delta \Phi}{\delta G} = \Sigma$,
and $\frac{\delta^2 \Phi}{\delta G^2} = -I_t$.
Since the functional derivative cannot be evaluated numerically,
it
mostly serves as a formal tool.
However, by choosing a specific variation $\delta G$ in the 
functional WI, one can derive more practical relations 
(as necessary but not sufficient conditions of the functional WIs).
For instance, one can easily deduce Eq.~\eqref{eq:WI} in the limit
$\omega \to 0$
by varying $G$ w.r.t.\ frequency
(see Ref.~\cite{Kopietz2010a} for a related treatment).
Moreover, one can derive the standard fRG self-energy $\dot{\Sigma}_{\mathrm{std}}$
by varying $G$ through the scale parameter \cite{Kugler2018b}.

\begin{figure}[tb!]
\centering
{{\resizebox{8.3cm}{!}{\includegraphics {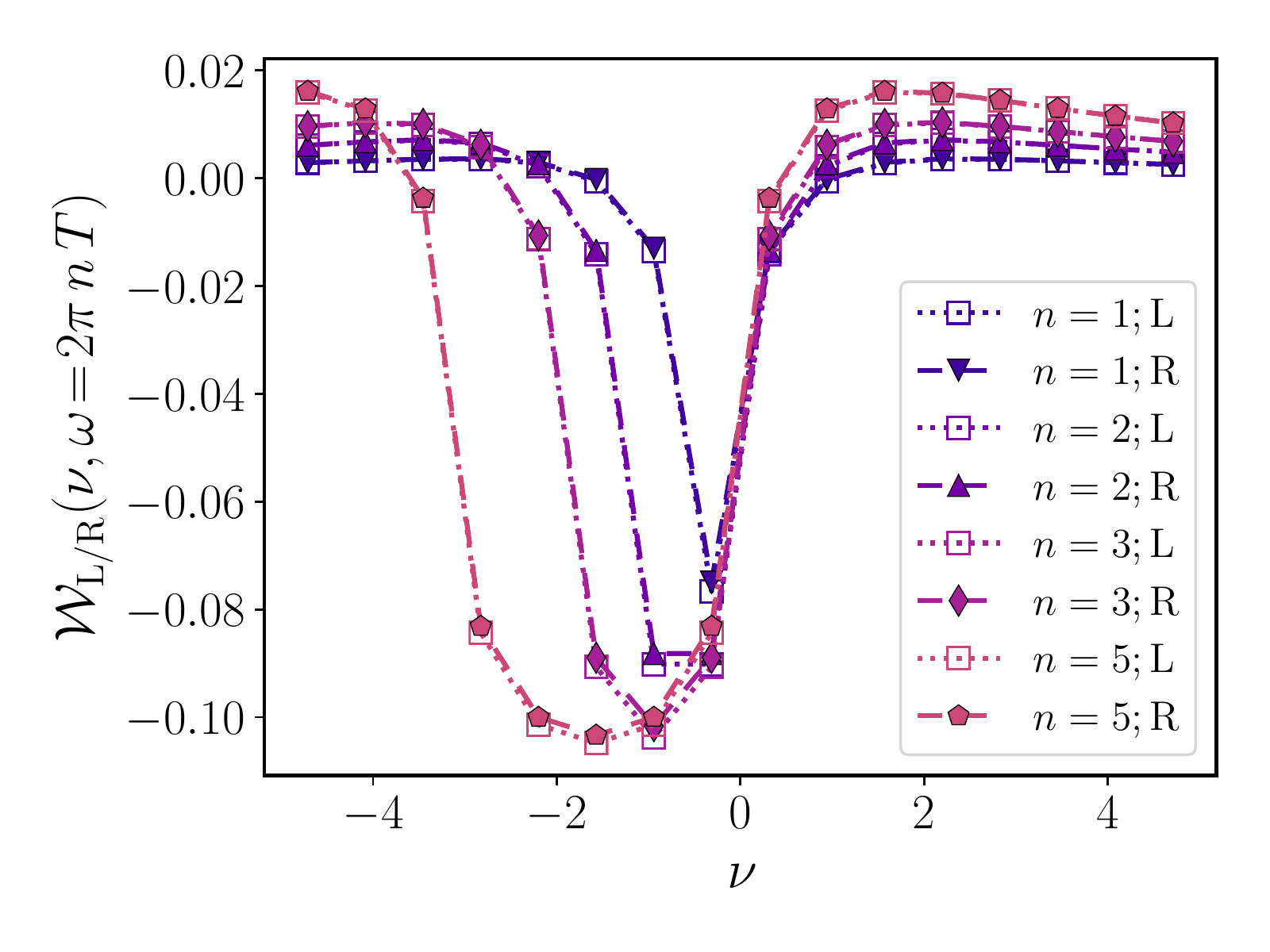}}} 
\caption{$\mathcal{W}_{\mathrm{L}}$ (dotted lines, empty squares) and $\mathcal{W}_{\mathrm{R}}$ (dashed lines, filled symbols) 
of the loop-converged mfRG solution, as a function of $\nu$, at $U\!=\!1$ and for different values of $\omega\! = \!2\pi n T$. 
The absolute discrepancy is largest for $\nu$ around $-\omega/2$.
}
\label{fig:Ward-mfRG-W}  
}
\end{figure} 

\textit{Numerical results}---%
Since the Pauli principle is preserved in the PA as well as (loop-converged) mfRG, one expects---on general grounds---these approximate schemes to violate the WIs to a certain extent. 
Arguably, the size of such violation should increase for increasing interaction strength, driven by the leading terms of the exact solution (where all fundamental relations are fulfilled) which are neglected in either 
approximate approach.
Furthermore, it is known \cite{Metzner2012,EnssThesis} that the $1\ell$ truncation leads to violations of the WIs. 
Katanin \cite{Katanin2004} proposed schemes to mitigate this deficiency. 
In particular, the $1\ell_K$ flow is widely used and often argued to better fulfill WIs. However, no explicit numerical studies were presented thus far. 
Here, we intend to fill this gap and investigate \textit{quantitatively} the fulfillment of WIs in fRG using our numerical results for the AIM. 
We focus on flowing (m)fRG results obtained with the $\Omega$-flow, in order to avoid the frequency extrapolation required for the $U$-flow (see Appendix~\ref{subsec:APP-tech-frequencies}). 

\begin{figure*}
\centering
{{\resizebox{17.8cm}{!}{\includegraphics {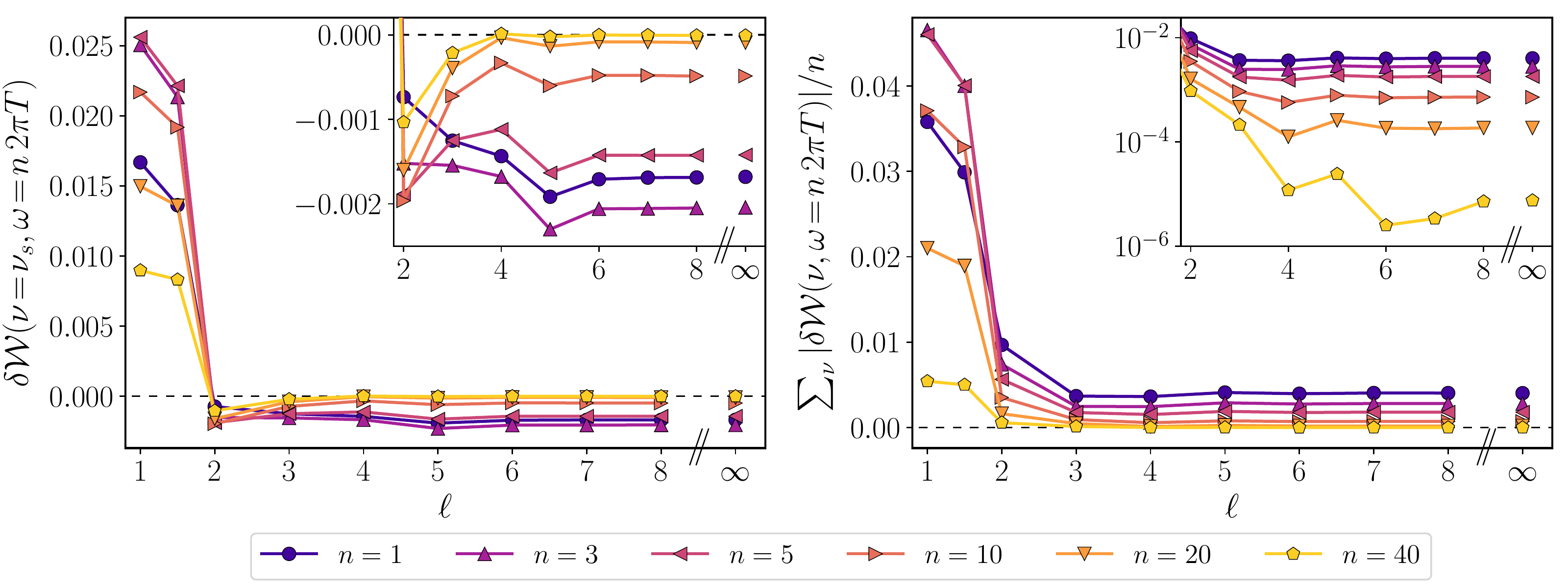}}} 
\caption{
$\delta\mathcal{W}$ in mfRG as a function of loop order at $U=1$ for different values of $\omega = 2\pi n T$ and different choices for $\nu$.
Left: $\nu = \nu_s = -\lceil n/2 \rceil 2\pi T + \pi T$, which gives the fermionic frequency closest to the symmetry axis $\nu = -\omega/2$ where the largest absolute deviation is found.
Right: $\nu$ is summed over a finite box (see text).
Larger values of $\omega$ are shown in yellow, smaller ones in violet. 
The insets show a zoom starting at $\ell=2$, using a linear (logarithmic) scale for the left (right) panel.
}
\label{fig:Ward-loop}  
}
\end{figure*}  

We start with Fig.~\ref{fig:Ward-1}, where the top row shows $\mathcal{W_{\rm L}(\nu,\omega)}$ (squares) and $\mathcal{W_{\rm R}(\nu,\omega)}$ (hexagons) for $\omega=2\pi T$ as a function of $\nu$ for $U=1$, as obtained from the flow.
We find that the $1\ell$ result exhibits the \textit{strongest deviation} in the WI for all $\nu$;  $1\ell_K$ yields already a visible 
improvement at the lowest Matsubara frequency.
However, the $2\ell$ and mfRG/PA results show an overall much more accurate description of the WI for all frequencies. In particular, we note that while, at the lowest Matsubara frequency, the deviation in $2\ell$ is smaller than in mfRG/PA, 
the trend 
is reversed for larger frequencies.

To better quantify the deviations between both sides of the WI, 
we focus on the quantity $\delta\mathcal{W}(\nu,\omega) = \mathcal{W}_{\mathrm{L}}(\nu,\omega)-\mathcal{W}_{\mathrm{R}}(\nu,\omega)$
at $\omega = 2\pi n T$ ($n \in \mathbb{N}$) for two different choices for $\nu$:
In the first case, we fix $\nu$ to $\nu_s\!=\!-\lceil n/2 \rceil 2\pi T + \pi T$, which gives the fermionic frequency closest to the symmetry axis $\nu = -\omega/2$, where the largest absolute deviations are found (e.g.\ $\nu_s\!=\!-\pi T$ for $\omega\!=\!2\pi T$ in Fig.~\ref{fig:Ward-1}, see also Fig.~\ref{fig:Ward-mfRG-W} discussed below). 
In the second case,
we sum $|\delta\mathcal{W}|$ for $\nu$ in a finite frequency box.
Specifically, we sum over $11$ frequencies to the left and $11$ frequencies to the right of the symmetry axis, adding also the contribution right at $\nu = -\omega/2$ if $n$ is odd. 
In this way, we incorporate the behavior at larger frequencies, while avoiding numerical inaccuracies from
the finite-frequency box effect of the high-frequency parametrization in our implementation \cite{Wentzell2020} (see Appendix~\ref{subsec:APP-tech-frequencies}).
When comparing results for different transfer frequency $\omega = 2\pi n T$, we divide by $n$ to obtain more comparable results.
The bottom row of Fig.~\ref{fig:Ward-1} shows 
$\delta\mathcal{W}$ for the $n=1$ data reported at the top.
The plot confirms that, at weak-coupling, already the first multiloop corrections strongly improve the fulfillment of the WI.
In particular, the 
minimal value for $\delta\mathcal{W}$ at $\nu=-\pi T$ (left panel) is found at $\ell=2$ 
and for $|\delta\mathcal{W}|$ summed over $\nu$ (right panel) at $\ell=3$.
Hence, our $U\!=\!1$ calculations show that the finite deviation from the exact fulfillment of the WI expected to occur in the loop-converged mfRG/PA results is notably smaller in comparison to $1\ell$ or $1\ell_K$, and that it quantitatively represents 
a marginal effect in the weak-coupling regime.
This trend is also confirmed regarding relative deviations $|\delta_r\mathcal{W}| = |\delta\mathcal{W}/\mathcal{W}_{\rm L}|$, as we explicitly show in Fig.~\ref{fig:APP-Ward-U-W1-rel-left} in Appendix~\ref{sec:APP-add}.

Next, we extend the analysis to larger values of $\omega = 2\pi n T$
and show in Fig.~\ref{fig:Ward-mfRG-W} 
loop-converged mfRG results for $1 \leq n \leq 5$. 
The plot
demonstrates that 
the mfRG data 
provide 
satisfactory agreement between
$\mathcal{W}_{\mathrm{L}}$ (empty squares) and $\mathcal{W}_{\mathrm{R}}$ (filled symbols) for all values of $\omega$ and $\nu$,
and that the largest absolute deviation indeed occurs for $\nu$ around $\nu_s$, i.e., the frequency closest to the symmetry axis $\nu=\omega/2$ (see above). 
Figure~\ref{fig:Ward-loop} presents $\delta\mathcal{W}$ as a function of $\ell$  for $n$ up to $40$.
Again, the fulfillment of the WI is slightly improved when going from $1\ell$ to $1\ell_K$ and strongly improved
starting from $2\ell$, for all values of $\omega$ (confirmed also by Fig.~\ref{fig:APP-Ward-U-W1-rel-left} in Appendix~\ref{sec:APP-add}).
However, the details in the change from $\ell=2$ to $\infty$ depend on $\omega$.
In general, we observe that the WI is better fulfilled for larger values of $\omega$. In fact, a perfect match is given for $\omega \to \infty$ and $\ell \to \infty$, since 
the WI reproduces the SDE for $\omega \to \infty$ (see Appendix~\ref{sec:WI_SDE}), which is exactly fulfilled in mfRG and the PA. 
This can be clearly seen in both insets of Fig.~\ref{fig:Ward-loop}. The inset of the right panel uses a logarithmic scale, where one can also spot the onset of 
oscillations in the multiloop convergence, in spite of their small amplitude.

\begin{figure*}
\centering
{{\resizebox{17.0cm}{!}{\includegraphics {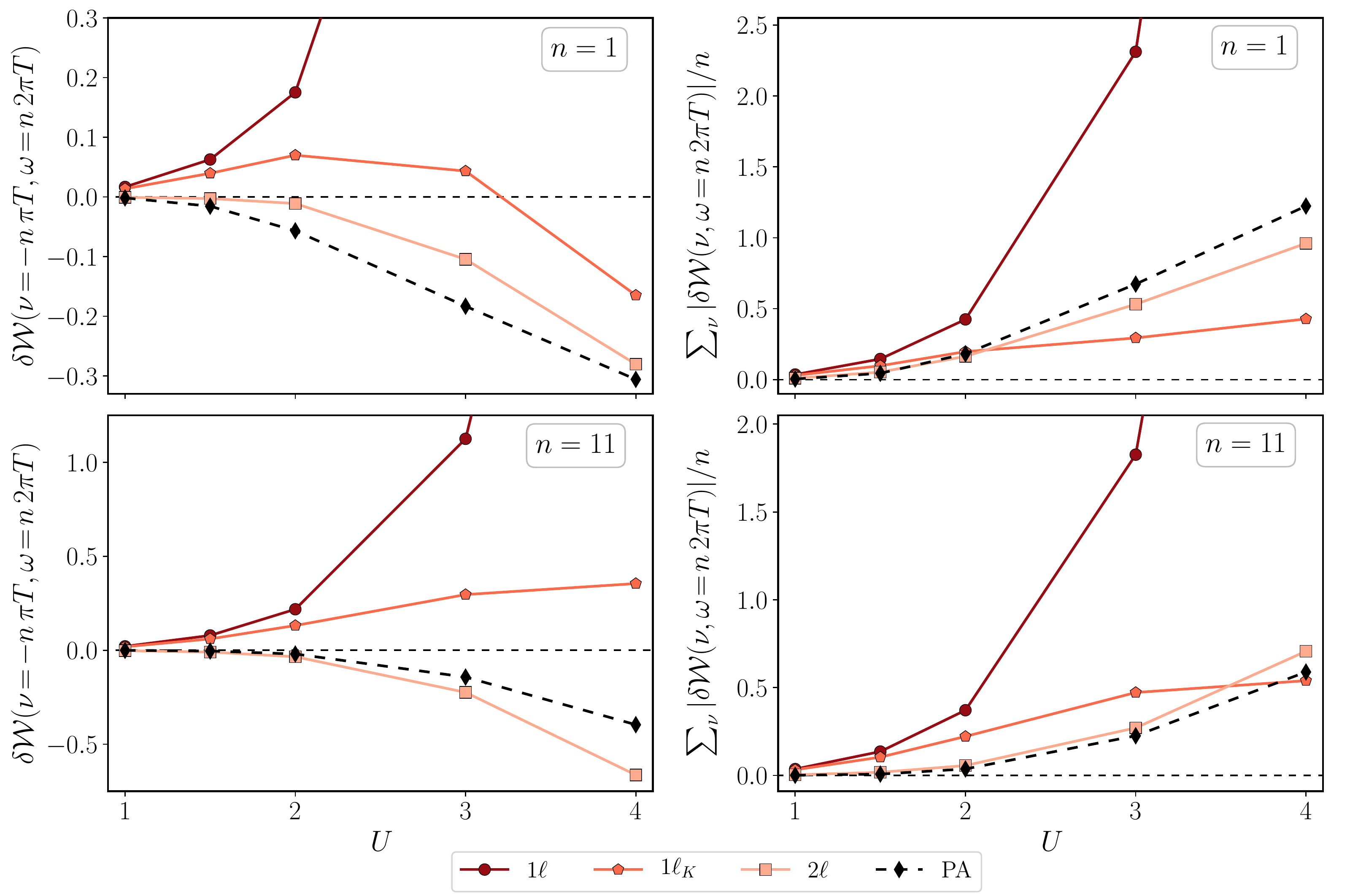}}} 
\caption{%
$\delta\mathcal{W}$ for increasing $U$ obtained with different methods.
Solid lines in shades of red denote (m)fRG schemes ($\Omega$-flow) at low loop order; 
the PA solution is shown in dashed black.
In the top (bottom) panels, $\omega = 2\pi n T$ is fixed at $n=1$ ($n=11$).
In the left panels, we use $\nu = -\pi T$ ($\nu = -11\pi T$). In the right panels $\nu$ is summed over a finite box (see text).}
\label{fig:Ward-U-W1}  
}
\end{figure*}

Finally, we analyze the effects of the interaction strength, by progressively increasing its value up to $U\!=\!4$.
In Fig.~\ref{fig:Ward-U-W1}, we examine
$\delta\mathcal{W}$ for $\omega=2\pi n T$ at $n=1$ and $n=11$,
comparing results of (m)fRG flows at low loop order with the PA.
At large interaction, the pure $1\ell$ flow is 
evidently unreliable,
violating the WI with very large values of $\delta\mathcal{W}$.
The situation visibly improves in $1\ell_K$, $2\ell$, and PA.
In particular, for $U \leq 2$, $1\ell_K$ is farther off than $2\ell$ and PA.
Interestingly, however, the $1\ell_K$ deviations display a highly non-trivial behavior with increasing $U$---they are non-monotonous in the top left panel and have a decreasing slope in the other panels---and thereby yield comparatively small values of $\delta\mathcal{W}$ at larger $U$.
By contrast, for the PA results, $|\delta\mathcal{W}|$ starts rather small but increases monotonously with increasing $U$.
Overall, for intermediate to large values of $U$, it seems that $1\ell_K$ provides the most accurate description of the WI at small frequencies ($n=1$),
while mfRG and the PA lead to a 
smaller violation of the WI for larger frequencies (here $n=11$). 
Further details on the individual deviations of $\mathcal{W}_{\mathrm{L}}$ 
and $\mathcal{W}_{\mathrm{R}}$ are given in Appendix~\ref{sec:APP-add}.

\begin{figure*}
\centering
{{\resizebox{17.0cm}{!}{\includegraphics {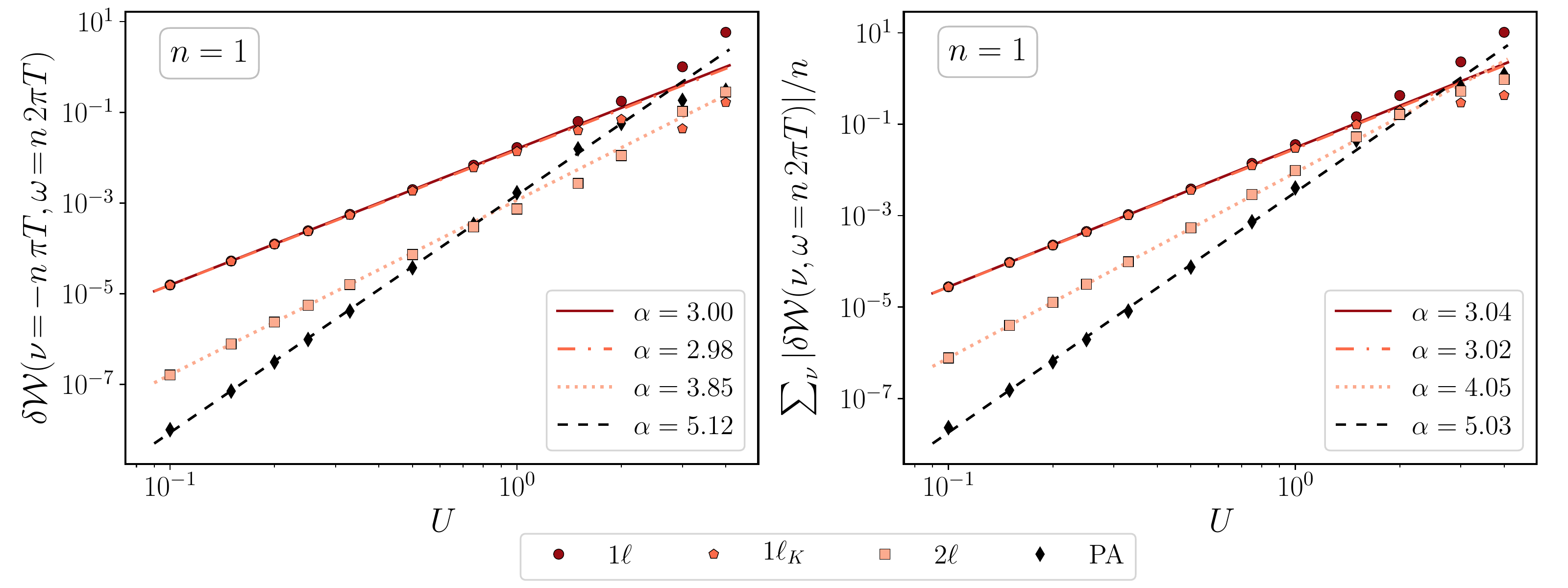}}} 
\caption{%
Same as Fig.~\ref{fig:Ward-U-W1} for $n=1$, but including data for very small interaction values $U<1$, using a log-log scale. The straight lines correspond to $f(x)=\alpha x + d$ fits using the first data points between $U=0.1$ and $U=0.5$, yielding the exponents of the $\sim U^\alpha$ behavior.
}
\label{fig:Ward-LOWU-W}  
}
\end{figure*}

As a last step, we compare the numerical deviations $\delta\mathcal{W}$ as a function of $U$ focusing on small interaction values $U<1$. 
Figure~\ref{fig:Ward-LOWU-W} shows $\delta\mathcal{W}$, similarly as in Fig.~\ref{fig:Ward-U-W1}, but on a log-log scale. Using a $f(x)=\alpha\,x + d$ fit, we extract the exponents of the deviations of the WI, 
$\delta \mathcal{W} \sim U^\alpha$, for the (m)fRG flow and PA scheme. Our analysis shows perfect agreement with the theoretical predictions of Ref.~\cite{Katanin2004}: the $1\ell$ scheme displays deviations that grow with the third power of $U$ ($\alpha \approx 3$, solid lines), and the $2\ell$ results are in agreement with a $U^4$ growth ($\alpha \approx 4$, dotted lines). The $1\ell_K$ results at small $U$ also manifest $\mathit{O}(U^3)$ deviations. 
This is in agreement with the analytic arguments of Ref.~\cite{Katanin2004} since, for the commonly used $1\ell_K$ scheme, only part of the $2\ell$ corrections are included by substituting $S\rightarrow\dot{G}$ (as described in Sec.~\ref{sec:formal}). Hence, some terms violating the WI at $\mathit{O}(U^3)$ remain, as seen in our numerical data in Fig.~\ref{fig:Ward-LOWU-W} ($\alpha \approx 3$, dashed-dotted lines). Note that the behavior at larger interaction values, as discussed above, is beyond the reach of the present analysis applicable at small values of $U$. 

Further, concerning the loop-converged mfRG/PA results, we find deviations of the WI, which behave as $\mathit{O}(U^5)$ (dashed lines). In general, one expects the PA/mfRG schemes to deviate from the exact solution as $\mathit{O}(U^4)$. 
However, at half filling, the combination of the particle-hole symmetry and spin symmetry of our problem causes the contributions to the WI from the forth-order ``envelope'' diagrams to exactly cancel, as we show explicitly in Appendix~\ref{subsec:APP-WI-envelope}. For completeness, we also note that the same behavior as in Fig.~\ref{fig:Ward-LOWU-W} is found for other frequency choices as well (e.g.~for $n=11$ used in the lower panel of Fig.~\ref{fig:Ward-U-W1}).

%%%%%%%%%%%%%%%%%%%%%%%%%%%%%%%%%%%%%%%%%%%%%%%%%%%%%%%%%%%%%%%%%%%%%%%%%%%%%%%%%%%%%%%%%%%%%%%%%%%%%%%%%%%%%%%%%%%%

\section{Conclusion and Outlook}
\label{sec:concl}

We investigated several essential features of the recently introduced mfRG approach
by performing a quantitative study of the particle-hole symmetric AIM for different coupling strengths.
As the numerical implementation of the mfRG applied to the AIM does not require additional algorithmic approximations (such as the form factor expansion used for the Hubbard model \cite{Tagliavini2019,Hille2020}), we were able to demonstrate how the 
precise convergence of the mfRG series to the corresponding PA results is readily obtained in the entire weak- to intermediate-coupling regime. A thorough inspection further confirmed the pivotal features of a converged mfRG solution, i.e., its independence of the specific RG cutoff adopted as well as the equivalence between flowing and post-processed results.
Hence, in the parameter regimes where a fast loop convergence of the mfRG is found, the application of this method 
offers potential advantages w.r.t.~to the full iterative solution of the PA  
through the intrinsic flexibility of the underlying fRG framework.

By increasing the value of the electronic interaction, we studied the oscillatory behavior emerging in the loop dependence of the mfRG series, which eventually hinders the convergence 
to the PA solution in the strong-coupling regime. Interestingly, the parameter region where a multiloop convergence could not be achieved appears roughly to match the one in which previous Quantum Monte Carlo studies \cite{Chalupa2018,Chalupa2021} have shown an explicit  breakdown of perturbative resummations to occur at the two-particle level. In this respect, the strong oscillatory behavior of the non-converging mfRG series could be plausibly regarded as a further hallmark of the nonperturbative \cite{Schaefer2013,Kozik2015,Gunnarsson2017,Chalupa2021} parameter regime, where significant physical differences between the PA and the exact solution of the AIM are found \cite{Chalupa2021}. 

The numerical data obtained in the region of proper convergence of the mfRG algorithm 
were then used for a quantitative investigation of the fulfillment of fundamental features of the many-electron problem, namely those linked to (i) the Pauli principle and (ii) the WIs.
For (i) the Pauli principle, we observed a sizable violation of sum rules in the conventional $1\ell$ fRG results, which gets systematically reduced by increasing the loop order. This is consistent with the fact that mfRG converges to the PA solution,
and that the PA obeys the Pauli principle by construction,
realized 
through the crossing symmetry 
and two-particle self-consistency. 
We also note that the indirect effects of the Pauli principle on the high-frequency asymptotic behavior of one-particle quantities are only recovered by including the multiloop additions to the self-energy flow, which start from the third loop onwards.
For (ii) the WIs, these are generally neither fulfilled in fRG nor in the PA. For weak to intermediate coupling, our results demonstrated that adding higher-loop terms systematically reduces the overall violation of WIs. In particular, while a first improvement can be already observed by including the one-loop Katanin ($1\ell_K$) substitution, higher loop orders and the PA yield quantitatively much smaller deviations. By increasing the interaction, however, the situation becomes more complex. 
Going beyond the $1\ell$ fRG level, 
whose description of the WIs is largely unreliable, 
we find that $1\ell_K$ mitigates most efficiently the WI violations at low frequencies,
while higher-loop mfRG and the PA yield better results for large frequencies. This is consistent with our observation that the WI reproduces the SDE for $\omega\to\infty$.
Additionally, we confirmed the predictions of Ref.~\cite{Katanin2004} for the asymptotic weak-coupling behavior of the WI deviations as a function of $U$ for the $1\ell$ and $2\ell$ scheme. Our numerical results for the mfRG/PA scheme revealed a $\mathit{O}(U^5)$ deviation, smaller than the expected $\mathit{O}(U^4)$, which we showed to be related to the particle-hole and spin symmetry used in our computations.

The insights gained in our study, which might be extended in the future to other regimes (e.g., out of half filling, and/or in the presence of a magnetic field) and more complex systems, are important for several reasons. On the one hand, they improve the understanding of the convergence of the mfRG procedure, whose relevance extends to more complex contexts than the basic AIM considered here. Such insights may be particularly important if the mfRG is used to include {\sl nonlocal correlations} on top of the DMFT solution of strongly correlated lattice problems, thus extending the DMF$^2$RG algorithms beyond the 
conventional ($1\ell$) fRG 
used so far \cite{Vilardi2019,Bonetti2022}. 
In that context, the mfRG might offer important advantages over corresponding parquet-based implementations. In contrast to the  latter, the mfRG flow does \textit{not} rely on the numerical manipulation of two-particle irreducible vertex functions, which display multiple divergences in the intermediate-to-strong coupling regime of different many-electron models \cite{Schaefer2013,Janis2014,Schaefer2016c,Ribic2016,Vucicevic2018, Thunstroem2018,Chalupa2018,Springer2019,Kotliar2020,Chalupa2021}. This should allow the circumvention of several of the problems faced by parquet-based DMFT extensions \cite{Rohringer2018} constructed upon such potentially diverging irreducible vertices, such as parquet D$\Gamma$A \cite{Toschi2007,Valli2015} or QUADRILEX \cite{Ayral2016}.

On the other hand, 
the possible relation of the loop convergence properties in mfRG with the breakdown of the perturbation expansion might have interesting theoretical and algorithmic implications, calling for an extension of our study to more complex physical situations than those considered here.
Together with our precise analysis of the fulfillment or violation of sum rules and WIs, this might shed new light on fundamental aspects of the many-electron theory and help to further develop refined calculation strategies for the most challenging parameter regimes.

\section{Acknowledgments}
The authors thank C.~Eckhardt,  S.~Heinzelmann, A.~Kauch, F.~Krien, S.~Jakobs, V.~Meden, G.~Rohringer, T.~Sch\"afer, A.~Tagliavini, and N.~Wentzell for valuable discussions. We acknowledge financial support from the Deutsche Forschungsgemeinschaft (DFG) through Project No.\ AN 815/6-1 (S.A.)
and through Germany's Excellence Strategy EXC-2111 (Project No.\ 390814868) (J.v.D.), 
as well as from the Austrian Science Fund (FWF) through Project No.\ I 2794-N35 (P.C.\ and A.T.). Calculations were done in part on the Vienna Scientific Cluster (VSC). 
F.B.K.\ acknowledges support by the Alexander von Humboldt Foundation through the Feodor Lynen Fellowship.

\appendix
\section*{APPENDIX}

\begin{figure*}[t]
\centering
{{\resizebox{17.8cm}{!}{\includegraphics {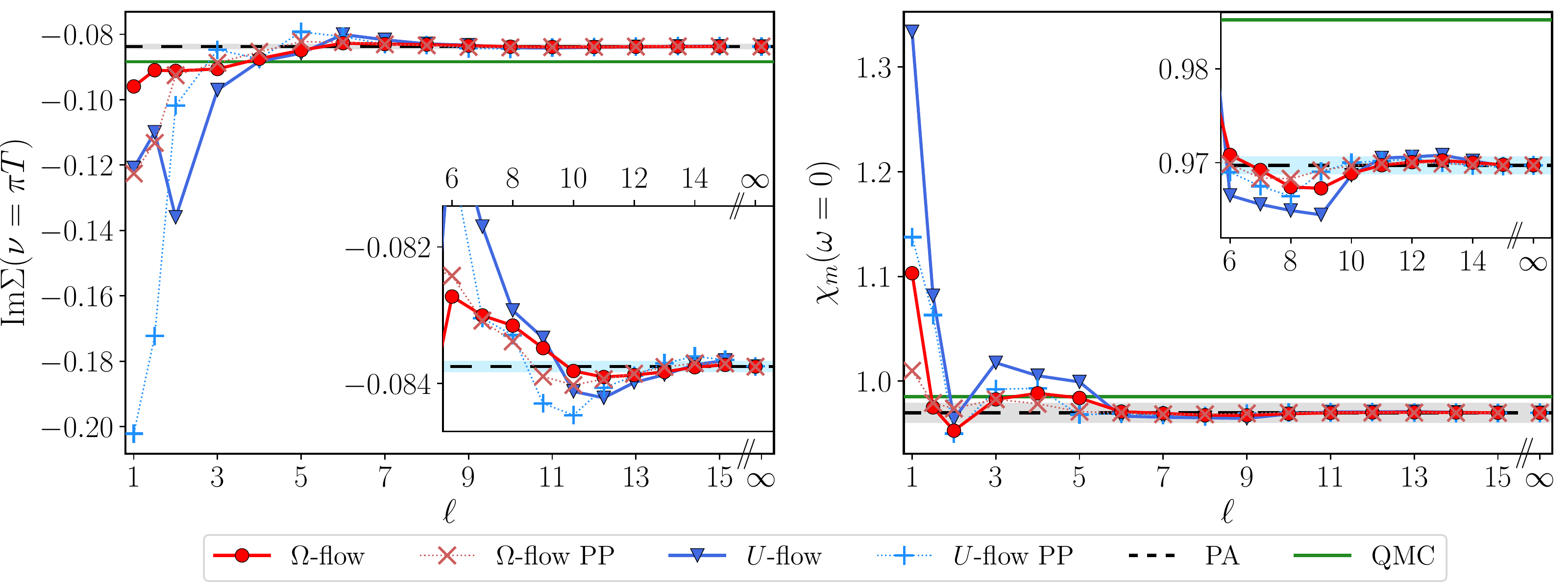}}} 
\caption{
\text{Im}$\Sigma(\nu = \pi T)$ and $\chi_m(\omega = 0)$ as in Fig.\,\ref{fig:mloop-convergence} but for $U=1.5$. Insets show a zoom for $\ell\ge6$.
The gray (blue) area indicates $1\%$ ($0.1\%$) deviation from the PA.
}
\label{fig:mloop-convergenceU15}  
}
\end{figure*} 

\begin{figure*}[t]
\centering
{{\resizebox{17.8cm}{!}{\includegraphics {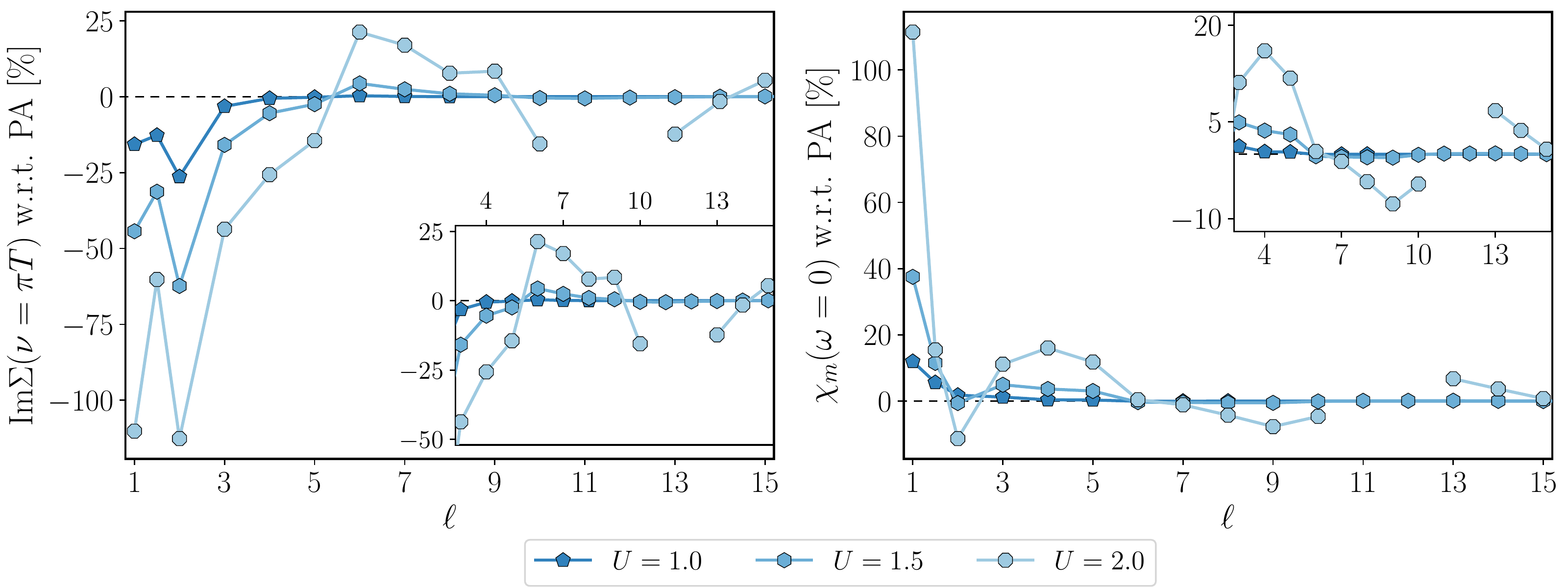}}} 
\caption{
Relative difference between $U$-flow mfRG calculations and the corresponding PA solutions for Im$\Sigma(\nu=\pi T)$ (left) and $\chi_m(\omega=0)$ (right), as a function of loop order $\ell$ and different values of the interaction $U$, as in Fig.~\ref{fig:mloop-Ucomparison}. Insets show a zoom for $\ell\ge3$.
}
\label{fig:APP_UFlow_UCOMP}  
}
\end{figure*} 

In the Appendix, we provide additional results, details on the numerical treatment as well as diagrammatic derivations, in order to specify our approach and further support the messages of the main part. The additional results are in Appendix~\ref{sec:APP-add}, mainly focused on the $U$-flow and the fulfillment of the WI. Details on our numerical approach, especially the dependence of different quantities on the number of Matsubara frequencies included in the computations, are discussed in Appendix~\ref{sec:APP-technicalities}. Finally, we give the diagrammatic derivations of several relations used in Sec.~\ref{sec:PPandWI} in Appendix~\ref{sec:diagr-derivations}.

\section{Additional results}
\label{sec:APP-add}

In Fig.~\ref{fig:mloop-convergenceU15}, we report the results for $U=1.5$ ($\beta=10$, half filling), which were anticipated in Sec.~\ref{subsec:mfrg-params}. 
For this parameter set, too, the mfRG scheme converges perfectly in loop order. 
For $\ell\ge 15$, both regulators lead to identical results for all quantities, and the PP 
(dotted lines with `$\times$' or `$+$' symbols) and flowing data coincide. As stated in the main text, no qualitative difference in the convergence behavior is observed, apart from the fact that, for $U=1.5$, more loop orders are necessary to reach it. 

In Fig.~\ref{fig:APP_UFlow_UCOMP}, the relative comparison between $U$-flow results and the PA for $U\!=\!1,1.5,2$ is shown in the same fashion as in Fig.~\ref{fig:mloop-Ucomparison} for the $\Omega$-flow. 
While there is no qualitative difference, quantitatively the $U$-flow shows larger relative differences with respect to the PA. Note that we were unable to converge the $U$-flow calculation for $\ell=11,12$; see also 
Appendix~\ref{subsec:APP-mfRGcalc-loop}.

\begin{figure*}[t!]
\centering
{{\resizebox{17.0cm}{!}{\includegraphics {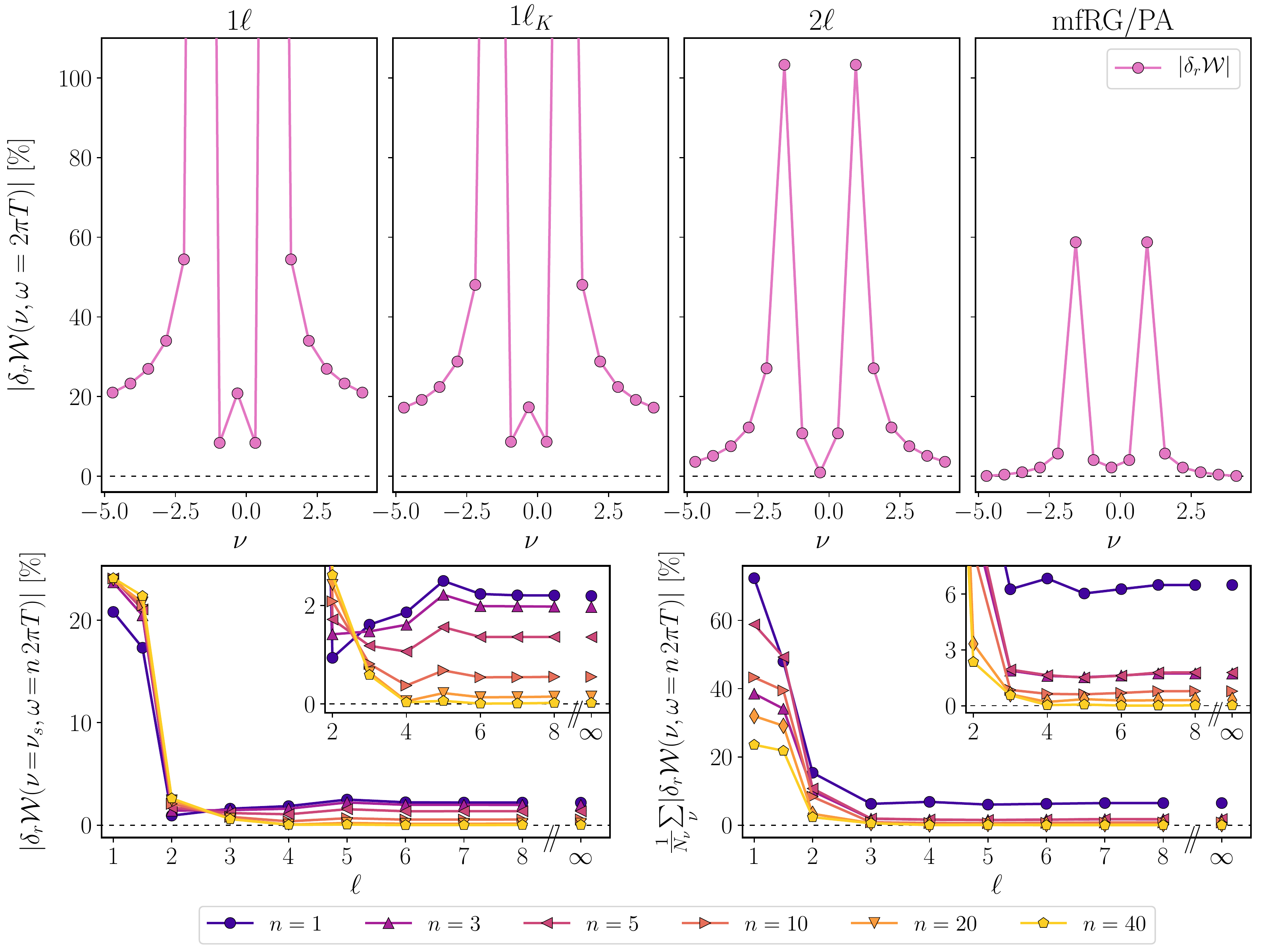}}} 
\caption{Top: Relative deviation of the WI \eqref{eq:WI} $|\delta_r\mathcal{W}(\nu,\omega)|\!=\!|\delta\mathcal{W}(\nu,\omega)/\mathcal{W}_{\rm L}(\nu,\omega)|$ for $\Omega$-flow mfRG calculations at $\omega\!=\!2\pi T$, $U\!=\!1$, similarly as the top row panels of Fig.~\ref{fig:Ward-1}, as a function of $\nu$. The $y$-axis is cut at $100\%$ to provide enough resolution for $|\delta_r\mathcal{W}(\nu,\omega=2\pi T)|$ at the various values of $\nu$.
Bottom: As Fig.~\ref{fig:Ward-loop} but showing $|\delta_r\mathcal{W}|$ instead. In the right panel, the normalizing factor $1/n$ of the main text is replaced by $1/N_{\nu}$, where $N_{\nu}$ is the number of frequencies summed over (see text).
All quantities are given in percent [\%]. 
}
\label{fig:APP-Ward-U-W1-rel-left}  
}
\end{figure*}

\begin{figure}[t!]
\centering
{{\resizebox{8.3cm}{!}{\includegraphics {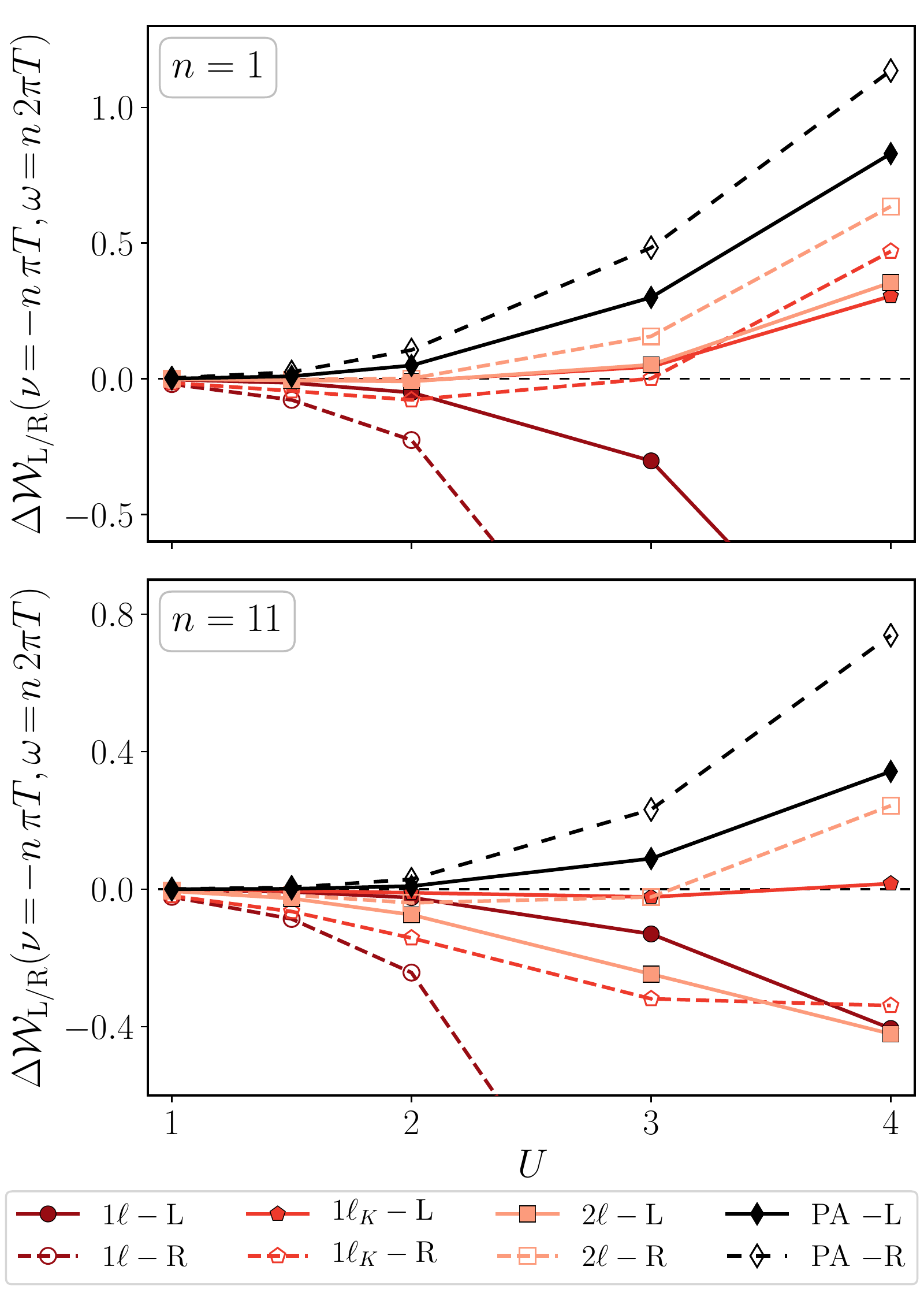}}} 
\caption{$\Delta\mathcal{W}_{\rm L/R}$ as a function of $U$ for different approaches where $\omega=2\pi n T$ is fixed at $n=1$ ($n=11$) in the top (bottom) panel. The solid (dashed) lines represent the 
left (right) side of the WI in comparison with the QMC result, see text.}
\label{fig:APP-Ward-U-W1-QMC}  
}
\end{figure} 

Finally, we add further analyses on the fulfillment of the WI, namely (i) on the relative deviations for the cases discussed in the main text, and (ii) more details on the deviations as a function of $U$ in the different approaches.
Concerning (i), Fig.~\ref{fig:APP-Ward-U-W1-rel-left} is a combined plot of Figs.~\ref{fig:Ward-1} and \ref{fig:Ward-loop} of the main text, but instead of $\delta\mathcal{W}$, we show $|\delta_r\mathcal{W}(\nu,\omega)| = {|\mathcal{W}_{\rm L}(\nu,\omega) - \mathcal{W}_{\rm R} (\nu,\omega) }/{\mathcal{W}_{\rm L}(\nu,\omega)}|$. 
In the top row, $|\delta_r\mathcal{W}|$ is shown for $\omega=2\pi T$, similarly as in the top row of Fig.~\ref{fig:Ward-1}. 
Note that the $y$-axis is cut at  $|\delta_r\mathcal{W}| \!=\! 100\%$ in order to present the behavior of $|\delta_r\mathcal{W}(\nu,\omega \!=\! 2\pi T)|$ for the various values of $\nu$ with sufficient resolution. 
The reason for the peak of $|\delta_r\mathcal{W}|$ at one specific Matsubara frequency is the sign change (and hence the closeness to zero) of $|\mathcal{W}_{\rm L}|$. The bottom panels and the corresponding insets show the relative deviation $|\delta_r\mathcal{W}(\nu,\omega \!=\! n2\pi T)|$ for $\nu\!=\!\nu_s$ (left) as well as for an averaged sum over a finite frequency box (see main text for both).  
Due to the averaging effect of the factor ${1}/{N_\nu}$ in $\frac{1}{N_\nu}\sum_\nu |\delta_r\mathcal{W}(\nu,\omega\!=\!n\,2\pi T)| $, where $N_\nu$ is the number of elements summed over, the factor ${1}/{n}$ used in Fig.~\ref{fig:Ward-loop} is omitted. 
In general, Fig.~\ref{fig:APP-Ward-U-W1-rel-left} confirms the trend described in the main text. One notices how the increase of the loop order $\ell$ leads to a reduction of the relative deviations for all frequencies $\omega$ and $\nu$. As pointed out in Sec.~\ref{sec:PPandWI}, the WI is exactly fulfilled for the mfRG/PA solution at $n\rightarrow \infty$, which is also confirmed in Fig.~\ref{fig:APP-Ward-U-W1-rel-left} (see insets). 
An important difference to Fig.~\ref{fig:Ward-loop} is that for the $1\ell$, $1\ell_K$ and $2\ell$ scheme, $\delta_r\mathcal{W}(\nu=\nu_s,\omega)$ is roughly constant, or even grows as $n$ is increased. This reflects the fact that these approaches do not respect the SDE, and hence do not fulfill the WI exactly for $n\to\infty$. 

Regarding (ii), we use the numerically exact QMC solution (fulfilling the WI) as a reference and compare $\mathcal{W}_{\rm L}$ and $\mathcal{W}_{\rm R}$ obtained by fRG/PA for $\nu=\nu_s$ (see main text) individually with the QMC result.
Figure~\ref{fig:APP-Ward-U-W1-QMC} shows this analysis for different values of $U$, in a similar fashion as Fig.~\ref{fig:Ward-U-W1}. 
The comparison of the left side, $\Delta \mathcal{W}_{\rm L}\!=\!\mathcal{W}_{\rm L}^{\rm x}(\nu\!=\!\nu_s,\omega)-\mathcal{W}_{\rm L}^{\rm QMC}(\nu\!=\!\nu_s,\omega)$, where ${\rm x}$ represents the given approach, is shown as full symbols with solid lines; the one of the right side, $\Delta\mathcal{W}_{\rm R}\!=\!\mathcal{W}_{\rm R}^{\rm x}(\nu\!=\!\nu_s,\omega)-\mathcal{W}_{\rm R}^{\rm QMC}(\nu\!=\!\nu_s,\omega)$, as empty symbols with dashed lines, where
${W}_{\rm L}^{\rm QMC}(\nu,\omega)\!=\!{W}_{\rm R}^{\rm QMC}(\nu,\omega)$. 
Let us point out that two distinct effects need to be distinguished in Fig.~\ref{fig:APP-Ward-U-W1-QMC}: on the one hand, there are the deviations of the fRG/PA results from the numerically exact QMC results, on the other hand, the fact that the fRG/PA results do not fulfill the WIs, and are hence not conserving. As discussed in the main part in Fig.~\ref{fig:phys-Ularge}, the deviations between PA/fRG calculations and the QMC results grow with $U$, which can also be seen in Fig.~\ref{fig:APP-Ward-U-W1-QMC}. The solution of a conserving approximation would show this deviation, but would not show a difference between the left and right side, i.e., the full and the dashed lines would coincide. 
Hence, it is not the value on the $y$-axis itself, but the difference in the deviation of $\Delta\mathcal{W}_{\rm L}$ and $\Delta\mathcal{W}_{\rm R}$, which turns out to be instructive. 
As can be seen in Fig.~\ref{fig:APP-Ward-U-W1-QMC}, for most cases, it is the right side of the WI that deviates more from the QMC solution, the $\Omega$-flow $1\ell_K$-results for $n=11$ represent the extreme case.
While in the PA, the $1\ell$ and $2\ell$ results show a steadily growing difference between the solid and the dashed line, 
the situation is less monotonous for the $1\ell_K$ approach. From its data for $n\!=\!1$ (top), one clearly notices the change in behavior as 
$\Delta\mathcal{W}_{\rm R}$ changes sign, leading to the sign change of $\delta\mathcal{W}_{\rm L}^{1\ell_K}(\nu\!=\!-\pi T, \omega\!=\! 2\pi T)$ seen in Fig.~\ref{fig:Ward-U-W1}.

\section{Details on the numerical approach}
\label{sec:APP-technicalities}

\subsection{fRG and mfRG calculations}
\label{subsec:APP-tech-frequencies}

Our fRG, mfRG, and PA computations for the AIM are based on the implementation used in Refs.~\cite{Wentzell2020,Tagliavini2019}. As stated in the main text, we employ the following parametrization of the reducible vertex functions \cite{Wentzell2020} $\gamma_r = K_{1r}\! +\! K_{2r}\! +\! K_{2'r} \!+\! K_{3r}$. The high-frequency asymptotics are included in the $K_{1r}$ and $K_{2^{(\prime)}r}$ functions with one and two 
frequency arguments, respectively. The remaining full frequency dependence, which has a relevant contribution at 
low Matsubara frequencies, is contained in $K_{3r}$. 
These contributions increase with increasing interaction values, and it is hence necessary to extend the frequency box, i.e., the number of frequencies where the full frequency dependence of $K_{3r}$ is taken into account. In Table~\ref{tab:APP_numpam}, we provide the number of positive fermionic frequencies of $K_{3r}$, $N_{f_+}$, for different approaches and values of $U$. The parameter $N_{f_+}$ also dictates all other frequency ranges in the same way as detailed in Ref.~\cite{Wentzell2020}.
Outside the finite frequency box, the $K_{3r}$ functions are set to zero, which is the core of the high-frequency asymptotics approximation. While this affects all quantities calculated with the different approaches, the difference in the results observed by comparing computations with different box sizes is negligible for the $\Omega$-flow and PA. By contrast, for the $U$-flow, an extrapolation in $N_{f_+}$ is necessary, as detailed in the following subsection.  

\begin{table}[tb]
    \centering
    \begin{tabular}{m{2.2cm} m{2.2cm} m{2.2cm}}
        \hline
        \hline
         $U$  &   flow      &   $N_{f_+}$  \\
         \hline
         1.0&   $\Omega$    &   32 \\
            &   $U$         &   32, 40, 64, 82 \\
        1.5 &   $\Omega$    &   36 \\
            &   $U$         &   32, 36, 40, 44 \\
        2.0 &   $\Omega$    &   40 \\
            &   $U$         &   36, 40, 44 \\
        3.0 &   $\Omega$    &   52 \\
        4.0 &   $\Omega$    &   52 \\
        \hline
        \hline
    \end{tabular}
    \caption{Number of positive fermionic Matsubara frequencies used in the calculations of the full frequency dependence ($K_{3r}$). For the PA calculations, we used the same number as for the $\Omega$-flow. }
    \label{tab:APP_numpam}
\end{table}

\subsubsection{Frequency extrapolation for the $U$-flow}
\label{subsec:APP-Uflow-freqextrap}
 
\begin{figure}[t!]
\centering
{{\resizebox{8.3cm}{!}{\includegraphics {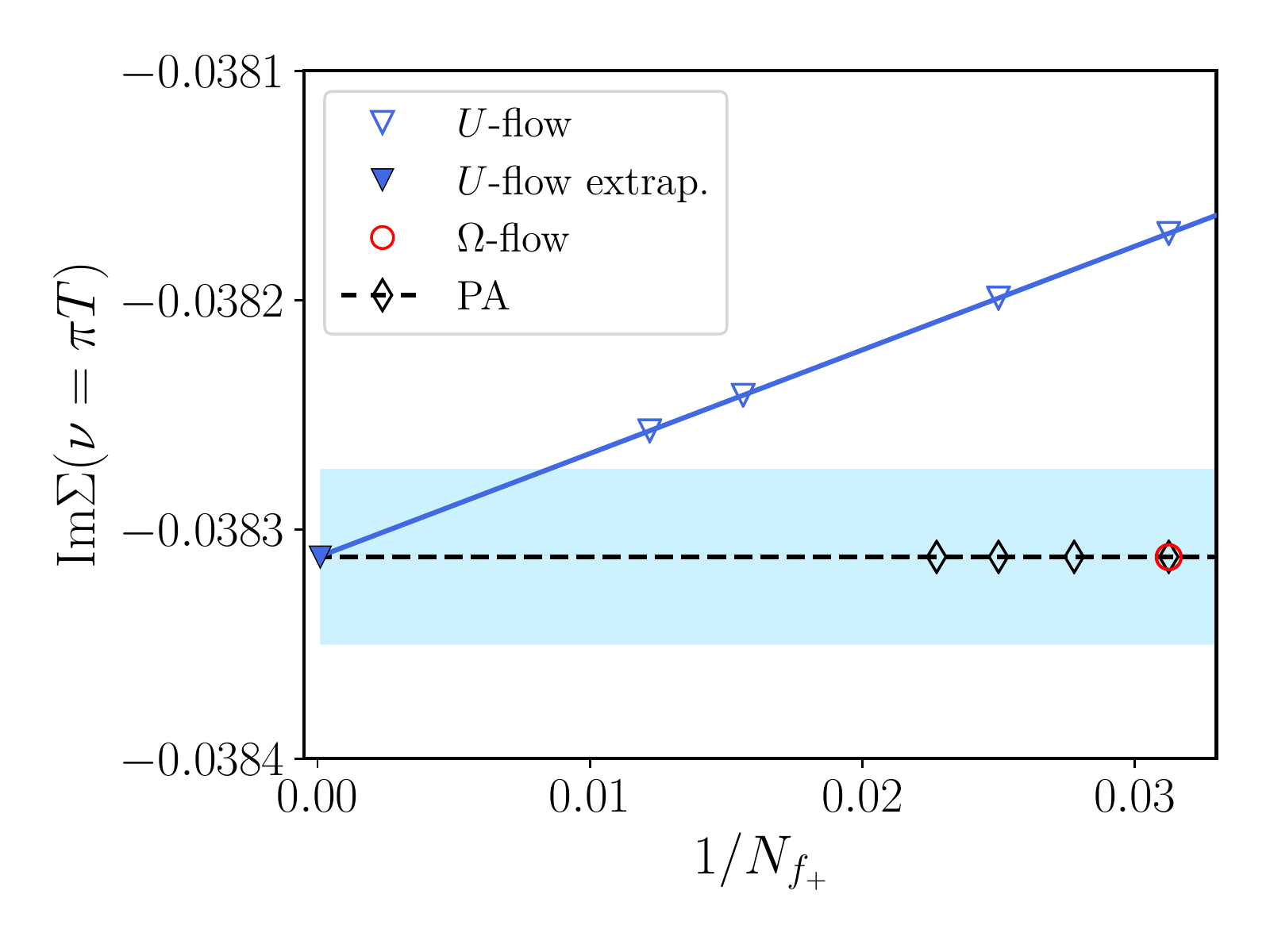}}} 
\caption{Im$\Sigma(\nu=\pi T)$ as obtained by $\infty$-loop $U$-flow calculations 
using different sizes of $N_{f_+}$, see Table \ref{tab:APP_numpam} (open blue triangles). 
The extrapolated value (filled blue triangle) is obtained using a $f(x)=A+B/x$ fit (blue line), 
which matches perfectly the PA solution for $N_{f_+}=32$ (black dashed line). 
The results of PA computations using different box sizes ($N_{f_+}=32,36,40,44$) are shown as open black diamonds. The blue-shaded area represents $0.1\%$ 
deviation from the PA solution for $N_{f_+}=32$. For a comparison a $\infty$-loop 
$\Omega$-flow result using $N_{f_+}=32$ (red open circle) is also shown.
}
\label{fig:APP_Uflow_extrap}  
}
\end{figure} 

In order to achieve agreement between the $\infty$-loop 
mfRG solution using the $U$-flow and the corresponding PA result to the precision chosen in the main part of the paper ($0.1\%$ in the insets), 
it is necessary to perform a frequency extrapolation.
To this end, several calculations for the same parameter set are performed with different sizes of $N_{f_+}$ (see Table \ref{tab:APP_numpam}). 
In Fig.~\ref{fig:APP_Uflow_extrap}, 
we showcase this for $U\!=\!1$ and $\beta\!=\!10$, i.e., the case discussed in Sec.~\ref{subsec:mfrg-conv}. 
The open blue symbols represent the results for Im$\Sigma(\nu=\pi T)$ 
as obtained by different $\infty$-loop $U$-flow calculations, plotted as a function of $1/N_{f_+}$. For comparison, the results 
of corresponding PA calculations with different box sizes are shown as open black symbols, which hardly display any dependence on $N_{f_+}$ at this scale.
Using a $f(x)=A+B/x$ fit (blue line), we obtain the extrapolated value (filled blue triangle), which lies on-top of the PA result for $N_{f_+}=32$ (dashed black line). 
For comparison, we also plot the result of an $\infty$-loop $\Omega$-flow calculation (open red circle) using $N_{f_+}=32$, which highlights that, for the $\Omega$-flow, no frequency extrapolation is necessary to reach agreement with PA at this precision, as stated above. 

All $U$-flow results for all loop orders shown in the main text and the Appendix are obtained in this way. For all quantities, a $f(x) = A+B/x$ fit proved to work best, except for the high-frequency value of $\Sigma$ discussed in Sec.~\ref{subsec:sumrules} (no $U$-flow results shown), where a $A+B/x+C/x^2$ fit turned out to be the best choice. 

\subsection{mfRG calculations}
\label{subsec:APP-mfRGcalc-loop}

In this part of the Appendix, we provide further details on our multiloop calculations. In particular, we specify how the $\infty$-loop mfRG solution is obtained and concisely discuss the iteration of $\Sigma$.

\subsubsection{$\infty$-loop mfRG solution}
\label{subsec:APP-inftyloop-mfrg}

At each step of the fRG flow, the changes in all quantities for all Matsubara frequencies when going from $\ell$ to $\ell + 1$ are measured. 
As soon as the relative (absolute) changes are lower than a given $\epsilon$, in our case $10^{-5}$ ($10^{-7}$), the calculation of higher loop orders is stopped. 
This speeds up the computation especially at the beginning of the flow, where usually a low loop order is sufficient; for more details on this, see Ref.~\cite{Hille2020Thesis}. 
While for obtaining the solution of loop order $\ell$, the multiloop calculation is stopped at this specific $\ell$, 
it is continued until the changes are smaller than $\epsilon$ to calculate the $\infty$-loop order solution. In Table \ref{tab:APP_loopsandits}, we provide the 
actual number of loops needed ($\ell_{\rm max}$) to obtain the $\infty$-loop order solution for the different flows and parameter sets. 

\begin{table}[tb]
    \centering
    \begin{tabular}{m{1.1cm} m{1.8cm} m{1.2cm} m{1.2cm} m{1.2cm} m{1.2cm}}
        \hline
        \hline
         U  &   flow      & $\ell_{\rm max}$ & $ N_{\Sigma\text{-iter}}$ & $N_{\text{step}}$ & $N_{\text{PA-iter}}$  \\
         \hline
         1.0&   $\Omega$    &   15 & 3  & 54 & \\
            &   $U$         &   23 & 4  & 9 & \\
            &   PA          &  &  & & 27 \\
         1.5&   $\Omega$    &   44 & 5  & 61 & \\
            &   $U$         &   61 & 5  & 14 & \\
            &   PA          &     &   &  & 43 \\
        2.0 &   $\Omega$ ($\ell\!=\!15$)    &   -  & 8 & 69 & \\
            &   $U$ ($\ell\!=\!15$)         &   -  & 9  & 23 &\\
            &   PA          &    &   & &  56 \\
        3.0 &   $\Omega$ ($\ell\!=\!15$)    &   -  & 3$^\ast$  & 98 & \\ 
            &   PA          &    &   & & 129 \\
        \hline
        \hline
    \end{tabular}
    \caption{Maximum number of loops ($\ell_{\rm max}$) and iterations of $\Sigma$ ($N_{\Sigma\text{-iter}}$) needed for the $\infty$-loop mfRG solution. In addition, for all interaction values, we list the number of Runge-Kutta integration steps in $\Lambda$ during the fRG flows ($N_{\text{step}}$), and for the PA, the number of iterations need to reach convergence ($N_{\text{PA-iter}}$). Where no  $\infty$-loop mfRG solution was obtained, we list  $N_{\Sigma\text{-iter}}$ and $N_{\text{step}}$ of the calculations with $\ell=15$. We reduced $ N_{\Sigma\text{-iter}}$ for $U\!=\!3$ for the $\Omega$ ($\ell\!=\!15$) calculation, see text.}
    \label{tab:APP_loopsandits}
\end{table}

\subsubsection{Iteration of\, $\Sigma$}
\label{subsec:APP-mfrg-Sigiter}

Part of the mfRG scheme is also the iteration of $\Sigma$ at each step of the flow \cite{Kugler2018,Kugler2018a,Kugler2018b}. The 
effect of these self-energy iterations was analyzed in great detail in Ref.~\cite{Hille2020}. Throughout our calculations, their impact proved to be small, e.g., comparing $\chi_m(\omega=0)$ with and without the 
iteration of $\Sigma$ for the $\Omega$-flow at $U=1$ leads to a difference of $\mathcal{O}(10^{-4})$. 
In Table \ref{tab:APP_loopsandits}, we provide the necessary number of iterations for the $\infty$-loop mfRG solution ($N_{\Sigma\text{-iter}}$ ) to arrive at 
differences smaller than $\epsilon$ (given above, see Appendix \ref{subsec:APP-inftyloop-mfrg}) when comparing 
iteration $i$ with $i+1$. For all other loop orders, the same condition was used. As it turns out, the number of necessary iterations proved to be very similar, except for the $\ell=11,12$ $U$-flow calculations for $U=2$.  
There, the number of required $\Sigma$ iterations increased considerably, preventing our numerical calculation from converging in a reasonable amount of time. 
Lowering the maximum number of $\Sigma$ iterations did not allow for obtaining a converged result, 
as the adaptive solver used for our computations did no longer converge in this case.

For completeness, Table \ref{tab:APP_loopsandits} also lists the number of  Runge-Kutta integration steps in $\Lambda$ during the fRG flow for both regulators ($N_{\text{step}}$), as well as the number of PA iterations ($N_{\text{PA-iter}}$). Note that, since the calculations for $U=3$ were numerically very costly, as they required a large frequency box for $K_{3r}$, we restricted the number of iterations for the $\Omega$-flow computations shown in the main part to $3$.

\subsection{QMC calculations}
\label{sec:APP-tech-QMC}

As stated in the main part, we employed the \textsc{w2dynamics} \cite{w2dynamics} package (version 1.0.0) as a continuous-time QMC \cite{Gull2011a} solver. 
We used the default sampling method for all calculations shown apart from the data for Fig.~\ref{fig:asymptote}. There, we performed 
Worm sampling \cite{Gunacker15,Gunacker2016} computations with symmetric improved estimators \cite{Kaufmann2019}
instead, which reduces the high-frequency noise. While we used about 2000 CPU hours for the former computations, the Worm sampling calculations were done using up to 25000 CPU hours.

%%%%%%%%%%%%%%%%%%%%%%%%%%%%%%%%%%%%%%%%%%%%%%%%%%%%%%%%%%%%%%%%%%%%%%%%%%%%%%%%%%%%%%%%%%%%%%%

\section{Diagrammatic derivations}
\label{sec:diagr-derivations}

\subsection{Relations between the self-energy asymptote and the susceptibility sum rule}
\label{sec:chi-sum_Sigma-asymptote}

\begin{figure}[t!]
\centering
\includegraphics[width=0.483\textwidth]{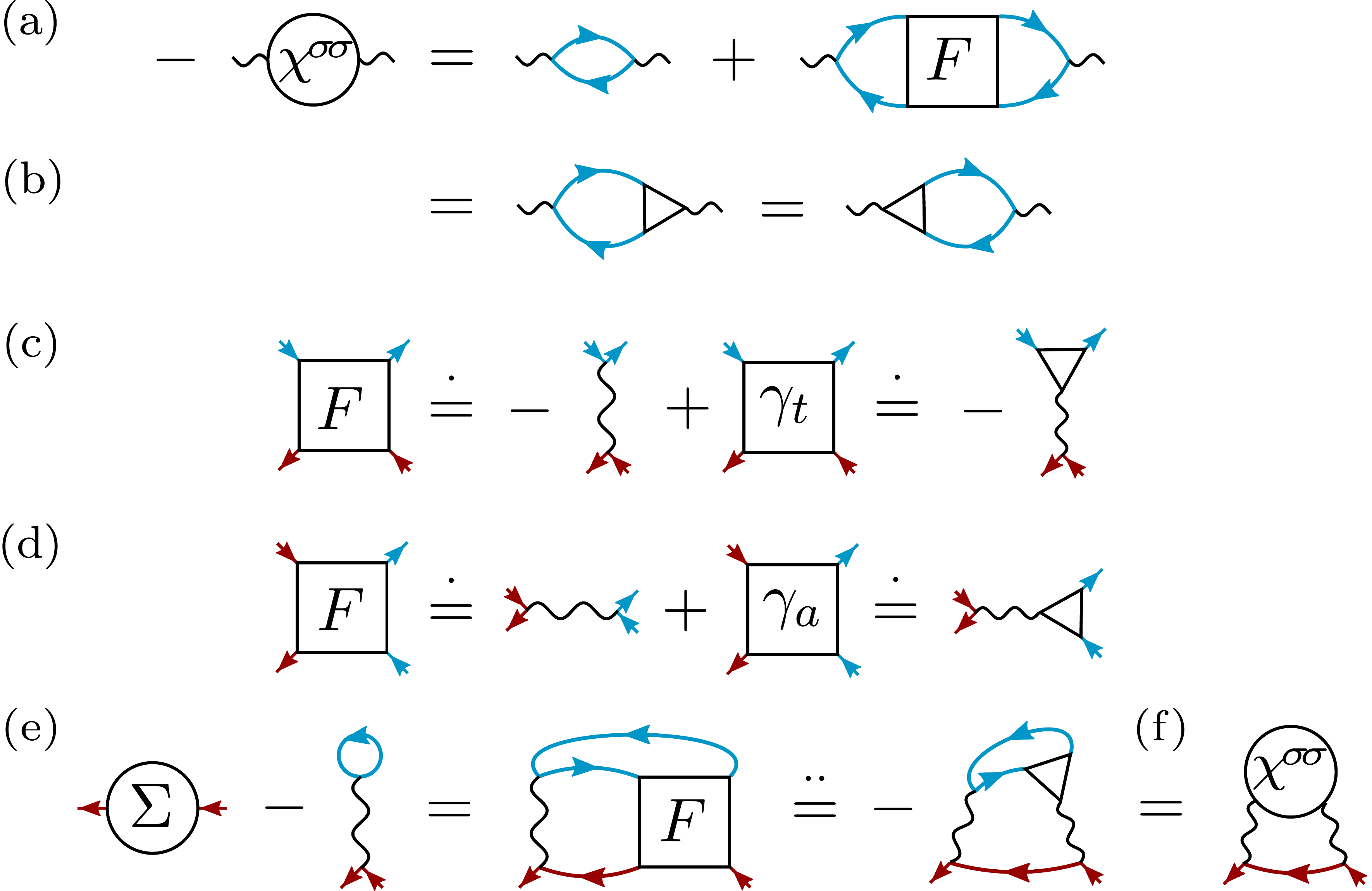}
\caption{(a,b) The susceptibility $\chi^{\sigma\sigma}$ can be expressed through the four-point vertex $F$ or the three-point vertex $\lambda$. 
(c,d) In the limit $\nu\!\to\infty$, the vertex, carrying $\nu$ on the external legs marked in red, collapses to a subset of diagrams up to corrections $\mathit{O}(1/\nu)$ (signified by `$\dot{=}$').
These can also be expressed through $\lambda$.
(e,f) In this limit, we can deduce
the self-energy $\Sigma_\nu$ 
up to corrections $\mathit{O}(1/\nu^2)$ (signified by `$\ddot{=}$')
from the SDE and express the result through $\lambda$ or $\chi^{\sigma\sigma}$.}
\label{fig:SDE-asymp}
\end{figure}

In this section, we will derive relations between the high-frequency asymptote 
of $\Sigma$ and the sum rule of $\chi^{\sigma\sigma}$.
First, we will show that the two are directly related through the SDE in parquet-type approaches.
Then, we move on to fRG flows.
We will show that the standard self-energy flow
also relates the $\Sigma$ asymptote to the susceptibility sum rule,
with $\chi^{\sigma\sigma}$ given by its one-loop flow.
Since the latter does not fulfill the sum rule, the former violates the exact asymptote.
Both the sum rule and the asymptote \textit{are} fulfilled in multiloop fRG.
We will show which terms of the multiloop corrections to $\dot{\Sigma}$
complete the relation, so that the $\Sigma$ asymptote 
is determined by $\chi^{\sigma\sigma}$ obtained in a multiloop flow.
The entire derivation will proceed diagrammatically.

\subsubsection*{Connection through the SDE}

In Fig.~\ref{fig:SDE-asymp}(a)
we recall Eq.~\eqref{eq:chi-GG-vtx},
which expresses $\chi^{\sigma\sigma}$ through a $GG$ bubble and 
corrections in terms of the full four-point vertex $F$.
The vertex $F$ is contracted by pairs of propagators on both sides.
Therefore, one can also express $\chi^{\sigma\sigma}$
through a (full) three-point vertex $\lambda$ on either 
the left or the right side,
as illustrated in Fig.~\ref{fig:SDE-asymp}(b).
The vertex $\lambda$ is particularly useful when considering $F$
in the limit of large fermionic frequencies.

Indeed, to find the self-energy asymptote, we will consider a large fermionic frequency $\nu$.
In Figs.~\ref{fig:SDE-asymp}(c,d), we show which diagrams of the vertex $F$,
carrying $\nu$ on the external legs marked in red,
remain nonzero in the limit $\nu \!\to\! \infty$, 
i.e., which diagrams are independent of $\nu$.
We use the symbol `$\dot{=}$' for that purpose, signifying equality up to $\mathit{O}(1/\nu)$.
To have nonzero contributions when $\nu \!\to\! \infty$, 
the red (amputated) external legs must directly meet at a bare interaction vertex. 
This is clearly fulfilled for $F \!=\! F_0$, but there can also be arbitrary vertex corrections
after the two red legs have met.
If $\nu$ is on the lower two legs
[Fig.~\ref{fig:SDE-asymp}(c)],
such corrections are a subset of the vertex $\gamma_t$,
reducible in \textit{t}ransverse (vertical) particle-hole lines.
If it is on the left two legs
[Fig.~\ref{fig:SDE-asymp}(d)], 
the corrections belong to $\gamma_a$,
reducible in \textit{a}ntiparallel (horizontal) lines.
The bare vertex and and the corrections are summarized by the three-point vertex $\lambda$.
To see this, one may insert the BSEs for $\gamma_{t/a}$, connecting the
irreducible vertices $I_{t/a}$ to the full vertex $F$.
Since $I_{t/a}$ are irreducible in their respective channels, they collapse to $F_0$
in the limit $\nu \!\to\! \infty$, and one obtains $\lambda$ similarly as in going from Fig.~\ref{fig:SDE-asymp}(a) to  Fig.~\ref{fig:SDE-asymp}(b).

Now, by means of the SDE~\eqref{eq:SDE}, the self-energy 
(minus its static Hartree part) is determined by the vertex $F$
connected to three propagators,
as we recall in Fig.~\ref{fig:SDE-asymp}(e).
To find $\Sigma$ to first order in $1/\nu$,
we need $F$ to zeroth order.
We choose to transport $\nu$ through the propagator at the bottom.
Then, we can directly use the relation in Fig.~\ref{fig:SDE-asymp}(c)
to replace $F$ by $\lambda$
up to corrections $\mathit{O}(1/\nu^2)$
(signified by the symbol `$\ddot{=}$').
Using Fig.~\ref{fig:SDE-asymp}(a), we obtain $\chi^{\sigma\sigma}$
through $\lambda$.
The last step is similar to Eq.~\eqref{eq:SE_asymptote_2ndOrder}:
Take the red propagator as $G^{\bar{\sigma}}_{\nu+\omega}$.
For $\nu \!\gg\! \omega$, we can replace
$G^{\bar{\sigma}}_{\nu+\omega}$ by $1/(i\nu)$ up to corrections
$\mathit{O}(1/\nu^2)$.
This leaves $\chi^{\sigma\sigma}_\omega$ summed over all $\omega$,
and, with a prefactor $U^2$ from the two interaction lines,
we obtain Eq.~\eqref{eq:SE-asymp_chi} for $\Sigma^{\bar{\sigma}}$.

\subsubsection*{Standard self-energy flow}

\begin{figure}[t!]
\centering
\includegraphics[width=0.483\textwidth]{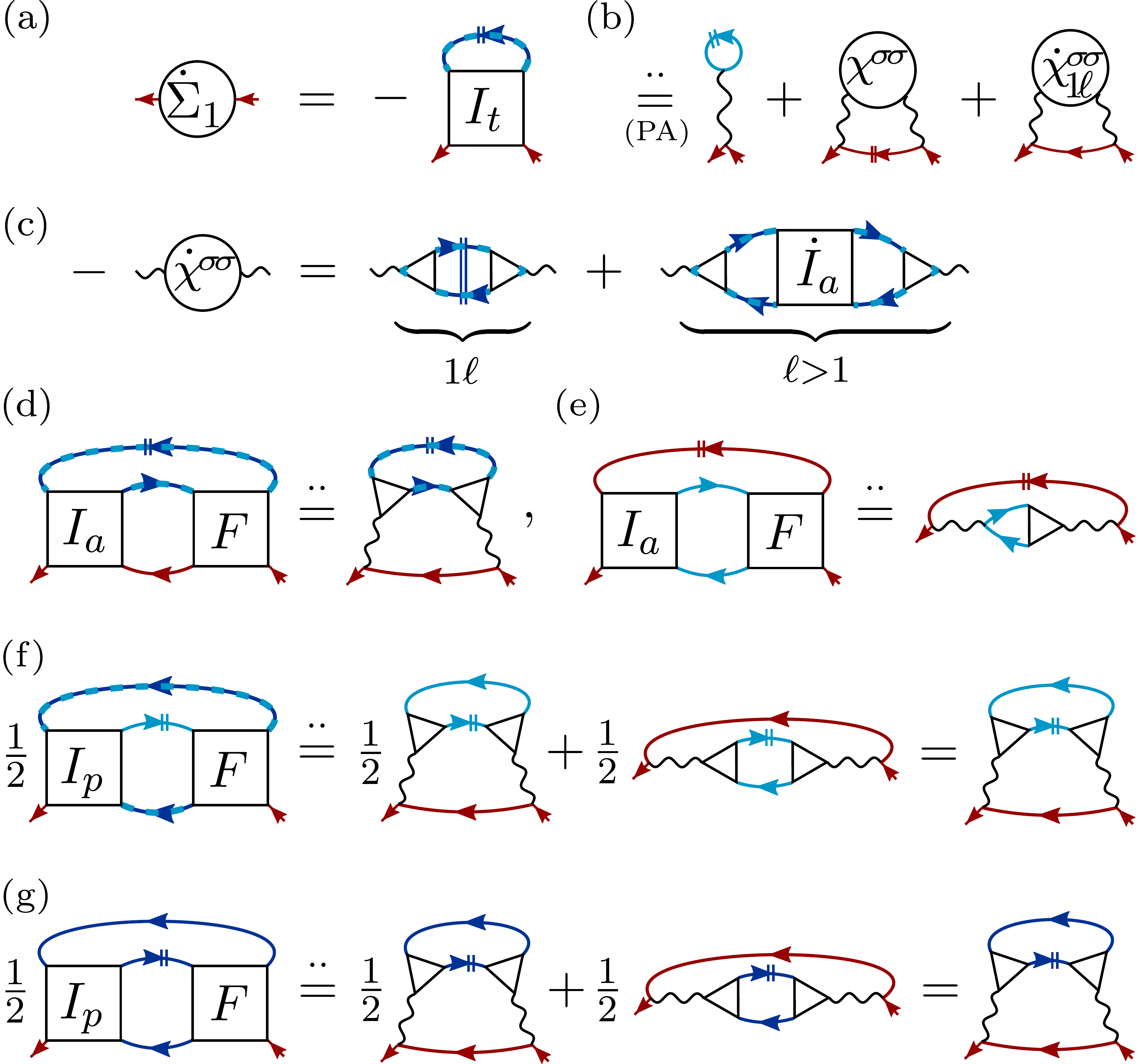}
\caption{(a) The standard self-energy flow 
$\dot{\Sigma}_{\mathrm{std}} \!\equiv\! \dot{\Sigma}_1$
in skeleton form \cite{Kugler2018b}.
(b) In the PA, it yields asymptotic contributions
of the same structure as Fig.~\ref{fig:SDE-asymp}(f),
where $\dot{\chi}^{\sigma\sigma}$ is approximated by its one-loop flow.
(c) One-loop ($\ell \!=\! 1$) and multiloop terms 
($\ell \!>\! 1$) for $\dot{\chi}^{\sigma\sigma}$ \cite{Kugler2018b}.
To derive the relation in (b),
we insert the BSEs for $\gamma_a$, $\gamma_p$ 
as part of $I_t$;
(d,e) concerns $\gamma_a$,
while (f,g) concerns $\gamma_p$.}
\label{fig:SE1-asymp}
\end{figure}

Next, we turn to fRG flows.
The standard self-energy flow is given by
$\dot{\Sigma}^\sigma_{\mathrm{std}} 
\!=-\! F^{\sigma\sigma'} \!\cdot\! S^{\sigma'}$,
where `$\cdot$' denotes the contraction of the top two vertex legs by the following propagator,
$S$ is the single-scale propagator,
and a sum over $\sigma'$ is understood.
For formal derivations, it is helpful to analyze 
$\dot{\Sigma}^\sigma_{\mathrm{std}}$ by means of its equivalent
skeleton version \cite{Kugler2018b},
$\dot{\Sigma}^\sigma_1 \!=\! - I_t^{\sigma\sigma'} \!\cdot\! \dot{G}^{\sigma'}$,
illustrated in Fig.~\ref{fig:SE1-asymp}(a).
As before,
a line with a doubled orthogonal slash denotes $\dot{G}$,
and dashed dark and light colors indicate a summation over spin.

As mentioned previously, $\dot{\Sigma}_1$
is exact only for an exact vertex, which is not available in practice.
Instead, we will consider the much more relevant case
of a vertex obtained in the PA
or, equivalently, a multiloop flow.
In this case, $\dot{\Sigma}_1$ is approximate.
We will show that it generates a high-frequency asymptote
of similar type as the exact relation Fig.~\ref{fig:SDE-asymp}(f),
but with $\chi^{\sigma\sigma}$ obtained by its (approximate) one-loop flow.
The connection from the general, $\Lambda$-independent statement 
Fig.~\ref{fig:SDE-asymp}(f)
to an fRG flow is made by taking the
scale derivative $\partial_\Lambda$ on the entire equation.
In this way, $\partial_\Lambda$ is subsequently applied to the trivial Hartree part,
to the red propagator alongside $\chi^{\sigma\sigma}$,
and finally to $\chi^{\sigma\sigma}$ itself.
Indeed, we will precisely find such a structure,
where the derivative $\partial_\Lambda \chi^{\sigma\sigma}$
is approximated by $\dot{\chi}^{\sigma\sigma}_{1\ell}$,
see Fig.~\ref{fig:SE1-asymp}(b).
The one-loop flow $\dot{\chi}^{\sigma\sigma}_{1\ell}$ 
is given by the first summand of Fig.~\ref{fig:SE1-asymp}(c).
(The long double slash denotes a 
differentiated two-particle propagator, $\dot{\Pi} \!=\! \dot{G} G \!+\! G \dot{G}$.)
The multiloop corrections to $\dot{\chi}^{\sigma\sigma}$,
which are compactly encoded in the 
second summand of Fig.~\ref{fig:SE1-asymp}(c)
and will be considered more closely in the next part,
restore equivalence to the general susceptibility--vertex relation
shown in Fig.~\ref{fig:SDE-asymp}(a).

To derive Fig.~\ref{fig:SE1-asymp}(b),
we start from $I_t \!=\! F_0 \!+\! \gamma_a \!+\! \gamma_p$
in the PA.
The bare vertex $F_0$ immediately gives the differentiated Hartree part
as the first summand of Fig.~\ref{fig:SE1-asymp}(b).
From Fig.~\ref{fig:SE1-asymp}(d) onward, 
we analyze the effect of $\gamma_{a/p}$ using their BSEs.
The analysis is slightly more complicated than in Fig.~\ref{fig:SDE-asymp}(e):
There, we had just a single vertical interaction line;
now, we have two spin-dependent vertices, 
where same-spin propagators can meet both 
vertically and horizontally.

In Figs.~\ref{fig:SE1-asymp}(d,e),
we insert the BSE of $\gamma_a^{\sigma\sigma'}$, 
with a summation on the spin carried by $\dot{G}^{\sigma'}$.
This gives three terms: 
(i) $\gamma_a^{\sigma\bar{\sigma}}$, where the 
antiparallel two-particle propagator $\Pi_a$
necessarily has two opposite spins;
$\gamma_a^{\sigma\sigma}$, where $\Pi_a$ has
(ii) both spins equal to $\sigma$ and 
(iii) both spins equal to $\bar{\sigma}$.
Cases (i) and (ii) are contained in Fig.~\ref{fig:SE1-asymp}(d),
with a spin sum encoded in the dashed colors.
Regarding $\sim\!\nu^{-1}$ contributions, Fig.~\ref{fig:SE1-asymp}(d)
contains all diagrams where 
the lower two legs of $I_a$ and $F$ directly meet at vertical interaction lines
and the large frequency is transported through the bottom propagator.
Since both $I_a$ and $F$ contain $F_0 \!+\! \gamma_t$, 
their $\sim\!\nu^{0}$ contributions are expressed through $\lambda$
according to Fig.~\ref{fig:SDE-asymp}(c).
Proceeding with case (iii), Fig.~\ref{fig:SE1-asymp}(e) contains all diagrams where
the left (right) legs of $I_a$ ($F$) directly meet at horizontal interaction lines
and the large frequency is transported through the top propagator.
While $F$ contains both $F_0$ and $\gamma_a$, $I_a$ contains only $F_0$.
Hence, their $\sim\!\nu^{0}$ contributions are expressed through $\lambda$
and a bare interaction line, respectively,
according to Fig.~\ref{fig:SDE-asymp}(d).

We continue with $\gamma_p$
and insert in Fig.~\ref{fig:SE1-asymp}(f) 
the BSE of $\gamma_p^{\sigma\bar{\sigma}}$
($\dot{G}$ is in light color),
where the parallel two-particle propagator $\Pi_p$ is summed over both spins
(and thus the typical prefactor $1/2$ is kept).
Since both $I_p$ and $F$ are crossing symmetric, 
contributions stemming from vertical and horizontal interaction lines enter equivalently.
Indeed, in the first (second) summand of Fig.~\ref{fig:SE1-asymp}(f),
the red propagator passes by vertical (horizontal) interaction lines.
Since both $I_p$ and $F$ contain $F_0 \!+\! \gamma_a \!+\! \gamma_t$,
we replace their $\sim\!\nu^{0}$ contributions by $\lambda$
using Fig.~\ref{fig:SDE-asymp}(c,d),
and we end up with two equivalent terms.
In Fig.~\ref{fig:SE1-asymp}(g), we insert the BSE of $\gamma_p^{\sigma\sigma}$,
where $\Pi_p$ must also carry spins $\sigma$
(the prefactor $1/2$ remains).
Again, the red propagator can pass by vertical and horizontal interaction lines,
and we get two equivalent terms expressed through $\lambda$.

Finally, we see that 
Fig.~\ref{fig:SE1-asymp}(e) gives the second summand of
Fig.~\ref{fig:SE1-asymp}(b)
[by means of Fig.~\ref{fig:SDE-asymp}(b)],
and the sum of Figs.~\ref{fig:SE1-asymp}(d,f,g) 
reproduces the first summand of Fig.~\ref{fig:SE1-asymp}(c).
This yields the last part of Fig.~\ref{fig:SE1-asymp}(b),
thus concluding the derivation.

\subsubsection*{Multiloop corrections to the self-energy flow}

\begin{figure}[t!]
\centering
\includegraphics[width=0.483\textwidth]{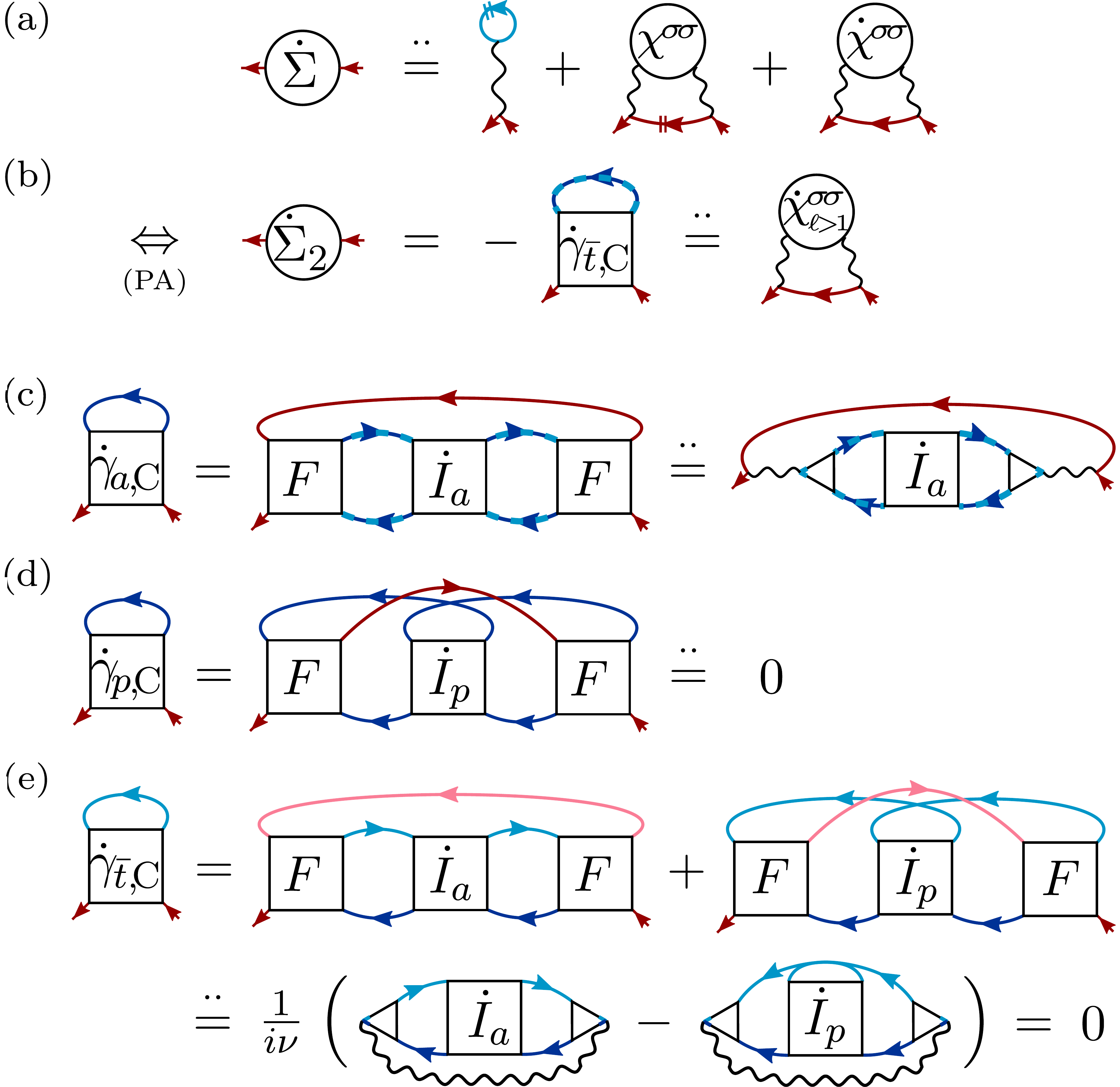}
\caption{(a) The multiloop self-energy flow is equivalent to the SDE
and thus generates a high-frequency asymptote in direct correspondence to Fig.~\ref{fig:SDE-asymp}(f). 
(b) Using $\dot{\Sigma} \!=\! \dot{\Sigma}_1 \!+\! \dot{\Sigma}_2$ and the result of Fig.~\ref{fig:SE1-asymp}(b), the asymptote of $\dot{\Sigma}_2$ must be related to the multiloop corrections of $\dot{\chi}^{\sigma\sigma}$.
To show this, we split the contraction of $\gamma_{\bar{t},\mathrm{C}}$ into four summands:
(c) $\gamma_{a,\mathrm{C}}^{\sigma\sigma} \cdot G^{\sigma}$ already yields the desired expression;
(d) $\gamma_{p,\mathrm{C}}^{\sigma\sigma} \cdot G^{\sigma}$ vanishes up to corrections $\mathit{O}(1/\nu^2)$;
(e) $(\gamma_{a,\mathrm{C}}^{\sigma\bar{\sigma}} + \gamma_{p,\mathrm{C}}^{\sigma\bar{\sigma}}) \cdot G^{\bar{\sigma}}$ cancel to that order,
as can be seen after factoring out $G^{\bar{\sigma}} \!\sim\! 1/(i\nu)$ for the first and
$G^{\bar{\sigma}} \!\sim\! 1/(-i\nu)$ for the second summand.}
\label{fig:SE2-asymp}
\end{figure}

The multiloop corrections to the self-energy flow
provide equivalence to the SDE while working in the PA \cite{Kugler2018b}.
Thereby, the multiloop self-energy flow is guaranteed 
to generate the correct high-frequency asymptote.
Its $\sim\!\nu^{-1}$ contribution
must be equal to the scale derivative
of Fig.~\ref{fig:SDE-asymp}(f),
shown in Fig.~\ref{fig:SE2-asymp}(a).
The multiloop self-energy flow can be written \cite{Kugler2018b} as 
$\dot{\Sigma} \!=\! \dot{\Sigma}_1 + \dot{\Sigma}_2$,
with $\dot{\Sigma}^\sigma_1 \!=\! - I_t^{\sigma\sigma'} \!\cdot\! \dot{G}^{\sigma'}$
from before and 
$\dot{\Sigma}_2^\sigma \!=\! 
-\dot{\gamma}_{\bar{t},C}^{\sigma\sigma'} \cdot G^{\sigma'}$.
Hence, Fig.~\ref{fig:SE2-asymp}(a)
and Fig.~\ref{fig:SE1-asymp}(b)
imply that Fig.~\ref{fig:SE2-asymp}(b) must hold.

It is interesting to analyze how Fig.~\ref{fig:SE2-asymp}(b) comes about.
Through the spin sum and the composite nature of $\dot{\gamma}_{\bar{t},\mathrm{C}}$,
$\dot{\Sigma}_2$ has four contributions, stemming from 
$\dot{\gamma}_{a,\mathrm{C}}^{\sigma\sigma}$,
$\dot{\gamma}_{p,\mathrm{C}}^{\sigma\sigma}$,
$\dot{\gamma}_{a,\mathrm{C}}^{\sigma\bar{\sigma}}$,
and $\dot{\gamma}_{p,\mathrm{C}}^{\sigma\bar{\sigma}}$.
We will show that the first term already gives the desired
result in Fig.~\ref{fig:SE2-asymp}(b).
Up to corrections $\mathit{O}(1/\nu^2)$,
the second term vanishes while the last two terms cancel.

Inserting $\dot{\gamma}_{a,\mathrm{C}}^{\sigma\sigma}$,
the only way to get $\sim\!\nu^{-1}$ contributions
is to transport the large frequency through the loop propagator at the top,
marked red in Fig.~\ref{fig:SE2-asymp}(c)
(note that both two-particle propagators $\Pi_a$ are summed over spin).
Further, all red lines must directly meet at (horizontal) interaction lines.
Hence, the four-point vertices $F$ at the left and right can be replaced by
three-point vertices $\lambda$.
The combination of $\lambda$, $\dot{I}_a$, $\lambda$ 
comprises the multiloop corrections to the flow of $\chi^{\sigma\sigma}$,
see Fig.~\ref{fig:SE1-asymp}(c),
thus yielding Fig.~\ref{fig:SE2-asymp}(b).

For the remaining terms,
one immediately sees in Fig.~\ref{fig:SE2-asymp}(d) that
$\dot{\gamma}_{p,\mathrm{C}}^{\sigma\sigma}$
has no $\sim\!\nu^{-1}$ contribution:
The external legs and the loop propagator 
would need to directly meet as two out-going (in-going) lines 
at a bare interaction line of the left (right) vertex.
However, they all have the same spin, and the bare interaction 
requires in- and out-going lines to have opposite spin.
Next, the opposite-spin contribution
$\dot{\gamma}_{a,\mathrm{C}}^{\sigma\bar{\sigma}} \!+\!
\dot{\gamma}_{p,\mathrm{C}}^{\sigma\bar{\sigma}}$ is shown in
Fig.~\ref{fig:SE2-asymp}(e).
By choosing fixed spin labels for the two $\Pi_p$
entering $\dot{\gamma}_{p,\mathrm{C}}$,
we eliminate the typical prefactor $(1/2)^2$.
The upper loop propagator carrying the $\nu$ dependence
goes in opposite directions for the first compared to second summand.
Hence, after factoring out the dominant $1/(i\nu)$, we get opposite signs for the
$\sim\!\nu^{-1}$ contributions between the $a$ and $p$ channel.
The remaining part for both is summed over all internal frequencies,
including $\omega$, as indicated by the closed wiggly line.
Their sum cancels, as can be checked explicitly at low orders.
Note that, for this to work, one needs the same number of diagrams in
$\gamma_a^{\sigma\bar{\sigma}}$ and $\gamma_p^{\sigma\bar{\sigma}}$ 
at each interaction order, as is indeed the case \cite{Kugler2018c}.

\subsection{Deriving the SDE from the WI}
\label{sec:WI_SDE}

The WI~\eqref{eq:WI} relates a difference of self-energies,
$\Sigma^\sigma_{\nu+\omega} - \Sigma^\sigma_{\nu}$,
to a vertex contracted by a combination of propagators.
For infinitely large $\omega$, while $\nu$ remains finite,
the first self-energy simplifies to its static value,
$\Sigma^\sigma_{\nu+\omega} \to U n_{\bar{\sigma}}$,
and we thus obtain a relation for $\Sigma^\sigma_{\nu}$ alone.
This relation is precisely the SDE~\eqref{eq:SDE}, as we show now.

\begin{figure}[t!]
\centering
\includegraphics[width=0.483\textwidth]{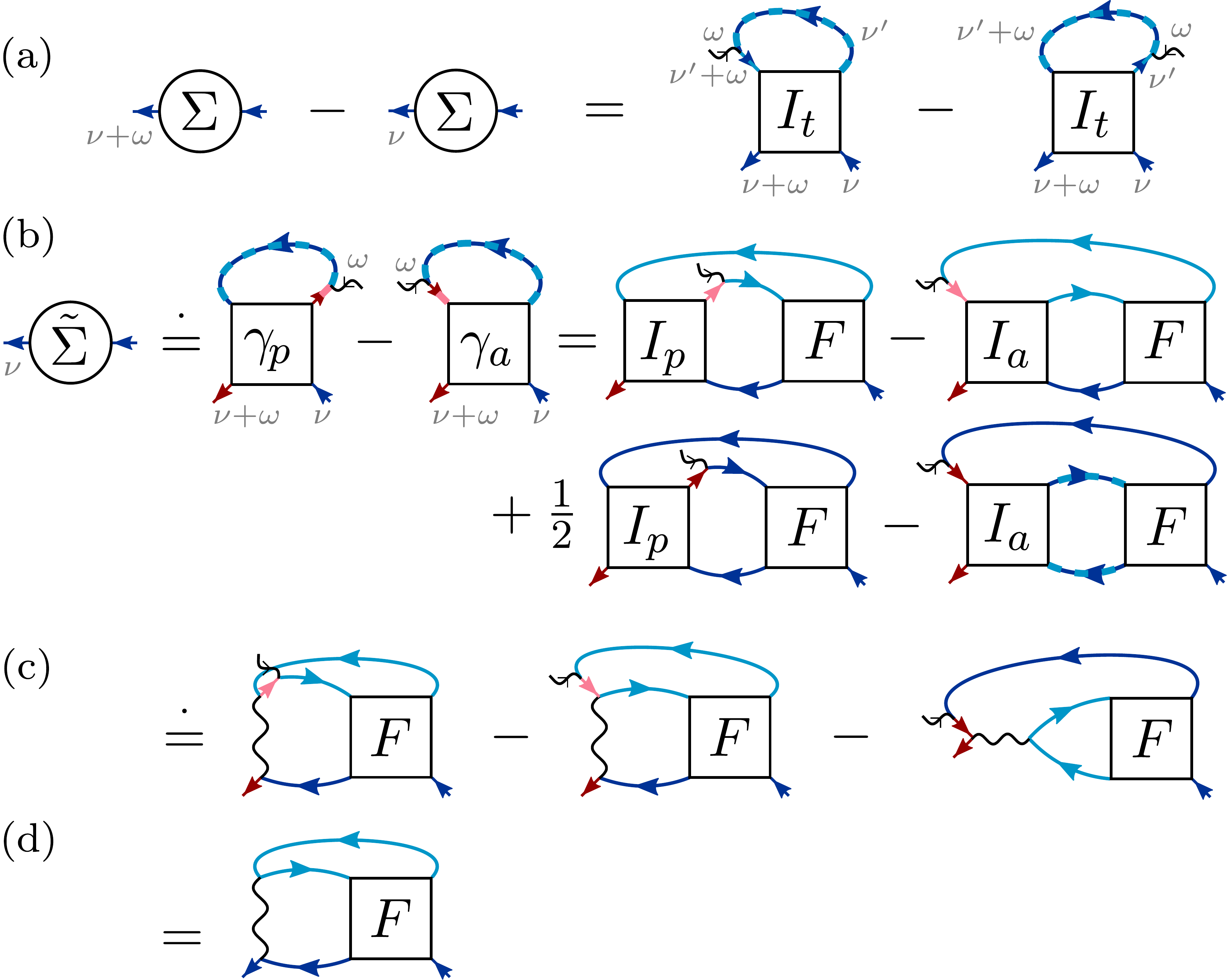}
\caption{(a) Illustration of the WI~\eqref{eq:WI_natural-param},
in a slightly different but equivalent form to Fig.~\ref{fig:chi-vtx_SDE_WI}(c).
Note that short lines denote amputated legs which are not part of the equation.
(b) Derivation of the SDE from the WI by taking the limit $\omega\!\to\!\infty$,
see text. Lines in red colors carry the large frequency $\omega$.}
\label{fig:WI_SDE}
\end{figure}

We start by restating Eq.~\eqref{eq:WI} in the form
\begin{align}
\Sigma^\sigma_{\nu+\omega}
-
\Sigma^\sigma_\nu
& =
-
\frac{1}{\beta}
\sum_{\sigma'\nu'}
I^{\sigma\sigma';\omega}_{t;\nu,\nu'}
( G^{\sigma'}_{\nu'+\omega} - G^{\sigma'}_{\nu'} )
.
\label{eq:WI_natural-param}
\end{align}
Here, we labeled $I_t$ by only three frequencies,
chosen in the natural parametrization of the $t$ channel,
with the bosonic frequency $\omega$ as a superscript and the
two fermionic frequencies $\nu$ and $\nu'$ as subscripts.
We also introduce a diagrammatic representation of the WI
that is slightly different from Fig.~\ref{fig:chi-vtx_SDE_WI}(c):
In Fig.~\ref{fig:WI_SDE}(a), we have the difference in self-energies on the left
and a difference of the vertices, each contracted by a different propagator on the right.
Indeed, each vertex is contracted by only the propagator corresponding to the long line.
All the short, external legs are amputated; they do not contribute to the diagram.
In particular, the short wavy line only serves to ensure energy conservation for each vertex;
it does not enter the equation itself.
We recall that dark and light colors distinguish the two spin species;
dashed lines with dark and light colors symbolize a sum over spin.

If we take the limit $\omega\to\infty$ in Eq.~\eqref{eq:WI_natural-param}
or Fig.~\ref{fig:WI_SDE}(a), the l.h.s.\ simplifies to $-\tilde{\Sigma}$,
where $\tilde{\Sigma}^\sigma_\nu \!=\! \Sigma^\sigma_\nu \!-\! U n_{\bar{\sigma}}$
is the self-energy without its static Hartree part.
In this limit, the r.h.s.\ simplifies as well.
First, we express $I_t$, the vertex irreducible in the $t$ channel, 
as a sum of the fully irreducible vertex $R_{\mathrm{2PI}}$ and the vertices reducible in the complementary 
channels, $\gamma_a$ and $\gamma_p$.
Fully irreducible vertex diagrams beyond the bare vertex, $F_0$, decay in all frequency arguments;
therefore, $\lim_{\omega\to\infty} R = F_0$.
However, $F_0$ makes no contribution to Eq.~\eqref{eq:WI_natural-param}, 
as it is frequency independent and thus leads to cancellation in the $\nu'$ sum.
In contrast to $R_{\mathrm{2PI}}$, the reducible vertices $\gamma_r$ have specific contributions 
that are independent of certain (fermionic) frequencies.
By substituting $\gamma_a \!+\! \gamma_p$ for $I_t$ in Eq.~\eqref{eq:WI_natural-param},
we get
\begin{align}
\tilde{\Sigma}^\sigma_\nu
& =
\lim_{\omega\to\infty}
\frac{1}{\beta}
\sum_{\sigma'\nu'}
I^{\sigma\sigma';\omega}_{t;\nu+\omega,\nu'+\omega}
( G^{\sigma'}_{\nu'+\omega} - G^{\sigma'}_{\nu'} )
\label{eq:Sigmatilde1}
\\
& =
\lim_{\omega\to\infty}
\frac{1}{\beta}
\sum_{\sigma'\nu'}
(
\gamma^{\sigma\sigma';\nu'-\nu}_{a;\nu+\omega,\nu}
\!+\!
\gamma^{\sigma\sigma';\nu+\nu'+\omega}_{p;\nu+\omega,\nu}
)
( G^{\sigma'}_{\nu'+\omega} \!-\! G^{\sigma'}_{\nu'} )
.
\nonumber
\end{align}
Here, we expressed $\gamma_a$ and $\gamma_p$ each in their natural frequency 
parametrization. As fermionic frequencies, we chose the two lower vertex legs in Fig.~\ref{fig:WI_SDE}(a) for both $\gamma_a$ and $\gamma_p$.
The transfer frequency is $\nu' \!-\! \nu$ w.r.t.\ to the $a$ channel
and $\nu \!+\! \nu' \!+\! \omega$ w.r.t.\ to the $p$ channel.

Next, we use the fact that a reducible vertex always decays with its bosonic transfer argument,
$\lim_{\omega\to\infty} \gamma^{\sigma\sigma';\omega}_{r;\nu,\nu'} \!=\! 0$,
and that a propagator $G^\sigma_\nu$ decays as $1/\nu$.
It follows that
$\lim_{\omega\to\infty} \sum_{\nu'} \gamma^{\sigma\sigma';\nu'-\nu}_{a;\nu+\omega,\nu} G^{\sigma'}_{\nu'+\omega} \!=\! 0$,
since nonzero values of $\gamma_a$ require $\nu' \!\sim\! \nu$,
i.e.\ finite $\nu'$,
so that $G^{\sigma'}_{\nu'+\omega} \!\to\! 0$.
Similarly, 
$\lim_{\omega\to\infty} \sum_{\nu'} \gamma^{\sigma\sigma';\nu+\nu'+\omega}_{p;\nu+\omega,\nu} G^{\sigma'}_{\nu'} = 0$,
since nonzero values of $\gamma_p$ require $\nu' \sim -\omega$,
increasing in magnitude with $\omega$,
so that $G^{\sigma'}_{\nu'} \to 0$.
By contrast, the remaining two terms in Eq.~\eqref{eq:Sigmatilde1} give finite contributions,
\begin{align*}
\tilde{\Sigma}^\sigma_\nu
=
\lim_{\omega\to\infty}
\Big(
\frac{1}{\beta}
\sum_{\sigma'\hat{\nu}}
\gamma^{\sigma\sigma';\nu+\hat{\nu}}_{p;\nu+\omega,\nu}
G^{\sigma'}_{\hat{\nu}}
-
\frac{1}{\beta}
\sum_{\sigma'\nu'}
\gamma^{\sigma\sigma';\nu'-\nu}_{a;\nu+\omega,\nu}
G^{\sigma'}_{\nu'}
\Big)
,
\end{align*}
where we relabeled $\hat{\nu} \!=\! \nu' \!+\! \omega$ in the $p$ channel.
This relation is the first equality in Fig.~\ref{fig:WI_SDE}(b).
The symbol `$\dot{=}$' here means that both sides agree up to $\mathit{O}(1/\omega)$,
i.e., they are equal in the limit $\omega \!\to\! \infty$.

\begin{figure}[t!]
\centering
\includegraphics[width=0.483\textwidth]{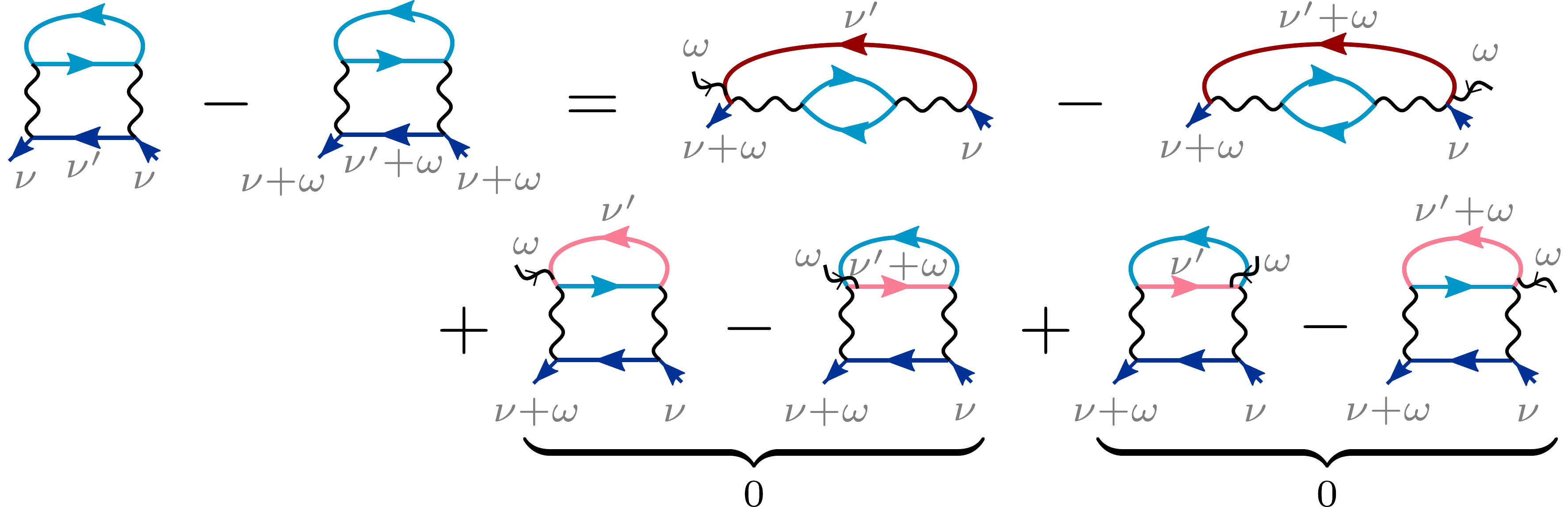}
\caption{The WI as in Fig.~\ref{fig:WI_SDE}(a) evaluated in second-order (bold) perturbation theory. The l.h.s., $\Sigma^\sigma_{\nu+\omega} \!-\! \Sigma^\sigma_\nu$, matches 
the $\gamma_a^{\sigma\sigma}$ vertex contributions on the r.h.s., top row.
The bottom row on the r.h.s., coming from $\gamma_a^{\sigma\bar{\sigma}}$
and $\gamma_p^{\sigma\bar{\sigma}}$ cancels.
Note that $\gamma_p^{\sigma\sigma}$ has no second-order contribution;
see, e.g., Fig.~5 in Ref.~\cite{Kugler2018c} for a collection of all second-order diagrams.
We give some frequency labels for clarity.
On the r.h.s., red colors mark the loop propagator
contracting the vertex $I_t$ in Fig.~\ref{fig:WI_SDE}(a).}
\label{fig:WI_2ndOrder}
\end{figure}

For the rest of the analysis, we refrain from spelling out the equations and proceed diagrammatically.
In Fig.~\ref{fig:WI_SDE}(b),
lines in red colors (dark and light for the two spins) carry the large frequency $\omega$.
All of these are amputated external legs, for, otherwise, the result would vanish
in the limit $\omega \!\to\! \infty$.
This means that only those diagrams of $\gamma_p$ and $\gamma_a$ contribute
where the red legs directly meet at the same bare vertex.
Thereby, $\omega$ is transferred without entering an actual propagator,
and the result is completely independent of $\omega$.
We can gather all those diagrams by inserting the BSEs for the reducible vertices.
This is done in the second equality of Fig.~\ref{fig:WI_SDE}(b).
The first two and last two summands per row differ by the choice of spin
in the propagator loop on top of $\gamma_{a/p}$.
In the BSE for $\gamma_p^{\sigma\bar{\sigma}}$, we fixed the spin $\sigma$ at the bottom propagator, thus eliminating the prefactor $1/2$.

By virtue of the BSEs, we have two out-going red legs attached to $I_p$,
and an in- and an out-going red leg attached to the left of $I_a$.
Since $I_p$ and $I_a$ are irreducible in parallel and antiparallel lines, respectively,
the only diagram for each that allows the red legs to meet directly is the bare vertex $F_0$.
Furthermore, as $F_0$ is only nonzero between different spins, 
the result collapses to the three contributions
(without any spin summation) shown in Fig.~\ref{fig:WI_SDE}(c).
At this point, the red lines meet at a bare vertex, 
and the $\omega$ dependence (and thus the wiggly line) can be simply removed.
We see that the first two terms in Fig.~\ref{fig:WI_SDE}(c) cancel.
It remains to use the crossing symmetry of $F$ to transform the last summand
of Fig.~\ref{fig:WI_SDE}(c)
into the expression of Fig.~\ref{fig:WI_SDE}(d).
The latter is precisely the SDE (the Hartree term is absorbed in $\tilde{\Sigma}$)
in the form known from Eq.~\eqref{eq:SDE} and Fig.~\ref{fig:chi-vtx_SDE_WI}(b).

It is no coincidence that the first two summands of Fig.~\ref{fig:chi-vtx_SDE_WI}(c)
canceled, and the nonzero contribution to $\tilde{\Sigma}$ is the one 
from the equal-spin vertex $\gamma_a^{\sigma\sigma}$ coming from
$I_t^{\sigma\sigma}$. 
In fact, the WI~\eqref{eq:WI_natural-param} also holds without spin sum,
\begin{align}
\Sigma^\sigma_{\nu+\omega}
-
\Sigma^\sigma_\nu
& =
-
\frac{1}{\beta}
\sum_{\nu'}
I^{\sigma\sigma;\omega}_{t;\nu,\nu'}
( G^{\sigma}_{\nu'+\omega} - G^{\sigma}_{\nu'} )
.
\label{eq:WI_equal-spin}
\end{align}
For convenience, we check this explicitly at second-order in $U$ in Fig.~\ref{fig:WI_2ndOrder}.
Equation~\eqref{eq:WI_equal-spin} 
can be found by deriving the WI not only 
using the local charge operator,
$\hat{\rho}_1 = \sum_\sigma \hat{n}_\sigma$,
but also the local spin operator
$\hat{\rho}_2 = \sum_\sigma \tau^z_{\sigma\sigma} \hat{n}_\sigma$,
where $\tau^z$ is the third Pauli matrix.
For the latter, the WI reads
\begin{align}
\Sigma^\sigma_{\nu+\omega}
-
\Sigma^\sigma_\nu
& =
-
\frac{1}{\beta}
\sum_{\sigma'\nu'}
\tau^z_{\sigma'\sigma'}
I^{\sigma\sigma';\omega}_{t;\nu,\nu'}
( G^{\sigma'}_{\nu'+\omega} - G^{\sigma'}_{\nu'} )
.
\label{eq:WI_spin-op}
\end{align}
Summing Eqs.~\eqref{eq:WI_natural-param} and \eqref{eq:WI_spin-op},
one obtains
Eq.~\eqref{eq:WI_equal-spin}.

\subsection{Envelope diagrams in the WI}
\label{subsec:APP-WI-envelope}

\begin{figure}[t!]
\centering
\includegraphics[width=0.485\textwidth]{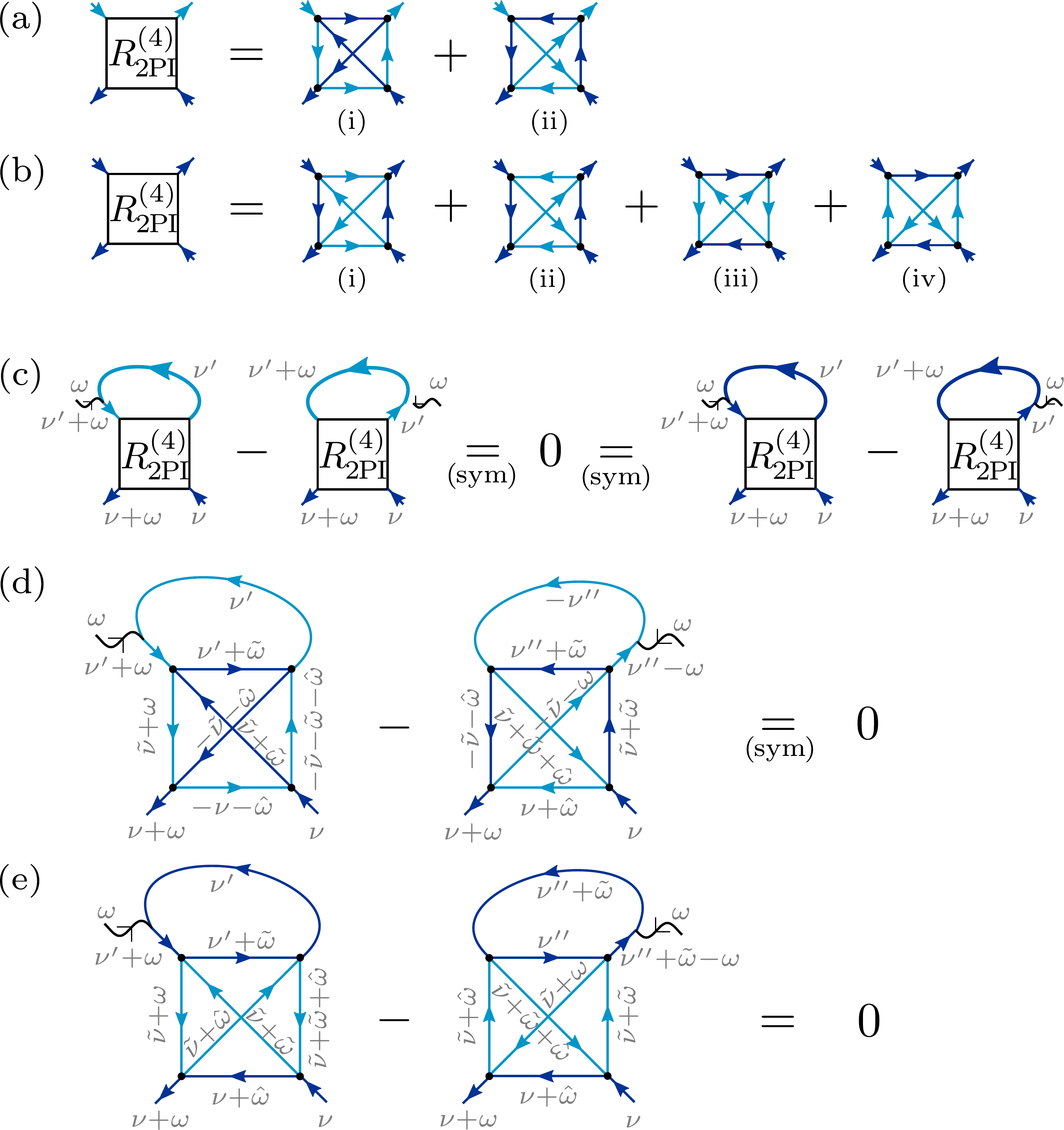}
\caption{%
$R_{\mathrm{2PI}}$ at order $U^4$ (envelope diagrams) with (a) different and (b) equal spins
on the external legs.
(c) Under particle-hole and spin symmetry, 
$R_{\mathrm{2PI}}^{(4)}$ does not contribute to the WI.
This is because pairs of diagrams cancel in the difference of Eq.~\eqref{eq:WI},
as apparent after a suitable transformation $\nu' \to \nu''$ in, say, the second term.
(d) Cancellation of diagrams (i) and (ii) from (a) inserted in the first and second term of the WI.
(e) Same for diagrams (iii) and (iv) from (b).}
\label{fig:diagr_envelope}
\end{figure}

The vertex in the PA deviates from the exact vertex starting at fourth order in the bare interaction $U$.
Through the SDE, the self-energy is exact up to order $U^4$, while errors start at order $U^5$.
\textit{A priori}, one thus expects the PA to violate the WI to order $U^4$,
as induced by the missing $U^4$ vertex diagrams---the so-called envelope diagrams.
However, it could also happen that this specific class of vertex diagrams does not contribute to the WI, i.e.,
that the envelope diagrams cancel out in Eq.~\eqref{eq:WI}.
Indeed, numerically, we found the PA to violate the WI to order $U^5$ instead of $U^4$.
In this section, we show analytically that 
in the special case of particle-hole and spin symmetry,
the envelope diagrams cancel in the WI.

Figures~\ref{fig:diagr_envelope}(a) and \ref{fig:diagr_envelope}(b) show the envelope diagrams
with different and equal spins (distinguished by light and dark colors) on the external legs, respectively.
(For brevity, we collapsed the interaction line to a dot.)
These diagrams can also be found in Figs.~14 and 15 of Ref.~\cite{Rohringer2012}.
Together, they form the fully irreducible vertex $R_{\mathrm{2PI}}$ at fourth order in $U$;
$R_{\mathrm{2PI}}^{(4)\sigma\bar{\sigma}}$ has two 
and $R_{\mathrm{2PI}}^{(4)\sigma\sigma}$ four diagrams,
as enumerated by Roman numbers.
Figure~\ref{fig:diagr_envelope}(c) states that
neither $R_{\mathrm{2PI}}^{(4)\sigma\bar{\sigma}}$ nor
$R_{\mathrm{2PI}}^{(4)\sigma\sigma}$ contribute to the WI
given particle-hole and spin symmetry.

Indeed, using $G^\sigma_\nu = - G^\sigma_{-\nu}$ and $G^\uparrow_\nu = G^\downarrow_\nu$,
one can always find pairs of envelope diagrams that cancel in the subtraction
inherent to the WI.
This cancellation becomes apparent after suitably transforming the summation frequency in,
say, the second term of the WI (thus changing $\nu'$ to $\nu''$).
In Figs.~\ref{fig:diagr_envelope}(d) and \ref{fig:diagr_envelope}(e), we establish the cancellation
by explicitly writing frequency labels on all internal lines.

Figure~\ref{fig:diagr_envelope}(d) considers the case of $R_{\mathrm{2PI}}^{(4)\sigma\bar{\sigma}}$
with diagram (i) and (ii) in the first and second term of the WI, respectively.
The same set of frequency labels occurs in both terms,
differing at most by minus signs. 
Both terms have a total of three global minus signs in their frequency labels;
using $G^\sigma_\nu = - G^\sigma_{-\nu}$, these minus signs can be pulled out of the equation.
One obtains a mathematically identical expression for both terms and thus a vanishing difference.
The case of $R_{\mathrm{2PI}}^{(4)\sigma\bar{\sigma}}$
with diagram (ii) in the first and diagram (i) in the second term proceeds analogously.
Indeed, one can transform one case into the other by flipping the arrows on 
the two horizontal and the two diagonal lines
(accordingly changing $\nu \to -\nu$ on their frequency labels)
and suitably changing the spin labels. The cancellation works just as before.
Further, the cancellation also works for diagrams (i) and (ii) of $R_{\mathrm{2PI}}^{(4)\sigma\sigma}$
(again for both orders). 
In this case, one must further invoke spin symmetry, $G^\uparrow_\nu = G^\downarrow_\nu$.

Finally, Fig.~\ref{fig:diagr_envelope}(e) treats the case of $R_{\mathrm{2PI}}^{(4)\sigma\sigma}$
with diagram (iii) and (iv) in the first and second term of the WI, respectively.
The argumentation is the same as before. 
Thanks to the transformation $\nu' \!\to\! \nu''$ in the second term, one has a
mathematically identical expression for both terms and thus a vanishing difference.
Again, interchanging the role of diagrams (iii) and (iv) in the two terms of the WI
merely amounts to flipping the arrows;
here, this affects the two vertical and the two diagonal lines, all of which have spin $\bar{\sigma}$.
As apparent from Fig.~\ref{fig:diagr_envelope}(e), no minus signs are involved,
and matching pairs of frequency labels also have the same spins.
The cancellation of diagrams (iii) and (iv) thus works also in the absence of particle-hole or spin symmetry.

\bibliography{RevisedManuscript}

\begin{thebibliography}{101}%
\makeatletter
\providecommand \@ifxundefined [1]{%
 \@ifx{#1\undefined}
}%
\providecommand \@ifnum [1]{%
 \ifnum #1\expandafter \@firstoftwo
 \else \expandafter \@secondoftwo
 \fi
}%
\providecommand \@ifx [1]{%
 \ifx #1\expandafter \@firstoftwo
 \else \expandafter \@secondoftwo
 \fi
}%
\providecommand \natexlab [1]{#1}%
\providecommand \enquote  [1]{``#1''}%
\providecommand \bibnamefont  [1]{#1}%
\providecommand \bibfnamefont [1]{#1}%
\providecommand \citenamefont [1]{#1}%
\providecommand \href@noop [0]{\@secondoftwo}%
\providecommand \href [0]{\begingroup \@sanitize@url \@href}%
\providecommand \@href[1]{\@@startlink{#1}\@@href}%
\providecommand \@@href[1]{\endgroup#1\@@endlink}%
\providecommand \@sanitize@url [0]{\catcode `\\12\catcode `\$12\catcode
  `\&12\catcode `\#12\catcode `\^12\catcode `\_12\catcode `\%12\relax}%
\providecommand \@@startlink[1]{}%
\providecommand \@@endlink[0]{}%
\providecommand \url  [0]{\begingroup\@sanitize@url \@url }%
\providecommand \@url [1]{\endgroup\@href {#1}{\urlprefix }}%
\providecommand \urlprefix  [0]{URL }%
\providecommand \Eprint [0]{\href }%
\providecommand \doibase [0]{https://doi.org/}%
\providecommand \selectlanguage [0]{\@gobble}%
\providecommand \bibinfo  [0]{\@secondoftwo}%
\providecommand \bibfield  [0]{\@secondoftwo}%
\providecommand \translation [1]{[#1]}%
\providecommand \BibitemOpen [0]{}%
\providecommand \bibitemStop [0]{}%
\providecommand \bibitemNoStop [0]{.\EOS\space}%
\providecommand \EOS [0]{\spacefactor3000\relax}%
\providecommand \BibitemShut  [1]{\csname bibitem#1\endcsname}%
\let\auto@bib@innerbib\@empty
%</preamble>
\bibitem [{\citenamefont {S\'en\'echal}\ \emph {et~al.}(2004)\citenamefont
  {S\'en\'echal}, \citenamefont {Tremblay},\ and\ \citenamefont
  {Bourbonnais}}]{Bickersbook2004}%
  \BibitemOpen
  \bibinfo {editor} {\bibfnamefont {D.}~\bibnamefont {S\'en\'echal}}, \bibinfo
  {editor} {\bibfnamefont {A.-M.}\ \bibnamefont {Tremblay}},\ and\ \bibinfo
  {editor} {\bibfnamefont {C.}~\bibnamefont {Bourbonnais}},\ eds.,\ \bibinfo
  {title} {Theoretical methods for strongly correlated electrons}\ (\bibinfo
  {publisher} {Springer-Verlag New York Berlin Heidelberg},\ \bibinfo {year}
  {2004})\ pp.\ \bibinfo {pages} {237--296}\BibitemShut {NoStop}%
\bibitem [{\citenamefont {Yang}\ \emph {et~al.}(2009)\citenamefont {Yang},
  \citenamefont {Fotso}, \citenamefont {Liu}, \citenamefont {Maier},
  \citenamefont {Tomko}, \citenamefont {D'Azevedo}, \citenamefont {Scalettar},
  \citenamefont {Pruschke},\ and\ \citenamefont {Jarrell}}]{Yang2009}%
  \BibitemOpen
  \bibfield  {author} {\bibinfo {author} {\bibfnamefont {S.~X.}\ \bibnamefont
  {Yang}}, \bibinfo {author} {\bibfnamefont {H.}~\bibnamefont {Fotso}},
  \bibinfo {author} {\bibfnamefont {J.}~\bibnamefont {Liu}}, \bibinfo {author}
  {\bibfnamefont {T.~A.}\ \bibnamefont {Maier}}, \bibinfo {author}
  {\bibfnamefont {K.}~\bibnamefont {Tomko}}, \bibinfo {author} {\bibfnamefont
  {E.~F.}\ \bibnamefont {D'Azevedo}}, \bibinfo {author} {\bibfnamefont {R.~T.}\
  \bibnamefont {Scalettar}}, \bibinfo {author} {\bibfnamefont {T.}~\bibnamefont
  {Pruschke}},\ and\ \bibinfo {author} {\bibfnamefont {M.}~\bibnamefont
  {Jarrell}},\ }\bibfield  {title} {\bibinfo {title} {Parquet approximation for
  the $4\ifmmode\times\else\texttimes\fi{}4$ {H}ubbard cluster},\ }\href
  {https://doi.org/10.1103/PhysRevE.80.046706} {\bibfield  {journal} {\bibinfo
  {journal} {Phys. Rev. E}\ }\textbf {\bibinfo {volume} {80}},\ \bibinfo
  {pages} {046706} (\bibinfo {year} {2009})}\BibitemShut {NoStop}%
\bibitem [{\citenamefont {Tam}\ \emph {et~al.}(2013)\citenamefont {Tam},
  \citenamefont {Fotso}, \citenamefont {Yang}, \citenamefont {Lee},
  \citenamefont {Moreno}, \citenamefont {Ramanujam},\ and\ \citenamefont
  {Jarrell}}]{Tam2013}%
  \BibitemOpen
  \bibfield  {author} {\bibinfo {author} {\bibfnamefont {K.-M.}\ \bibnamefont
  {Tam}}, \bibinfo {author} {\bibfnamefont {H.}~\bibnamefont {Fotso}}, \bibinfo
  {author} {\bibfnamefont {S.-X.}\ \bibnamefont {Yang}}, \bibinfo {author}
  {\bibfnamefont {T.-W.}\ \bibnamefont {Lee}}, \bibinfo {author} {\bibfnamefont
  {J.}~\bibnamefont {Moreno}}, \bibinfo {author} {\bibfnamefont
  {J.}~\bibnamefont {Ramanujam}},\ and\ \bibinfo {author} {\bibfnamefont
  {M.}~\bibnamefont {Jarrell}},\ }\bibfield  {title} {\bibinfo {title} {Solving
  the parquet equations for the {H}ubbard model beyond weak coupling},\ }\href
  {https://doi.org/10.1103/PhysRevE.87.013311} {\bibfield  {journal} {\bibinfo
  {journal} {Phys. Rev. E}\ }\textbf {\bibinfo {volume} {87}},\ \bibinfo
  {pages} {013311} (\bibinfo {year} {2013})}\BibitemShut {NoStop}%
\bibitem [{\citenamefont {Valli}\ \emph {et~al.}(2015)\citenamefont {Valli},
  \citenamefont {Sch\"afer}, \citenamefont {Thunstr\"om}, \citenamefont
  {Rohringer}, \citenamefont {Andergassen}, \citenamefont {Sangiovanni},
  \citenamefont {Held},\ and\ \citenamefont {Toschi}}]{Valli2015}%
  \BibitemOpen
  \bibfield  {author} {\bibinfo {author} {\bibfnamefont {A.}~\bibnamefont
  {Valli}}, \bibinfo {author} {\bibfnamefont {T.}~\bibnamefont {Sch\"afer}},
  \bibinfo {author} {\bibfnamefont {P.}~\bibnamefont {Thunstr\"om}}, \bibinfo
  {author} {\bibfnamefont {G.}~\bibnamefont {Rohringer}}, \bibinfo {author}
  {\bibfnamefont {S.}~\bibnamefont {Andergassen}}, \bibinfo {author}
  {\bibfnamefont {G.}~\bibnamefont {Sangiovanni}}, \bibinfo {author}
  {\bibfnamefont {K.}~\bibnamefont {Held}},\ and\ \bibinfo {author}
  {\bibfnamefont {A.}~\bibnamefont {Toschi}},\ }\bibfield  {title} {\bibinfo
  {title} {Dynamical vertex approximation in its parquet implementation:
  Application to {H}ubbard nanorings},\ }\href
  {https://doi.org/10.1103/PhysRevB.91.115115} {\bibfield  {journal} {\bibinfo
  {journal} {Phys. Rev. B}\ }\textbf {\bibinfo {volume} {91}},\ \bibinfo
  {pages} {115115} (\bibinfo {year} {2015})}\BibitemShut {NoStop}%
\bibitem [{\citenamefont {Li}\ \emph {et~al.}(2016)\citenamefont {Li},
  \citenamefont {Wentzell}, \citenamefont {Pudleiner}, \citenamefont
  {Thunstr\"om},\ and\ \citenamefont {Held}}]{Li2016}%
  \BibitemOpen
  \bibfield  {author} {\bibinfo {author} {\bibfnamefont {G.}~\bibnamefont
  {Li}}, \bibinfo {author} {\bibfnamefont {N.}~\bibnamefont {Wentzell}},
  \bibinfo {author} {\bibfnamefont {P.}~\bibnamefont {Pudleiner}}, \bibinfo
  {author} {\bibfnamefont {P.}~\bibnamefont {Thunstr\"om}},\ and\ \bibinfo
  {author} {\bibfnamefont {K.}~\bibnamefont {Held}},\ }\bibfield  {title}
  {\bibinfo {title} {Efficient implementation of the parquet equations: Role of
  the reducible vertex function and its kernel approximation},\ }\href
  {https://doi.org/10.1103/PhysRevB.93.165103} {\bibfield  {journal} {\bibinfo
  {journal} {Phys. Rev. B}\ }\textbf {\bibinfo {volume} {93}},\ \bibinfo
  {pages} {165103} (\bibinfo {year} {2016})}\BibitemShut {NoStop}%
\bibitem [{\citenamefont {Wentzell}\ \emph {et~al.}(2020)\citenamefont
  {Wentzell}, \citenamefont {Li}, \citenamefont {Tagliavini}, \citenamefont
  {Taranto}, \citenamefont {Rohringer}, \citenamefont {Held}, \citenamefont
  {Toschi},\ and\ \citenamefont {Andergassen}}]{Wentzell2020}%
  \BibitemOpen
  \bibfield  {author} {\bibinfo {author} {\bibfnamefont {N.}~\bibnamefont
  {Wentzell}}, \bibinfo {author} {\bibfnamefont {G.}~\bibnamefont {Li}},
  \bibinfo {author} {\bibfnamefont {A.}~\bibnamefont {Tagliavini}}, \bibinfo
  {author} {\bibfnamefont {C.}~\bibnamefont {Taranto}}, \bibinfo {author}
  {\bibfnamefont {G.}~\bibnamefont {Rohringer}}, \bibinfo {author}
  {\bibfnamefont {K.}~\bibnamefont {Held}}, \bibinfo {author} {\bibfnamefont
  {A.}~\bibnamefont {Toschi}},\ and\ \bibinfo {author} {\bibfnamefont
  {S.}~\bibnamefont {Andergassen}},\ }\bibfield  {title} {\bibinfo {title}
  {High-frequency asymptotics of the vertex function: Diagrammatic
  parametrization and algorithmic implementation},\ }\href
  {https://doi.org/10.1103/PhysRevB.102.085106} {\bibfield  {journal} {\bibinfo
   {journal} {Phys. Rev. B}\ }\textbf {\bibinfo {volume} {102}},\ \bibinfo
  {pages} {085106} (\bibinfo {year} {2020})}\BibitemShut {NoStop}%
\bibitem [{\citenamefont {Smith}(1992)}]{Smith1992}%
  \BibitemOpen
  \bibfield  {author} {\bibinfo {author} {\bibfnamefont {R.~A.}\ \bibnamefont
  {Smith}},\ }\bibfield  {title} {\bibinfo {title} {Planar version of
  {B}aym-{K}adanoff theory},\ }\href {https://doi.org/10.1103/PhysRevA.46.4586}
  {\bibfield  {journal} {\bibinfo  {journal} {Phys. Rev. A}\ }\textbf {\bibinfo
  {volume} {46}},\ \bibinfo {pages} {4586} (\bibinfo {year}
  {1992})}\BibitemShut {NoStop}%
\bibitem [{\citenamefont {Jani\v{s}}\ \emph {et~al.}(2017)\citenamefont
  {Jani\v{s}}, \citenamefont {Kauch},\ and\ \citenamefont
  {Pokorn\'y}}]{Janis2017}%
  \BibitemOpen
  \bibfield  {author} {\bibinfo {author} {\bibfnamefont {V.}~\bibnamefont
  {Jani\v{s}}}, \bibinfo {author} {\bibfnamefont {A.}~\bibnamefont {Kauch}},\
  and\ \bibinfo {author} {\bibfnamefont {V.}~\bibnamefont {Pokorn\'y}},\
  }\bibfield  {title} {\bibinfo {title} {Thermodynamically consistent
  description of criticality in models of correlated electrons},\ }\href
  {https://doi.org/10.1103/PhysRevB.95.045108} {\bibfield  {journal} {\bibinfo
  {journal} {Phys. Rev. B}\ }\textbf {\bibinfo {volume} {95}},\ \bibinfo
  {pages} {045108} (\bibinfo {year} {2017})}\BibitemShut {NoStop}%
\bibitem [{\citenamefont {Kugler}\ and\ \citenamefont {von
  Delft}(2018{\natexlab{a}})}]{Kugler2018b}%
  \BibitemOpen
  \bibfield  {author} {\bibinfo {author} {\bibfnamefont {F.~B.}\ \bibnamefont
  {Kugler}}\ and\ \bibinfo {author} {\bibfnamefont {J.}~\bibnamefont {von
  Delft}},\ }\bibfield  {title} {\bibinfo {title} {Derivation of exact flow
  equations from the self-consistent parquet relations},\ }\href
  {https://doi.org/10.1088/1367-2630/aaf65f} {\bibfield  {journal} {\bibinfo
  {journal} {New J. Phys.}\ }\textbf {\bibinfo {volume} {20}},\ \bibinfo
  {pages} {123029} (\bibinfo {year} {2018}{\natexlab{a}})}\BibitemShut
  {NoStop}%
\bibitem [{\citenamefont {Metzner}\ \emph {et~al.}(2012)\citenamefont
  {Metzner}, \citenamefont {Salmhofer}, \citenamefont {Honerkamp},
  \citenamefont {Meden},\ and\ \citenamefont {Sch\"onhammer}}]{Metzner2012}%
  \BibitemOpen
  \bibfield  {author} {\bibinfo {author} {\bibfnamefont {W.}~\bibnamefont
  {Metzner}}, \bibinfo {author} {\bibfnamefont {M.}~\bibnamefont {Salmhofer}},
  \bibinfo {author} {\bibfnamefont {C.}~\bibnamefont {Honerkamp}}, \bibinfo
  {author} {\bibfnamefont {V.}~\bibnamefont {Meden}},\ and\ \bibinfo {author}
  {\bibfnamefont {K.}~\bibnamefont {Sch\"onhammer}},\ }\bibfield  {title}
  {\bibinfo {title} {Functional renormalization group approach to correlated
  fermion systems},\ }\href {https://doi.org/10.1103/RevModPhys.84.299}
  {\bibfield  {journal} {\bibinfo  {journal} {Rev. Mod. Phys.}\ }\textbf
  {\bibinfo {volume} {84}},\ \bibinfo {pages} {299} (\bibinfo {year}
  {2012})}\BibitemShut {NoStop}%
\bibitem [{\citenamefont {Salmhofer}(1999)}]{Salmhofer1999}%
  \BibitemOpen
  \bibfield  {author} {\bibinfo {author} {\bibfnamefont {M.}~\bibnamefont
  {Salmhofer}},\ }\href@noop {} {\emph {\bibinfo {title} {Renormalization - An
  Introduction}}},\ edited by\ \bibinfo {editor} {\bibfnamefont
  {R.}~\bibnamefont {Balian}}, \bibinfo {editor} {\bibfnamefont
  {W.}~\bibnamefont {Beiglb\"ock}}, \bibinfo {editor} {\bibfnamefont
  {H.}~\bibnamefont {Grosse}}, \bibinfo {editor} {\bibfnamefont {E.~H.}\
  \bibnamefont {Lieb}}, \bibinfo {editor} {\bibfnamefont {N.}~\bibnamefont
  {Reshetikhin}}, \bibinfo {editor} {\bibfnamefont {H.}~\bibnamefont {Spohn}},\
  and\ \bibinfo {editor} {\bibfnamefont {W.}~\bibnamefont {Thirring}}\
  (\bibinfo  {publisher} {Springer-Verlag Berlin Heidelberg},\ \bibinfo {year}
  {1999})\BibitemShut {NoStop}%
\bibitem [{\citenamefont {Berges}\ \emph {et~al.}(2002)\citenamefont {Berges},
  \citenamefont {Tetradis},\ and\ \citenamefont {Wetterich}}]{Berges2002}%
  \BibitemOpen
  \bibfield  {author} {\bibinfo {author} {\bibfnamefont {J.}~\bibnamefont
  {Berges}}, \bibinfo {author} {\bibfnamefont {N.}~\bibnamefont {Tetradis}},\
  and\ \bibinfo {author} {\bibfnamefont {C.}~\bibnamefont {Wetterich}},\
  }\bibfield  {title} {\bibinfo {title} {Non-perturbative renormalization flow
  in quantum field theory and statistical physics},\ }\href
  {https://doi.org/10.1016/S0370-1573(01)00098-9} {\bibfield  {journal}
  {\bibinfo  {journal} {Phys. Rep.}\ }\textbf {\bibinfo {volume} {363}},\
  \bibinfo {pages} {223 } (\bibinfo {year} {2002})}\BibitemShut {NoStop}%
\bibitem [{\citenamefont {Kopietz}\ \emph
  {et~al.}(2010{\natexlab{a}})\citenamefont {Kopietz}, \citenamefont
  {Bartosch},\ and\ \citenamefont {Sch\"{u}tz}}]{Kopietz2010}%
  \BibitemOpen
  \bibfield  {author} {\bibinfo {author} {\bibfnamefont {P.}~\bibnamefont
  {Kopietz}}, \bibinfo {author} {\bibfnamefont {L.}~\bibnamefont {Bartosch}},\
  and\ \bibinfo {author} {\bibfnamefont {F.}~\bibnamefont {Sch\"{u}tz}},\
  }\href {https://doi.org/10.1007/978-3-642-05094-7} {\emph {\bibinfo {title}
  {Introduction to the Functional Renormalization Group}}}\ (\bibinfo
  {publisher} {Springer Berlin Heidelberg},\ \bibinfo {year}
  {2010})\BibitemShut {NoStop}%
\bibitem [{\citenamefont {Dupuis}\ \emph {et~al.}(2021)\citenamefont {Dupuis},
  \citenamefont {Canet}, \citenamefont {Eichhorn}, \citenamefont {Metzner},
  \citenamefont {Pawlowski}, \citenamefont {Tissier},\ and\ \citenamefont
  {Wschebor}}]{Dupuis2021}%
  \BibitemOpen
  \bibfield  {author} {\bibinfo {author} {\bibfnamefont {N.}~\bibnamefont
  {Dupuis}}, \bibinfo {author} {\bibfnamefont {L.}~\bibnamefont {Canet}},
  \bibinfo {author} {\bibfnamefont {A.}~\bibnamefont {Eichhorn}}, \bibinfo
  {author} {\bibfnamefont {W.}~\bibnamefont {Metzner}}, \bibinfo {author}
  {\bibfnamefont {J.}~\bibnamefont {Pawlowski}}, \bibinfo {author}
  {\bibfnamefont {M.}~\bibnamefont {Tissier}},\ and\ \bibinfo {author}
  {\bibfnamefont {N.}~\bibnamefont {Wschebor}},\ }\bibfield  {title} {\bibinfo
  {title} {The nonperturbative functional renormalization group and its
  applications},\ }\href {https://doi.org/10.1016/j.physrep.2021.01.001}
  {\bibfield  {journal} {\bibinfo  {journal} {Phys. Rep.}\ }\textbf {\bibinfo
  {volume} {910}},\ \bibinfo {pages} {1} (\bibinfo {year} {2021})}\BibitemShut
  {NoStop}%
\bibitem [{\citenamefont {Kugler}\ and\ \citenamefont {von
  Delft}(2018{\natexlab{b}})}]{Kugler2018}%
  \BibitemOpen
  \bibfield  {author} {\bibinfo {author} {\bibfnamefont {F.~B.}\ \bibnamefont
  {Kugler}}\ and\ \bibinfo {author} {\bibfnamefont {J.}~\bibnamefont {von
  Delft}},\ }\bibfield  {title} {\bibinfo {title} {Multiloop functional
  renormalization group for general models},\ }\href
  {https://doi.org/10.1103/PhysRevB.97.035162} {\bibfield  {journal} {\bibinfo
  {journal} {Phys. Rev. B}\ }\textbf {\bibinfo {volume} {97}},\ \bibinfo
  {pages} {035162} (\bibinfo {year} {2018}{\natexlab{b}})}\BibitemShut
  {NoStop}%
\bibitem [{\citenamefont {Kugler}\ and\ \citenamefont {von
  Delft}(2018{\natexlab{c}})}]{Kugler2018a}%
  \BibitemOpen
  \bibfield  {author} {\bibinfo {author} {\bibfnamefont {F.~B.}\ \bibnamefont
  {Kugler}}\ and\ \bibinfo {author} {\bibfnamefont {J.}~\bibnamefont {von
  Delft}},\ }\bibfield  {title} {\bibinfo {title} {Multiloop functional
  renormalization group that sums up all parquet diagrams},\ }\href
  {https://doi.org/10.1103/PhysRevLett.120.057403} {\bibfield  {journal}
  {\bibinfo  {journal} {Phys. Rev. Lett.}\ }\textbf {\bibinfo {volume} {120}},\
  \bibinfo {pages} {057403} (\bibinfo {year} {2018}{\natexlab{c}})}\BibitemShut
  {NoStop}%
\bibitem [{\citenamefont {Katanin}(2004)}]{Katanin2004}%
  \BibitemOpen
  \bibfield  {author} {\bibinfo {author} {\bibfnamefont {A.~A.}\ \bibnamefont
  {Katanin}},\ }\bibfield  {title} {\bibinfo {title} {Fulfillment of ward
  identities in the functional renormalization group approach},\ }\href
  {https://doi.org/10.1103/PhysRevB.70.115109} {\bibfield  {journal} {\bibinfo
  {journal} {Phys. Rev. B}\ }\textbf {\bibinfo {volume} {70}},\ \bibinfo
  {pages} {115109} (\bibinfo {year} {2004})}\BibitemShut {NoStop}%
\bibitem [{\citenamefont {Enss}(2016)}]{EnssThesis}%
  \BibitemOpen
  \bibfield  {author} {\bibinfo {author} {\bibfnamefont {T.}~\bibnamefont
  {Enss}},\ }\emph {\bibinfo {title} {Renormalization, Conservation Laws and
  Transport in Correlated Electron Systems}},\ \href@noop {} {Ph.D. thesis},\
  \bibinfo  {school} {University of Stuttgart} (\bibinfo {year}
  {2016})\BibitemShut {NoStop}%
\bibitem [{\citenamefont {Veschgini}\ and\ \citenamefont
  {Salmhofer}(2013)}]{Veschgini2013}%
  \BibitemOpen
  \bibfield  {author} {\bibinfo {author} {\bibfnamefont {K.}~\bibnamefont
  {Veschgini}}\ and\ \bibinfo {author} {\bibfnamefont {M.}~\bibnamefont
  {Salmhofer}},\ }\bibfield  {title} {\bibinfo {title} {{S}chwinger-{D}yson
  renormalization group},\ }\href {https://doi.org/10.1103/PhysRevB.88.155131}
  {\bibfield  {journal} {\bibinfo  {journal} {Phys. Rev. B}\ }\textbf {\bibinfo
  {volume} {88}},\ \bibinfo {pages} {155131} (\bibinfo {year}
  {2013})}\BibitemShut {NoStop}%
\bibitem [{\citenamefont {Caltapanides}\ \emph {et~al.}(2021)\citenamefont
  {Caltapanides}, \citenamefont {Kennes},\ and\ \citenamefont
  {Meden}}]{Caltapanides2021}%
  \BibitemOpen
  \bibfield  {author} {\bibinfo {author} {\bibfnamefont {M.}~\bibnamefont
  {Caltapanides}}, \bibinfo {author} {\bibfnamefont {D.~M.}\ \bibnamefont
  {Kennes}},\ and\ \bibinfo {author} {\bibfnamefont {V.}~\bibnamefont
  {Meden}},\ }\bibfield  {title} {\bibinfo {title} {Finite-bias transport
  through the interacting resonant level model coupled to a phonon mode: A
  functional renormalization group study},\ }\href
  {https://doi.org/10.1103/PhysRevB.104.085125} {\bibfield  {journal} {\bibinfo
   {journal} {Phys. Rev. B}\ }\textbf {\bibinfo {volume} {104}},\ \bibinfo
  {pages} {085125} (\bibinfo {year} {2021})}\BibitemShut {NoStop}%
\bibitem [{\citenamefont {Sch\"utz}\ \emph {et~al.}(2005)\citenamefont
  {Sch\"utz}, \citenamefont {Bartosch},\ and\ \citenamefont
  {Kopietz}}]{Schuetz2005}%
  \BibitemOpen
  \bibfield  {author} {\bibinfo {author} {\bibfnamefont {F.}~\bibnamefont
  {Sch\"utz}}, \bibinfo {author} {\bibfnamefont {L.}~\bibnamefont {Bartosch}},\
  and\ \bibinfo {author} {\bibfnamefont {P.}~\bibnamefont {Kopietz}},\
  }\bibfield  {title} {\bibinfo {title} {Collective fields in the functional
  renormalization group for fermions, {W}ard identities, and the exact solution
  of the {T}omonaga-{L}uttinger model},\ }\href
  {https://doi.org/10.1103/PhysRevB.72.035107} {\bibfield  {journal} {\bibinfo
  {journal} {Phys. Rev. B}\ }\textbf {\bibinfo {volume} {72}},\ \bibinfo
  {pages} {035107} (\bibinfo {year} {2005})}\BibitemShut {NoStop}%
\bibitem [{\citenamefont {Bartosch}\ \emph {et~al.}(2009)\citenamefont
  {Bartosch}, \citenamefont {Freire}, \citenamefont {Cardenas},\ and\
  \citenamefont {Kopietz}}]{Bartosch2009}%
  \BibitemOpen
  \bibfield  {author} {\bibinfo {author} {\bibfnamefont {L.}~\bibnamefont
  {Bartosch}}, \bibinfo {author} {\bibfnamefont {H.}~\bibnamefont {Freire}},
  \bibinfo {author} {\bibfnamefont {J.~J.~R.}\ \bibnamefont {Cardenas}},\ and\
  \bibinfo {author} {\bibfnamefont {P.}~\bibnamefont {Kopietz}},\ }\bibfield
  {title} {\bibinfo {title} {A functional renormalization group approach to the
  {A}nderson impurity model},\ }\href
  {https://doi.org/10.1088/0953-8984/21/30/305602} {\bibfield  {journal}
  {\bibinfo  {journal} {J. Phys. Condens. Matter}\ }\textbf {\bibinfo {volume}
  {21}},\ \bibinfo {pages} {305602} (\bibinfo {year} {2009})}\BibitemShut
  {NoStop}%
\bibitem [{\citenamefont {Streib}\ \emph {et~al.}(2013)\citenamefont {Streib},
  \citenamefont {Isidori},\ and\ \citenamefont {Kopietz}}]{Streib2013}%
  \BibitemOpen
  \bibfield  {author} {\bibinfo {author} {\bibfnamefont {S.}~\bibnamefont
  {Streib}}, \bibinfo {author} {\bibfnamefont {A.}~\bibnamefont {Isidori}},\
  and\ \bibinfo {author} {\bibfnamefont {P.}~\bibnamefont {Kopietz}},\
  }\bibfield  {title} {\bibinfo {title} {Solution of the {A}nderson impurity
  model via the functional renormalization group},\ }\href
  {https://doi.org/10.1103/PhysRevB.87.201107} {\bibfield  {journal} {\bibinfo
  {journal} {Phys. Rev. B}\ }\textbf {\bibinfo {volume} {87}},\ \bibinfo
  {pages} {201107(R)} (\bibinfo {year} {2013})}\BibitemShut {NoStop}%
\bibitem [{\citenamefont {Diekmann}\ and\ \citenamefont
  {Jakobs}(2021)}]{Diekmann2020}%
  \BibitemOpen
  \bibfield  {author} {\bibinfo {author} {\bibfnamefont {J.}~\bibnamefont
  {Diekmann}}\ and\ \bibinfo {author} {\bibfnamefont {S.~G.}\ \bibnamefont
  {Jakobs}},\ }\bibfield  {title} {\bibinfo {title} {Parquet approximation and
  one-loop renormalization group: Equivalence on the leading-logarithmic
  level},\ }\href {https://doi.org/10.1103/PhysRevB.103.155156} {\bibfield
  {journal} {\bibinfo  {journal} {Phys. Rev. B}\ }\textbf {\bibinfo {volume}
  {103}},\ \bibinfo {pages} {155156} (\bibinfo {year} {2021})}\BibitemShut
  {NoStop}%
\bibitem [{\citenamefont {Karrasch}\ \emph
  {et~al.}(2010{\natexlab{a}})\citenamefont {Karrasch}, \citenamefont
  {Pletyukhov}, \citenamefont {Borda},\ and\ \citenamefont
  {Meden}}]{Karrasch2010a}%
  \BibitemOpen
  \bibfield  {author} {\bibinfo {author} {\bibfnamefont {C.}~\bibnamefont
  {Karrasch}}, \bibinfo {author} {\bibfnamefont {M.}~\bibnamefont
  {Pletyukhov}}, \bibinfo {author} {\bibfnamefont {L.}~\bibnamefont {Borda}},\
  and\ \bibinfo {author} {\bibfnamefont {V.}~\bibnamefont {Meden}},\ }\bibfield
   {title} {\bibinfo {title} {Functional renormalization group study of the
  interacting resonant level model in and out of equilibrium},\ }\href
  {https://doi.org/10.1103/PhysRevB.81.125122} {\bibfield  {journal} {\bibinfo
  {journal} {Phys. Rev. B}\ }\textbf {\bibinfo {volume} {81}},\ \bibinfo
  {pages} {125122} (\bibinfo {year} {2010}{\natexlab{a}})}\BibitemShut
  {NoStop}%
\bibitem [{\citenamefont {Karrasch}\ \emph
  {et~al.}(2010{\natexlab{b}})\citenamefont {Karrasch}, \citenamefont
  {Andergassen}, \citenamefont {Pletyukhov}, \citenamefont {Schuricht},
  \citenamefont {Borda}, \citenamefont {Meden},\ and\ \citenamefont
  {Schoeller}}]{Karrasch2010b}%
  \BibitemOpen
  \bibfield  {author} {\bibinfo {author} {\bibfnamefont {C.}~\bibnamefont
  {Karrasch}}, \bibinfo {author} {\bibfnamefont {S.}~\bibnamefont
  {Andergassen}}, \bibinfo {author} {\bibfnamefont {M.}~\bibnamefont
  {Pletyukhov}}, \bibinfo {author} {\bibfnamefont {D.}~\bibnamefont
  {Schuricht}}, \bibinfo {author} {\bibfnamefont {L.}~\bibnamefont {Borda}},
  \bibinfo {author} {\bibfnamefont {V.}~\bibnamefont {Meden}},\ and\ \bibinfo
  {author} {\bibfnamefont {H.}~\bibnamefont {Schoeller}},\ }\bibfield  {title}
  {\bibinfo {title} {Non-equilibrium current and relaxation dynamics of a
  charge-fluctuating quantum dot},\ }\href
  {https://iopscience.iop.org/article/10.1209/0295-5075/90/30003} {\bibfield
  {journal} {\bibinfo  {journal} {Europhys. Lett.}\ }\textbf {\bibinfo {volume}
  {90}},\ \bibinfo {pages} {30003} (\bibinfo {year}
  {2010}{\natexlab{b}})}\BibitemShut {NoStop}%
\bibitem [{\citenamefont {Kennes}\ and\ \citenamefont
  {Meden}(2013)}]{Kennes2013}%
  \BibitemOpen
  \bibfield  {author} {\bibinfo {author} {\bibfnamefont {D.~M.}\ \bibnamefont
  {Kennes}}\ and\ \bibinfo {author} {\bibfnamefont {V.}~\bibnamefont {Meden}},\
  }\bibfield  {title} {\bibinfo {title} {Interacting resonant-level model in
  nonequilibrium: Finite-temperature effects},\ }\href
  {https://doi.org/10.1103/PhysRevB.87.075130} {\bibfield  {journal} {\bibinfo
  {journal} {Phys. Rev. B}\ }\textbf {\bibinfo {volume} {87}},\ \bibinfo
  {pages} {075130} (\bibinfo {year} {2013})}\BibitemShut {NoStop}%
\bibitem [{\citenamefont {Kennes}\ \emph {et~al.}(2013)\citenamefont {Kennes},
  \citenamefont {Schuricht},\ and\ \citenamefont {Meden}}]{Kennes2013a}%
  \BibitemOpen
  \bibfield  {author} {\bibinfo {author} {\bibfnamefont {D.~M.}\ \bibnamefont
  {Kennes}}, \bibinfo {author} {\bibfnamefont {D.}~\bibnamefont {Schuricht}},\
  and\ \bibinfo {author} {\bibfnamefont {V.}~\bibnamefont {Meden}},\ }\bibfield
   {title} {\bibinfo {title} {Efficiency and power of a thermoelectric quantum
  dot device},\ }\href {https://doi.org/10.1209/0295-5075/102/57003} {\bibfield
   {journal} {\bibinfo  {journal} {Europhys. Lett.}\ }\textbf {\bibinfo
  {volume} {102}},\ \bibinfo {pages} {57003} (\bibinfo {year}
  {2013})}\BibitemShut {NoStop}%
\bibitem [{\citenamefont {Meden}\ \emph {et~al.}(2002)\citenamefont {Meden},
  \citenamefont {Metzner}, \citenamefont {Schollw\"ock},\ and\ \citenamefont
  {Sch\"onhammer}}]{Meden2002}%
  \BibitemOpen
  \bibfield  {author} {\bibinfo {author} {\bibfnamefont {V.}~\bibnamefont
  {Meden}}, \bibinfo {author} {\bibfnamefont {W.}~\bibnamefont {Metzner}},
  \bibinfo {author} {\bibfnamefont {U.}~\bibnamefont {Schollw\"ock}},\ and\
  \bibinfo {author} {\bibfnamefont {K.}~\bibnamefont {Sch\"onhammer}},\
  }\bibfield  {title} {\bibinfo {title} {Scaling behavior of impurities in
  mesoscopic {L}uttinger liquids},\ }\href
  {https://doi.org/10.1103/PhysRevB.65.045318} {\bibfield  {journal} {\bibinfo
  {journal} {Phys. Rev. B}\ }\textbf {\bibinfo {volume} {65}},\ \bibinfo
  {pages} {045318} (\bibinfo {year} {2002})}\BibitemShut {NoStop}%
\bibitem [{\citenamefont {Meden}\ \emph {et~al.}(2003)\citenamefont {Meden},
  \citenamefont {Andergassen}, \citenamefont {Metzner}, \citenamefont
  {Schollw\"ock},\ and\ \citenamefont {Sch\"onhammer}}]{Meden2003}%
  \BibitemOpen
  \bibfield  {author} {\bibinfo {author} {\bibfnamefont {V.}~\bibnamefont
  {Meden}}, \bibinfo {author} {\bibfnamefont {S.}~\bibnamefont {Andergassen}},
  \bibinfo {author} {\bibfnamefont {W.}~\bibnamefont {Metzner}}, \bibinfo
  {author} {\bibfnamefont {U.}~\bibnamefont {Schollw\"ock}},\ and\ \bibinfo
  {author} {\bibfnamefont {K.}~\bibnamefont {Sch\"onhammer}},\ }\bibfield
  {title} {\bibinfo {title} {Scaling of the conductance in a quantum wire},\
  }\href {https://doi.org/10.1209/epl/i2003-00624-x} {\bibfield  {journal}
  {\bibinfo  {journal} {Europhysics Letters (EPL)}\ }\textbf {\bibinfo {volume}
  {64}},\ \bibinfo {pages} {769} (\bibinfo {year} {2003})}\BibitemShut
  {NoStop}%
\bibitem [{\citenamefont {Andergassen}\ \emph {et~al.}(2004)\citenamefont
  {Andergassen}, \citenamefont {Enss}, \citenamefont {Meden}, \citenamefont
  {Metzner}, \citenamefont {Schollw\"ock},\ and\ \citenamefont
  {Sch\"onhammer}}]{Andergassen2004}%
  \BibitemOpen
  \bibfield  {author} {\bibinfo {author} {\bibfnamefont {S.}~\bibnamefont
  {Andergassen}}, \bibinfo {author} {\bibfnamefont {T.}~\bibnamefont {Enss}},
  \bibinfo {author} {\bibfnamefont {V.}~\bibnamefont {Meden}}, \bibinfo
  {author} {\bibfnamefont {W.}~\bibnamefont {Metzner}}, \bibinfo {author}
  {\bibfnamefont {U.}~\bibnamefont {Schollw\"ock}},\ and\ \bibinfo {author}
  {\bibfnamefont {K.}~\bibnamefont {Sch\"onhammer}},\ }\bibfield  {title}
  {\bibinfo {title} {Functional renormalization group for {L}uttinger liquids
  with impurities},\ }\href {https://doi.org/10.1103/PhysRevB.70.075102}
  {\bibfield  {journal} {\bibinfo  {journal} {Phys. Rev. B}\ }\textbf {\bibinfo
  {volume} {70}},\ \bibinfo {pages} {075102} (\bibinfo {year}
  {2004})}\BibitemShut {NoStop}%
\bibitem [{\citenamefont {Meden}\ \emph {et~al.}(2005)\citenamefont {Meden},
  \citenamefont {Enss}, \citenamefont {Andergassen}, \citenamefont {Metzner},\
  and\ \citenamefont {Sch\"onhammer}}]{Meden2005}%
  \BibitemOpen
  \bibfield  {author} {\bibinfo {author} {\bibfnamefont {V.}~\bibnamefont
  {Meden}}, \bibinfo {author} {\bibfnamefont {T.}~\bibnamefont {Enss}},
  \bibinfo {author} {\bibfnamefont {S.}~\bibnamefont {Andergassen}}, \bibinfo
  {author} {\bibfnamefont {W.}~\bibnamefont {Metzner}},\ and\ \bibinfo {author}
  {\bibfnamefont {K.}~\bibnamefont {Sch\"onhammer}},\ }\bibfield  {title}
  {\bibinfo {title} {Correlation effects on resonant tunneling in
  one-dimensional quantum wires},\ }\href
  {https://doi.org/10.1103/PhysRevB.71.041302} {\bibfield  {journal} {\bibinfo
  {journal} {Phys. Rev. B}\ }\textbf {\bibinfo {volume} {71}},\ \bibinfo
  {pages} {041302(R)} (\bibinfo {year} {2005})}\BibitemShut {NoStop}%
\bibitem [{\citenamefont {Enss}\ \emph {et~al.}(2005)\citenamefont {Enss},
  \citenamefont {Meden}, \citenamefont {Andergassen}, \citenamefont
  {Barnab\'e-Th\'eriault}, \citenamefont {Metzner},\ and\ \citenamefont
  {Sch\"onhammer}}]{Enss2005}%
  \BibitemOpen
  \bibfield  {author} {\bibinfo {author} {\bibfnamefont {T.}~\bibnamefont
  {Enss}}, \bibinfo {author} {\bibfnamefont {V.}~\bibnamefont {Meden}},
  \bibinfo {author} {\bibfnamefont {S.}~\bibnamefont {Andergassen}}, \bibinfo
  {author} {\bibfnamefont {X.}~\bibnamefont {Barnab\'e-Th\'eriault}}, \bibinfo
  {author} {\bibfnamefont {W.}~\bibnamefont {Metzner}},\ and\ \bibinfo {author}
  {\bibfnamefont {K.}~\bibnamefont {Sch\"onhammer}},\ }\bibfield  {title}
  {\bibinfo {title} {Impurity and correlation effects on transport in
  one-dimensional quantum wires},\ }\href
  {https://doi.org/10.1103/PhysRevB.71.155401} {\bibfield  {journal} {\bibinfo
  {journal} {Phys. Rev. B}\ }\textbf {\bibinfo {volume} {71}},\ \bibinfo
  {pages} {155401} (\bibinfo {year} {2005})}\BibitemShut {NoStop}%
\bibitem [{\citenamefont {Andergassen}\ \emph {et~al.}(2006)\citenamefont
  {Andergassen}, \citenamefont {Enss}, \citenamefont {Meden}, \citenamefont
  {Metzner}, \citenamefont {Schollw\"ock},\ and\ \citenamefont
  {Sch\"onhammer}}]{Andergassen2006a}%
  \BibitemOpen
  \bibfield  {author} {\bibinfo {author} {\bibfnamefont {S.}~\bibnamefont
  {Andergassen}}, \bibinfo {author} {\bibfnamefont {T.}~\bibnamefont {Enss}},
  \bibinfo {author} {\bibfnamefont {V.}~\bibnamefont {Meden}}, \bibinfo
  {author} {\bibfnamefont {W.}~\bibnamefont {Metzner}}, \bibinfo {author}
  {\bibfnamefont {U.}~\bibnamefont {Schollw\"ock}},\ and\ \bibinfo {author}
  {\bibfnamefont {K.}~\bibnamefont {Sch\"onhammer}},\ }\bibfield  {title}
  {\bibinfo {title} {Renormalization-group analysis of the one-dimensional
  extended {H}ubbard model with a single impurity},\ }\href
  {https://doi.org/10.1103/PhysRevB.73.045125} {\bibfield  {journal} {\bibinfo
  {journal} {Phys. Rev. B}\ }\textbf {\bibinfo {volume} {73}},\ \bibinfo
  {pages} {045125} (\bibinfo {year} {2006})}\BibitemShut {NoStop}%
\bibitem [{\citenamefont {Meden}\ \emph {et~al.}(2008)\citenamefont {Meden},
  \citenamefont {Andergassen}, \citenamefont {Enss}, \citenamefont
  {Schoeller},\ and\ \citenamefont {Sch\"onhammer}}]{Meden2008}%
  \BibitemOpen
  \bibfield  {author} {\bibinfo {author} {\bibfnamefont {V.}~\bibnamefont
  {Meden}}, \bibinfo {author} {\bibfnamefont {S.}~\bibnamefont {Andergassen}},
  \bibinfo {author} {\bibfnamefont {T.}~\bibnamefont {Enss}}, \bibinfo {author}
  {\bibfnamefont {H.}~\bibnamefont {Schoeller}},\ and\ \bibinfo {author}
  {\bibfnamefont {K.}~\bibnamefont {Sch\"onhammer}},\ }\bibfield  {title}
  {\bibinfo {title} {Fermionic renormalization group methods for transport
  through inhomogeneous {L}uttinger liquids},\ }\href
  {https://doi.org/10.1088/1367-2630/10/4/045012} {\bibfield  {journal}
  {\bibinfo  {journal} {New Journal of Physics}\ }\textbf {\bibinfo {volume}
  {10}},\ \bibinfo {pages} {045012} (\bibinfo {year} {2008})}\BibitemShut
  {NoStop}%
\bibitem [{\citenamefont {Tagliavini}\ \emph {et~al.}(2019)\citenamefont
  {Tagliavini}, \citenamefont {Hille}, \citenamefont {Kugler}, \citenamefont
  {Andergassen}, \citenamefont {Toschi},\ and\ \citenamefont
  {Honerkamp}}]{Tagliavini2019}%
  \BibitemOpen
  \bibfield  {author} {\bibinfo {author} {\bibfnamefont {A.}~\bibnamefont
  {Tagliavini}}, \bibinfo {author} {\bibfnamefont {C.}~\bibnamefont {Hille}},
  \bibinfo {author} {\bibfnamefont {F.~B.}\ \bibnamefont {Kugler}}, \bibinfo
  {author} {\bibfnamefont {S.}~\bibnamefont {Andergassen}}, \bibinfo {author}
  {\bibfnamefont {A.}~\bibnamefont {Toschi}},\ and\ \bibinfo {author}
  {\bibfnamefont {C.}~\bibnamefont {Honerkamp}},\ }\bibfield  {title} {\bibinfo
  {title} {{Multiloop functional renormalization group for the two-dimensional
  {H}ubbard model: Loop convergence of the response functions}},\ }\href
  {https://doi.org/10.21468/SciPostPhys.6.1.009} {\bibfield  {journal}
  {\bibinfo  {journal} {SciPost Phys.}\ }\textbf {\bibinfo {volume} {6}},\
  \bibinfo {pages} {009} (\bibinfo {year} {2019})},\ \Eprint
  {https://arxiv.org/abs/1807.02697} {1807.02697} \BibitemShut {NoStop}%
\bibitem [{\citenamefont {Hille}\ \emph
  {et~al.}(2020{\natexlab{a}})\citenamefont {Hille}, \citenamefont {Kugler},
  \citenamefont {Eckhardt}, \citenamefont {He}, \citenamefont {Kauch},
  \citenamefont {Honerkamp}, \citenamefont {Toschi},\ and\ \citenamefont
  {Andergassen}}]{Hille2020}%
  \BibitemOpen
  \bibfield  {author} {\bibinfo {author} {\bibfnamefont {C.}~\bibnamefont
  {Hille}}, \bibinfo {author} {\bibfnamefont {F.~B.}\ \bibnamefont {Kugler}},
  \bibinfo {author} {\bibfnamefont {C.~J.}\ \bibnamefont {Eckhardt}}, \bibinfo
  {author} {\bibfnamefont {Y.-Y.}\ \bibnamefont {He}}, \bibinfo {author}
  {\bibfnamefont {A.}~\bibnamefont {Kauch}}, \bibinfo {author} {\bibfnamefont
  {C.}~\bibnamefont {Honerkamp}}, \bibinfo {author} {\bibfnamefont
  {A.}~\bibnamefont {Toschi}},\ and\ \bibinfo {author} {\bibfnamefont
  {S.}~\bibnamefont {Andergassen}},\ }\bibfield  {title} {\bibinfo {title}
  {Quantitative functional renormalization group description of the
  two-dimensional {H}ubbard model},\ }\href
  {https://doi.org/10.1103/PhysRevResearch.2.033372} {\bibfield  {journal}
  {\bibinfo  {journal} {Phys. Rev. Research}\ }\textbf {\bibinfo {volume}
  {2}},\ \bibinfo {pages} {033372} (\bibinfo {year}
  {2020}{\natexlab{a}})}\BibitemShut {NoStop}%
\bibitem [{\citenamefont {Sch\"afer}\ \emph {et~al.}(2021)\citenamefont
  {Sch\"afer}, \citenamefont {Wentzell}, \citenamefont {\ifmmode~\check{S}\else
  \v{S}\fi{}imkovic}, \citenamefont {He}, \citenamefont {Hille}, \citenamefont
  {Klett}, \citenamefont {Eckhardt}, \citenamefont {Arzhang}, \citenamefont
  {Harkov}, \citenamefont {Le~R\'egent}, \citenamefont {Kirsch}, \citenamefont
  {Wang}, \citenamefont {Kim}, \citenamefont {Kozik}, \citenamefont {Stepanov},
  \citenamefont {Kauch}, \citenamefont {Andergassen}, \citenamefont {Hansmann},
  \citenamefont {Rohe}, \citenamefont {Vilk}, \citenamefont {LeBlanc},
  \citenamefont {Zhang}, \citenamefont {Tremblay}, \citenamefont {Ferrero},
  \citenamefont {Parcollet},\ and\ \citenamefont {Georges}}]{Schaefer2021}%
  \BibitemOpen
  \bibfield  {author} {\bibinfo {author} {\bibfnamefont {T.}~\bibnamefont
  {Sch\"afer}}, \bibinfo {author} {\bibfnamefont {N.}~\bibnamefont {Wentzell}},
  \bibinfo {author} {\bibfnamefont {F.}~\bibnamefont {\ifmmode~\check{S}\else
  \v{S}\fi{}imkovic}}, \bibinfo {author} {\bibfnamefont {Y.-Y.}\ \bibnamefont
  {He}}, \bibinfo {author} {\bibfnamefont {C.}~\bibnamefont {Hille}}, \bibinfo
  {author} {\bibfnamefont {M.}~\bibnamefont {Klett}}, \bibinfo {author}
  {\bibfnamefont {C.~J.}\ \bibnamefont {Eckhardt}}, \bibinfo {author}
  {\bibfnamefont {B.}~\bibnamefont {Arzhang}}, \bibinfo {author} {\bibfnamefont
  {V.}~\bibnamefont {Harkov}}, \bibinfo {author} {\bibfnamefont {F.-M.}\
  \bibnamefont {Le~R\'egent}}, \bibinfo {author} {\bibfnamefont
  {A.}~\bibnamefont {Kirsch}}, \bibinfo {author} {\bibfnamefont
  {Y.}~\bibnamefont {Wang}}, \bibinfo {author} {\bibfnamefont {A.~J.}\
  \bibnamefont {Kim}}, \bibinfo {author} {\bibfnamefont {E.}~\bibnamefont
  {Kozik}}, \bibinfo {author} {\bibfnamefont {E.~A.}\ \bibnamefont {Stepanov}},
  \bibinfo {author} {\bibfnamefont {A.}~\bibnamefont {Kauch}}, \bibinfo
  {author} {\bibfnamefont {S.}~\bibnamefont {Andergassen}}, \bibinfo {author}
  {\bibfnamefont {P.}~\bibnamefont {Hansmann}}, \bibinfo {author}
  {\bibfnamefont {D.}~\bibnamefont {Rohe}}, \bibinfo {author} {\bibfnamefont
  {Y.~M.}\ \bibnamefont {Vilk}}, \bibinfo {author} {\bibfnamefont {J.~P.~F.}\
  \bibnamefont {LeBlanc}}, \bibinfo {author} {\bibfnamefont {S.}~\bibnamefont
  {Zhang}}, \bibinfo {author} {\bibfnamefont {A.~M.~S.}\ \bibnamefont
  {Tremblay}}, \bibinfo {author} {\bibfnamefont {M.}~\bibnamefont {Ferrero}},
  \bibinfo {author} {\bibfnamefont {O.}~\bibnamefont {Parcollet}},\ and\
  \bibinfo {author} {\bibfnamefont {A.}~\bibnamefont {Georges}},\ }\bibfield
  {title} {\bibinfo {title} {Tracking the footprints of spin fluctuations: A
  multimethod, multimessenger study of the two-dimensional {H}ubbard model},\
  }\href {https://doi.org/10.1103/PhysRevX.11.011058} {\bibfield  {journal}
  {\bibinfo  {journal} {Phys. Rev. X}\ }\textbf {\bibinfo {volume} {11}},\
  \bibinfo {pages} {011058} (\bibinfo {year} {2021})}\BibitemShut {NoStop}%
\bibitem [{\citenamefont {Gull}\ \emph {et~al.}(2011)\citenamefont {Gull},
  \citenamefont {Millis}, \citenamefont {Lichtenstein}, \citenamefont
  {Rubtsov}, \citenamefont {Troyer},\ and\ \citenamefont {Werner}}]{Gull2011a}%
  \BibitemOpen
  \bibfield  {author} {\bibinfo {author} {\bibfnamefont {E.}~\bibnamefont
  {Gull}}, \bibinfo {author} {\bibfnamefont {A.~J.}\ \bibnamefont {Millis}},
  \bibinfo {author} {\bibfnamefont {A.~I.}\ \bibnamefont {Lichtenstein}},
  \bibinfo {author} {\bibfnamefont {A.~N.}\ \bibnamefont {Rubtsov}}, \bibinfo
  {author} {\bibfnamefont {M.}~\bibnamefont {Troyer}},\ and\ \bibinfo {author}
  {\bibfnamefont {P.}~\bibnamefont {Werner}},\ }\bibfield  {title} {\bibinfo
  {title} {Continuous-time {M}onte {C}arlo methods for quantum impurity
  models},\ }\href {https://doi.org/10.1103/RevModPhys.83.349} {\bibfield
  {journal} {\bibinfo  {journal} {Rev. Mod. Phys.}\ }\textbf {\bibinfo {volume}
  {83}},\ \bibinfo {pages} {349} (\bibinfo {year} {2011})}\BibitemShut
  {NoStop}%
\bibitem [{\citenamefont {Rohringer}\ \emph {et~al.}(2012)\citenamefont
  {Rohringer}, \citenamefont {Valli},\ and\ \citenamefont
  {Toschi}}]{Rohringer2012}%
  \BibitemOpen
  \bibfield  {author} {\bibinfo {author} {\bibfnamefont {G.}~\bibnamefont
  {Rohringer}}, \bibinfo {author} {\bibfnamefont {A.}~\bibnamefont {Valli}},\
  and\ \bibinfo {author} {\bibfnamefont {A.}~\bibnamefont {Toschi}},\
  }\bibfield  {title} {\bibinfo {title} {Local electronic correlation at the
  two-particle level},\ }\href {https://doi.org/10.1103/PhysRevB.86.125114}
  {\bibfield  {journal} {\bibinfo  {journal} {Phys. Rev. B}\ }\textbf {\bibinfo
  {volume} {86}},\ \bibinfo {pages} {125114} (\bibinfo {year}
  {2012})}\BibitemShut {NoStop}%
\bibitem [{\citenamefont {Tagliavini}\ \emph {et~al.}(2018)\citenamefont
  {Tagliavini}, \citenamefont {Hummel}, \citenamefont {Wentzell}, \citenamefont
  {Andergassen}, \citenamefont {Toschi},\ and\ \citenamefont
  {Rohringer}}]{Tagliavini2018}%
  \BibitemOpen
  \bibfield  {author} {\bibinfo {author} {\bibfnamefont {A.}~\bibnamefont
  {Tagliavini}}, \bibinfo {author} {\bibfnamefont {S.}~\bibnamefont {Hummel}},
  \bibinfo {author} {\bibfnamefont {N.}~\bibnamefont {Wentzell}}, \bibinfo
  {author} {\bibfnamefont {S.}~\bibnamefont {Andergassen}}, \bibinfo {author}
  {\bibfnamefont {A.}~\bibnamefont {Toschi}},\ and\ \bibinfo {author}
  {\bibfnamefont {G.}~\bibnamefont {Rohringer}},\ }\bibfield  {title} {\bibinfo
  {title} {Efficient {B}ethe-{S}alpeter equation treatment in dynamical
  mean-field theory},\ }\href {https://doi.org/10.1103/PhysRevB.97.235140}
  {\bibfield  {journal} {\bibinfo  {journal} {Phys. Rev. B}\ }\textbf {\bibinfo
  {volume} {97}},\ \bibinfo {pages} {235140} (\bibinfo {year}
  {2018})}\BibitemShut {NoStop}%
\bibitem [{\citenamefont {Rohringer}\ \emph {et~al.}(2018)\citenamefont
  {Rohringer}, \citenamefont {Hafermann}, \citenamefont {Toschi}, \citenamefont
  {Katanin}, \citenamefont {Antipov}, \citenamefont {Katsnelson}, \citenamefont
  {Lichtenstein}, \citenamefont {Rubtsov},\ and\ \citenamefont
  {Held}}]{Rohringer2018}%
  \BibitemOpen
  \bibfield  {author} {\bibinfo {author} {\bibfnamefont {G.}~\bibnamefont
  {Rohringer}}, \bibinfo {author} {\bibfnamefont {H.}~\bibnamefont
  {Hafermann}}, \bibinfo {author} {\bibfnamefont {A.}~\bibnamefont {Toschi}},
  \bibinfo {author} {\bibfnamefont {A.~A.}\ \bibnamefont {Katanin}}, \bibinfo
  {author} {\bibfnamefont {A.~E.}\ \bibnamefont {Antipov}}, \bibinfo {author}
  {\bibfnamefont {M.~I.}\ \bibnamefont {Katsnelson}}, \bibinfo {author}
  {\bibfnamefont {A.~I.}\ \bibnamefont {Lichtenstein}}, \bibinfo {author}
  {\bibfnamefont {A.~N.}\ \bibnamefont {Rubtsov}},\ and\ \bibinfo {author}
  {\bibfnamefont {K.}~\bibnamefont {Held}},\ }\bibfield  {title} {\bibinfo
  {title} {Diagrammatic routes to nonlocal correlations beyond dynamical mean
  field theory},\ }\href {https://doi.org/10.1103/RevModPhys.90.025003}
  {\bibfield  {journal} {\bibinfo  {journal} {Rev. Mod. Phys.}\ }\textbf
  {\bibinfo {volume} {90}},\ \bibinfo {pages} {025003} (\bibinfo {year}
  {2018})}\BibitemShut {NoStop}%
\bibitem [{\citenamefont {Lichtenstein}\ \emph {et~al.}(2017)\citenamefont
  {Lichtenstein}, \citenamefont {{S\'anchez de la Pe\~na}}, \citenamefont
  {Rohe}, \citenamefont {{Di Napoli}}, \citenamefont {Honerkamp},\ and\
  \citenamefont {Maier}}]{Lichtenstein2017}%
  \BibitemOpen
  \bibfield  {author} {\bibinfo {author} {\bibfnamefont {J.}~\bibnamefont
  {Lichtenstein}}, \bibinfo {author} {\bibfnamefont {D.}~\bibnamefont
  {{S\'anchez de la Pe\~na}}}, \bibinfo {author} {\bibfnamefont
  {D.}~\bibnamefont {Rohe}}, \bibinfo {author} {\bibfnamefont {E.}~\bibnamefont
  {{Di Napoli}}}, \bibinfo {author} {\bibfnamefont {C.}~\bibnamefont
  {Honerkamp}},\ and\ \bibinfo {author} {\bibfnamefont {S.}~\bibnamefont
  {Maier}},\ }\bibfield  {title} {\bibinfo {title} {High-performance functional
  {R}enormalization {G}roup calculations for interacting fermions},\ }\href
  {https://doi.org/https://doi.org/10.1016/j.cpc.2016.12.013} {\bibfield
  {journal} {\bibinfo  {journal} {Computer Physics Communications}\ }\textbf
  {\bibinfo {volume} {213}},\ \bibinfo {pages} {100} (\bibinfo {year}
  {2017})}\BibitemShut {NoStop}%
\bibitem [{\citenamefont {Eckhardt}\ \emph {et~al.}(2020)\citenamefont
  {Eckhardt}, \citenamefont {Honerkamp}, \citenamefont {Held},\ and\
  \citenamefont {Kauch}}]{Eckhardt2020}%
  \BibitemOpen
  \bibfield  {author} {\bibinfo {author} {\bibfnamefont {C.~J.}\ \bibnamefont
  {Eckhardt}}, \bibinfo {author} {\bibfnamefont {C.}~\bibnamefont {Honerkamp}},
  \bibinfo {author} {\bibfnamefont {K.}~\bibnamefont {Held}},\ and\ \bibinfo
  {author} {\bibfnamefont {A.}~\bibnamefont {Kauch}},\ }\bibfield  {title}
  {\bibinfo {title} {Truncated unity parquet solver},\ }\href
  {https://doi.org/10.1103/PhysRevB.101.155104} {\bibfield  {journal} {\bibinfo
   {journal} {Phys. Rev. B}\ }\textbf {\bibinfo {volume} {101}},\ \bibinfo
  {pages} {155104} (\bibinfo {year} {2020})}\BibitemShut {NoStop}%
\bibitem [{\citenamefont {Hedden}\ \emph {et~al.}(2004)\citenamefont {Hedden},
  \citenamefont {Meden}, \citenamefont {Pruschke},\ and\ \citenamefont
  {Sch\"onhammer}}]{Hedden2004}%
  \BibitemOpen
  \bibfield  {author} {\bibinfo {author} {\bibfnamefont {R.}~\bibnamefont
  {Hedden}}, \bibinfo {author} {\bibfnamefont {V.}~\bibnamefont {Meden}},
  \bibinfo {author} {\bibfnamefont {T.}~\bibnamefont {Pruschke}},\ and\
  \bibinfo {author} {\bibfnamefont {K.}~\bibnamefont {Sch\"onhammer}},\
  }\bibfield  {title} {\bibinfo {title} {A functional renormalization group
  approach to zero-dimensional interacting systems},\ }\href
  {https://doi.org/10.1088/0953-8984/16/29/019} {\bibfield  {journal} {\bibinfo
   {journal} {J. Phys. Condens. Matter}\ }\textbf {\bibinfo {volume} {16}},\
  \bibinfo {pages} {5279} (\bibinfo {year} {2004})}\BibitemShut {NoStop}%
\bibitem [{\citenamefont {Karrasch}\ \emph {et~al.}(2008)\citenamefont
  {Karrasch}, \citenamefont {Hedden}, \citenamefont {Peters}, \citenamefont
  {Pruschke}, \citenamefont {Sch\"onhammer},\ and\ \citenamefont
  {Meden}}]{Karrasch2008}%
  \BibitemOpen
  \bibfield  {author} {\bibinfo {author} {\bibfnamefont {C.}~\bibnamefont
  {Karrasch}}, \bibinfo {author} {\bibfnamefont {R.}~\bibnamefont {Hedden}},
  \bibinfo {author} {\bibfnamefont {R.}~\bibnamefont {Peters}}, \bibinfo
  {author} {\bibfnamefont {T.}~\bibnamefont {Pruschke}}, \bibinfo {author}
  {\bibfnamefont {K.}~\bibnamefont {Sch\"onhammer}},\ and\ \bibinfo {author}
  {\bibfnamefont {V.}~\bibnamefont {Meden}},\ }\bibfield  {title} {\bibinfo
  {title} {A finite-frequency functional renormalization group approach to the
  single impurity {A}nderson model},\ }\href
  {https://doi.org/10.1088/0953-8984/20/34/345205} {\bibfield  {journal}
  {\bibinfo  {journal} {J. Phys. Condens. Matter}\ }\textbf {\bibinfo {volume}
  {20}},\ \bibinfo {pages} {345205} (\bibinfo {year} {2008})}\BibitemShut
  {NoStop}%
\bibitem [{\citenamefont {Jakobs}\ \emph {et~al.}(2010)\citenamefont {Jakobs},
  \citenamefont {Pletyukhov},\ and\ \citenamefont {Schoeller}}]{Jakobs2010}%
  \BibitemOpen
  \bibfield  {author} {\bibinfo {author} {\bibfnamefont {S.~G.}\ \bibnamefont
  {Jakobs}}, \bibinfo {author} {\bibfnamefont {M.}~\bibnamefont {Pletyukhov}},\
  and\ \bibinfo {author} {\bibfnamefont {H.}~\bibnamefont {Schoeller}},\
  }\bibfield  {title} {\bibinfo {title} {Nonequilibrium functional
  renormalization group with frequency-dependent vertex function: A study of
  the single-impurity {A}nderson model},\ }\href
  {https://doi.org/10.1103/PhysRevB.81.195109} {\bibfield  {journal} {\bibinfo
  {journal} {Phys. Rev. B}\ }\textbf {\bibinfo {volume} {81}},\ \bibinfo
  {pages} {195109} (\bibinfo {year} {2010})}\BibitemShut {NoStop}%
\bibitem [{\citenamefont {Isidori}\ \emph {et~al.}(2010)\citenamefont
  {Isidori}, \citenamefont {Roosen}, \citenamefont {Bartosch}, \citenamefont
  {Hofstetter},\ and\ \citenamefont {Kopietz}}]{Isidori2010}%
  \BibitemOpen
  \bibfield  {author} {\bibinfo {author} {\bibfnamefont {A.}~\bibnamefont
  {Isidori}}, \bibinfo {author} {\bibfnamefont {D.}~\bibnamefont {Roosen}},
  \bibinfo {author} {\bibfnamefont {L.}~\bibnamefont {Bartosch}}, \bibinfo
  {author} {\bibfnamefont {W.}~\bibnamefont {Hofstetter}},\ and\ \bibinfo
  {author} {\bibfnamefont {P.}~\bibnamefont {Kopietz}},\ }\bibfield  {title}
  {\bibinfo {title} {Spectral function of the {A}nderson impurity model at
  finite temperatures},\ }\href {https://doi.org/10.1103/PhysRevB.81.235120}
  {\bibfield  {journal} {\bibinfo  {journal} {Phys. Rev. B}\ }\textbf {\bibinfo
  {volume} {81}},\ \bibinfo {pages} {235120} (\bibinfo {year}
  {2010})}\BibitemShut {NoStop}%
\bibitem [{\citenamefont {Rentrop}\ \emph {et~al.}(2016)\citenamefont
  {Rentrop}, \citenamefont {Meden},\ and\ \citenamefont
  {Jakobs}}]{Rentrop2016}%
  \BibitemOpen
  \bibfield  {author} {\bibinfo {author} {\bibfnamefont {J.~F.}\ \bibnamefont
  {Rentrop}}, \bibinfo {author} {\bibfnamefont {V.}~\bibnamefont {Meden}},\
  and\ \bibinfo {author} {\bibfnamefont {S.~G.}\ \bibnamefont {Jakobs}},\
  }\bibfield  {title} {\bibinfo {title} {Renormalization group flow of the
  {L}uttinger-{W}ard functional: Conserving approximations and application to
  the {A}nderson impurity model},\ }\href
  {https://doi.org/10.1103/PhysRevB.93.195160} {\bibfield  {journal} {\bibinfo
  {journal} {Phys. Rev. B}\ }\textbf {\bibinfo {volume} {93}},\ \bibinfo
  {pages} {195160} (\bibinfo {year} {2016})}\BibitemShut {NoStop}%
\bibitem [{\citenamefont {Yirga}\ and\ \citenamefont
  {Campbell}(2021)}]{Yirga2021}%
  \BibitemOpen
  \bibfield  {author} {\bibinfo {author} {\bibfnamefont {N.~K.}\ \bibnamefont
  {Yirga}}\ and\ \bibinfo {author} {\bibfnamefont {D.~K.}\ \bibnamefont
  {Campbell}},\ }\bibfield  {title} {\bibinfo {title} {Frequency-dependent
  functional renormalization group for interacting fermionic systems},\ }\href
  {https://doi.org/10.1103/PhysRevB.103.235165} {\bibfield  {journal} {\bibinfo
   {journal} {Phys. Rev. B}\ }\textbf {\bibinfo {volume} {103}},\ \bibinfo
  {pages} {235165} (\bibinfo {year} {2021})}\BibitemShut {NoStop}%
\bibitem [{\citenamefont {Chalupa}\ \emph {et~al.}(2021)\citenamefont
  {Chalupa}, \citenamefont {Sch\"afer}, \citenamefont {Reitner}, \citenamefont
  {Springer}, \citenamefont {Andergassen},\ and\ \citenamefont
  {Toschi}}]{Chalupa2021}%
  \BibitemOpen
  \bibfield  {author} {\bibinfo {author} {\bibfnamefont {P.}~\bibnamefont
  {Chalupa}}, \bibinfo {author} {\bibfnamefont {T.}~\bibnamefont {Sch\"afer}},
  \bibinfo {author} {\bibfnamefont {M.}~\bibnamefont {Reitner}}, \bibinfo
  {author} {\bibfnamefont {D.}~\bibnamefont {Springer}}, \bibinfo {author}
  {\bibfnamefont {S.}~\bibnamefont {Andergassen}},\ and\ \bibinfo {author}
  {\bibfnamefont {A.}~\bibnamefont {Toschi}},\ }\bibfield  {title} {\bibinfo
  {title} {Fingerprints of the local moment formation and its {K}ondo screening
  in the generalized susceptibilities of many-electron problems},\ }\href
  {https://doi.org/10.1103/PhysRevLett.126.056403} {\bibfield  {journal}
  {\bibinfo  {journal} {Phys. Rev. Lett.}\ }\textbf {\bibinfo {volume} {126}},\
  \bibinfo {pages} {056403} (\bibinfo {year} {2021})}\BibitemShut {NoStop}%
\bibitem [{\citenamefont {Vilardi}\ \emph {et~al.}(2019)\citenamefont
  {Vilardi}, \citenamefont {Taranto},\ and\ \citenamefont
  {Metzner}}]{Vilardi2019}%
  \BibitemOpen
  \bibfield  {author} {\bibinfo {author} {\bibfnamefont {D.}~\bibnamefont
  {Vilardi}}, \bibinfo {author} {\bibfnamefont {C.}~\bibnamefont {Taranto}},\
  and\ \bibinfo {author} {\bibfnamefont {W.}~\bibnamefont {Metzner}},\
  }\bibfield  {title} {\bibinfo {title} {Antiferromagnetic and $d$-wave pairing
  correlations in the strongly interacting two-dimensional {H}ubbard model from
  the functional renormalization group},\ }\href
  {https://doi.org/10.1103/PhysRevB.99.104501} {\bibfield  {journal} {\bibinfo
  {journal} {Phys. Rev. B}\ }\textbf {\bibinfo {volume} {99}},\ \bibinfo
  {pages} {104501} (\bibinfo {year} {2019})}\BibitemShut {NoStop}%
\bibitem [{\citenamefont {Bonetti}\ \emph {et~al.}(2022)\citenamefont
  {Bonetti}, \citenamefont {Toschi}, \citenamefont {Hille}, \citenamefont
  {Andergassen},\ and\ \citenamefont {Vilardi}}]{Bonetti2022}%
  \BibitemOpen
  \bibfield  {author} {\bibinfo {author} {\bibfnamefont {P.~M.}\ \bibnamefont
  {Bonetti}}, \bibinfo {author} {\bibfnamefont {A.}~\bibnamefont {Toschi}},
  \bibinfo {author} {\bibfnamefont {C.}~\bibnamefont {Hille}}, \bibinfo
  {author} {\bibfnamefont {S.}~\bibnamefont {Andergassen}},\ and\ \bibinfo
  {author} {\bibfnamefont {D.}~\bibnamefont {Vilardi}},\ }\bibfield  {title}
  {\bibinfo {title} {Single-boson exchange representation of the functional
  renormalization group for strongly interacting many-electron systems},\
  }\href {https://doi.org/10.1103/PhysRevResearch.4.013034} {\bibfield
  {journal} {\bibinfo  {journal} {Phys. Rev. Research}\ }\textbf {\bibinfo
  {volume} {4}},\ \bibinfo {pages} {013034} (\bibinfo {year}
  {2022})}\BibitemShut {NoStop}%
\bibitem [{\citenamefont {Georges}\ \emph {et~al.}(1996)\citenamefont
  {Georges}, \citenamefont {Kotliar}, \citenamefont {Krauth},\ and\
  \citenamefont {Rozenberg}}]{Georges1996}%
  \BibitemOpen
  \bibfield  {author} {\bibinfo {author} {\bibfnamefont {A.}~\bibnamefont
  {Georges}}, \bibinfo {author} {\bibfnamefont {G.}~\bibnamefont {Kotliar}},
  \bibinfo {author} {\bibfnamefont {W.}~\bibnamefont {Krauth}},\ and\ \bibinfo
  {author} {\bibfnamefont {M.~J.}\ \bibnamefont {Rozenberg}},\ }\bibfield
  {title} {\bibinfo {title} {Dynamical mean-field theory of strongly correlated
  fermion systems and the limit of infinite dimensions},\ }\href
  {https://doi.org/10.1103/RevModPhys.68.13} {\bibfield  {journal} {\bibinfo
  {journal} {Rev. Mod. Phys.}\ }\textbf {\bibinfo {volume} {68}},\ \bibinfo
  {pages} {13} (\bibinfo {year} {1996})}\BibitemShut {NoStop}%
\bibitem [{\citenamefont {Bulla}\ \emph {et~al.}(2008)\citenamefont {Bulla},
  \citenamefont {Costi},\ and\ \citenamefont {Pruschke}}]{Bulla2008}%
  \BibitemOpen
  \bibfield  {author} {\bibinfo {author} {\bibfnamefont {R.}~\bibnamefont
  {Bulla}}, \bibinfo {author} {\bibfnamefont {T.~A.}\ \bibnamefont {Costi}},\
  and\ \bibinfo {author} {\bibfnamefont {T.}~\bibnamefont {Pruschke}},\
  }\bibfield  {title} {\bibinfo {title} {Numerical renormalization group method
  for quantum impurity systems},\ }\href
  {https://doi.org/10.1103/RevModPhys.80.395} {\bibfield  {journal} {\bibinfo
  {journal} {Rev. Mod. Phys.}\ }\textbf {\bibinfo {volume} {80}},\ \bibinfo
  {pages} {395} (\bibinfo {year} {2008})}\BibitemShut {NoStop}%
\bibitem [{\citenamefont {Taranto}\ \emph {et~al.}(2014)\citenamefont
  {Taranto}, \citenamefont {Andergassen}, \citenamefont {Bauer}, \citenamefont
  {Held}, \citenamefont {Katanin}, \citenamefont {Metzner}, \citenamefont
  {Rohringer},\ and\ \citenamefont {Toschi}}]{Taranto2014}%
  \BibitemOpen
  \bibfield  {author} {\bibinfo {author} {\bibfnamefont {C.}~\bibnamefont
  {Taranto}}, \bibinfo {author} {\bibfnamefont {S.}~\bibnamefont
  {Andergassen}}, \bibinfo {author} {\bibfnamefont {J.}~\bibnamefont {Bauer}},
  \bibinfo {author} {\bibfnamefont {K.}~\bibnamefont {Held}}, \bibinfo {author}
  {\bibfnamefont {A.}~\bibnamefont {Katanin}}, \bibinfo {author} {\bibfnamefont
  {W.}~\bibnamefont {Metzner}}, \bibinfo {author} {\bibfnamefont
  {G.}~\bibnamefont {Rohringer}},\ and\ \bibinfo {author} {\bibfnamefont
  {A.}~\bibnamefont {Toschi}},\ }\bibfield  {title} {\bibinfo {title} {From
  infinite to two dimensions through the functional renormalization group},\
  }\href {https://doi.org/10.1103/PhysRevLett.112.196402} {\bibfield  {journal}
  {\bibinfo  {journal} {Phys. Rev. Lett.}\ }\textbf {\bibinfo {volume} {112}},\
  \bibinfo {pages} {196402} (\bibinfo {year} {2014})}\BibitemShut {NoStop}%
\bibitem [{\citenamefont {Wentzell}\ \emph {et~al.}(2015)\citenamefont
  {Wentzell}, \citenamefont {Taranto}, \citenamefont {Katanin}, \citenamefont
  {Toschi},\ and\ \citenamefont {Andergassen}}]{Wentzell2015}%
  \BibitemOpen
  \bibfield  {author} {\bibinfo {author} {\bibfnamefont {N.}~\bibnamefont
  {Wentzell}}, \bibinfo {author} {\bibfnamefont {C.}~\bibnamefont {Taranto}},
  \bibinfo {author} {\bibfnamefont {A.}~\bibnamefont {Katanin}}, \bibinfo
  {author} {\bibfnamefont {A.}~\bibnamefont {Toschi}},\ and\ \bibinfo {author}
  {\bibfnamefont {S.}~\bibnamefont {Andergassen}},\ }\bibfield  {title}
  {\bibinfo {title} {Correlated starting points for the functional
  renormalization group},\ }\href {https://doi.org/10.1103/PhysRevB.91.045120}
  {\bibfield  {journal} {\bibinfo  {journal} {Phys. Rev. B}\ }\textbf {\bibinfo
  {volume} {91}},\ \bibinfo {pages} {045120} (\bibinfo {year}
  {2015})}\BibitemShut {NoStop}%
\bibitem [{\citenamefont {Anderson}(1961)}]{Anderson1961}%
  \BibitemOpen
  \bibfield  {author} {\bibinfo {author} {\bibfnamefont {P.~W.}\ \bibnamefont
  {Anderson}},\ }\bibfield  {title} {\bibinfo {title} {Localized magnetic
  states in metals},\ }\href {https://doi.org/10.1103/PhysRev.124.41}
  {\bibfield  {journal} {\bibinfo  {journal} {Phys. Rev.}\ }\textbf {\bibinfo
  {volume} {124}},\ \bibinfo {pages} {41} (\bibinfo {year} {1961})}\BibitemShut
  {NoStop}%
\bibitem [{\citenamefont {Hewson}(1993)}]{Hewson1993}%
  \BibitemOpen
  \bibfield  {author} {\bibinfo {author} {\bibfnamefont {A.}~\bibnamefont
  {Hewson}},\ }\href@noop {} {\emph {\bibinfo {title} {The Kondo Problem to
  Heavy Fermions}}}\ (\bibinfo  {publisher} {Cambridge University Press},\
  \bibinfo {year} {1993})\BibitemShut {NoStop}%
\bibitem [{\citenamefont {Chalupa}\ \emph {et~al.}(2018)\citenamefont
  {Chalupa}, \citenamefont {Gunacker}, \citenamefont {Sch\"afer}, \citenamefont
  {Held},\ and\ \citenamefont {Toschi}}]{Chalupa2018}%
  \BibitemOpen
  \bibfield  {author} {\bibinfo {author} {\bibfnamefont {P.}~\bibnamefont
  {Chalupa}}, \bibinfo {author} {\bibfnamefont {P.}~\bibnamefont {Gunacker}},
  \bibinfo {author} {\bibfnamefont {T.}~\bibnamefont {Sch\"afer}}, \bibinfo
  {author} {\bibfnamefont {K.}~\bibnamefont {Held}},\ and\ \bibinfo {author}
  {\bibfnamefont {A.}~\bibnamefont {Toschi}},\ }\bibfield  {title} {\bibinfo
  {title} {Divergences of the irreducible vertex functions in correlated
  metallic systems: Insights from the {A}nderson impurity model},\ }\href
  {https://doi.org/10.1103/PhysRevB.97.245136} {\bibfield  {journal} {\bibinfo
  {journal} {Phys. Rev. B}\ }\textbf {\bibinfo {volume} {97}},\ \bibinfo
  {pages} {245136} (\bibinfo {year} {2018})}\BibitemShut {NoStop}%
\bibitem [{\citenamefont {Coleman}(2015)}]{Coleman2015}%
  \BibitemOpen
  \bibfield  {author} {\bibinfo {author} {\bibfnamefont {P.}~\bibnamefont
  {Coleman}},\ }\href {https://doi.org/10.1017/CBO9781139020916} {\emph
  {\bibinfo {title} {Introduction to Many-Body Physics}}}\ (\bibinfo
  {publisher} {Cambridge University Press},\ \bibinfo {year}
  {2015})\BibitemShut {NoStop}%
\bibitem [{\citenamefont {Honerkamp}\ \emph {et~al.}(2004)\citenamefont
  {Honerkamp}, \citenamefont {Rohe}, \citenamefont {Andergassen},\ and\
  \citenamefont {Enss}}]{Honerkamp2004}%
  \BibitemOpen
  \bibfield  {author} {\bibinfo {author} {\bibfnamefont {C.}~\bibnamefont
  {Honerkamp}}, \bibinfo {author} {\bibfnamefont {D.}~\bibnamefont {Rohe}},
  \bibinfo {author} {\bibfnamefont {S.}~\bibnamefont {Andergassen}},\ and\
  \bibinfo {author} {\bibfnamefont {T.}~\bibnamefont {Enss}},\ }\bibfield
  {title} {\bibinfo {title} {Interaction flow method for many-fermion
  systems},\ }\href {https://doi.org/10.1103/PhysRevB.70.235115} {\bibfield
  {journal} {\bibinfo  {journal} {Phys. Rev. B}\ }\textbf {\bibinfo {volume}
  {70}},\ \bibinfo {pages} {235115} (\bibinfo {year} {2004})}\BibitemShut
  {NoStop}%
\bibitem [{\citenamefont {Eberlein}(2014)}]{Eberlein2014}%
  \BibitemOpen
  \bibfield  {author} {\bibinfo {author} {\bibfnamefont {A.}~\bibnamefont
  {Eberlein}},\ }\bibfield  {title} {\bibinfo {title} {Fermionic two-loop
  functional renormalization group for correlated fermions: Method and
  application to the attractive {H}ubbard model},\ }\href
  {https://doi.org/10.1103/PhysRevB.90.115125} {\bibfield  {journal} {\bibinfo
  {journal} {Phys. Rev. B}\ }\textbf {\bibinfo {volume} {90}},\ \bibinfo
  {pages} {115125} (\bibinfo {year} {2014})}\BibitemShut {NoStop}%
\bibitem [{\citenamefont {R\"uck}\ and\ \citenamefont
  {Reuther}(2018)}]{Rueck2018}%
  \BibitemOpen
  \bibfield  {author} {\bibinfo {author} {\bibfnamefont {M.}~\bibnamefont
  {R\"uck}}\ and\ \bibinfo {author} {\bibfnamefont {J.}~\bibnamefont
  {Reuther}},\ }\bibfield  {title} {\bibinfo {title} {Effects of two-loop
  contributions in the pseudofermion functional renormalization group method
  for quantum spin systems},\ }\href
  {https://doi.org/10.1103/PhysRevB.97.144404} {\bibfield  {journal} {\bibinfo
  {journal} {Phys. Rev. B}\ }\textbf {\bibinfo {volume} {97}},\ \bibinfo
  {pages} {144404} (\bibinfo {year} {2018})}\BibitemShut {NoStop}%
\bibitem [{\citenamefont {Thoenniss}\ \emph {et~al.}(2020)\citenamefont
  {Thoenniss}, \citenamefont {Ritter}, \citenamefont {Kugler}, \citenamefont
  {von Delft},\ and\ \citenamefont {Punk}}]{Thoenniss2020}%
  \BibitemOpen
  \bibfield  {author} {\bibinfo {author} {\bibfnamefont {J.}~\bibnamefont
  {Thoenniss}}, \bibinfo {author} {\bibfnamefont {M.~K.}\ \bibnamefont
  {Ritter}}, \bibinfo {author} {\bibfnamefont {F.~B.}\ \bibnamefont {Kugler}},
  \bibinfo {author} {\bibfnamefont {J.}~\bibnamefont {von Delft}},\ and\
  \bibinfo {author} {\bibfnamefont {M.}~\bibnamefont {Punk}},\ }\href@noop {}
  {\bibinfo {title} {Multiloop pseudofermion functional renormalization for
  quantum spin systems: Application to the spin-$\frac{1}{2}$ kagome
  {H}eisenberg model}} (\bibinfo {year} {2020}),\ \Eprint
  {https://arxiv.org/abs/2011.01268} {arXiv:2011.01268 [cond-mat.str-el]}
  \BibitemShut {NoStop}%
\bibitem [{\citenamefont {Kiese}\ \emph {et~al.}(2020)\citenamefont {Kiese},
  \citenamefont {Mueller}, \citenamefont {Iqbal}, \citenamefont {Thomale},\
  and\ \citenamefont {Trebst}}]{Kiese2020}%
  \BibitemOpen
  \bibfield  {author} {\bibinfo {author} {\bibfnamefont {D.}~\bibnamefont
  {Kiese}}, \bibinfo {author} {\bibfnamefont {T.}~\bibnamefont {Mueller}},
  \bibinfo {author} {\bibfnamefont {Y.}~\bibnamefont {Iqbal}}, \bibinfo
  {author} {\bibfnamefont {R.}~\bibnamefont {Thomale}},\ and\ \bibinfo {author}
  {\bibfnamefont {S.}~\bibnamefont {Trebst}},\ }\href@noop {} {\bibinfo {title}
  {Multiloop functional renormalization group approach to quantum spin
  systems}} (\bibinfo {year} {2020}),\ \Eprint
  {https://arxiv.org/abs/2011.01269} {arXiv:2011.01269 [cond-mat.str-el]}
  \BibitemShut {NoStop}%
\bibitem [{Note1()}]{Note1}%
  \BibitemOpen
  \bibinfo {note} {The translation of the notation used in this work to the
  notation used in many other works, among them Ref.~\cite {Rohringer2018}, is
  the following: The diagrammatic channels relate to one another as $a=\protect
  \overline {ph}$, $p=pp$, $t=ph$; the vertex two-particle reducible in channel
  $r$ is referred to as $\gamma _r = \Phi _r$, the vertex irreducible in
  channel $r$ as $I_r = \Gamma _r$ and the fully two-particle irreducible
  vertex as $R_{\protect \mathrm {2PI}} = \Lambda _{\protect \mathrm
  {irr}}$.}\BibitemShut {Stop}%
\bibitem [{\citenamefont {Hille}\ \emph
  {et~al.}(2020{\natexlab{b}})\citenamefont {Hille}, \citenamefont {Rohe},
  \citenamefont {Honerkamp},\ and\ \citenamefont {Andergassen}}]{Hille2020a}%
  \BibitemOpen
  \bibfield  {author} {\bibinfo {author} {\bibfnamefont {C.}~\bibnamefont
  {Hille}}, \bibinfo {author} {\bibfnamefont {D.}~\bibnamefont {Rohe}},
  \bibinfo {author} {\bibfnamefont {C.}~\bibnamefont {Honerkamp}},\ and\
  \bibinfo {author} {\bibfnamefont {S.}~\bibnamefont {Andergassen}},\
  }\bibfield  {title} {\bibinfo {title} {Pseudogap opening in the
  two-dimensional {H}ubbard model: A functional renormalization group
  analysis},\ }\href {https://doi.org/10.1103/PhysRevResearch.2.033068}
  {\bibfield  {journal} {\bibinfo  {journal} {Phys. Rev. Research}\ }\textbf
  {\bibinfo {volume} {2}},\ \bibinfo {pages} {033068} (\bibinfo {year}
  {2020}{\natexlab{b}})}\BibitemShut {NoStop}%
\bibitem [{\citenamefont {Wallerberger}\ \emph {et~al.}(2019)\citenamefont
  {Wallerberger}, \citenamefont {Hausoel}, \citenamefont {Gunacker},
  \citenamefont {Kowalski}, \citenamefont {Parragh}, \citenamefont {Goth},
  \citenamefont {Held},\ and\ \citenamefont {Sangiovanni}}]{w2dynamics}%
  \BibitemOpen
  \bibfield  {author} {\bibinfo {author} {\bibfnamefont {M.}~\bibnamefont
  {Wallerberger}}, \bibinfo {author} {\bibfnamefont {A.}~\bibnamefont
  {Hausoel}}, \bibinfo {author} {\bibfnamefont {P.}~\bibnamefont {Gunacker}},
  \bibinfo {author} {\bibfnamefont {A.}~\bibnamefont {Kowalski}}, \bibinfo
  {author} {\bibfnamefont {N.}~\bibnamefont {Parragh}}, \bibinfo {author}
  {\bibfnamefont {F.}~\bibnamefont {Goth}}, \bibinfo {author} {\bibfnamefont
  {K.}~\bibnamefont {Held}},\ and\ \bibinfo {author} {\bibfnamefont
  {G.}~\bibnamefont {Sangiovanni}},\ }\bibfield  {title} {\bibinfo {title}
  {w2dynamics: Local one- and two-particle quantities from dynamical mean field
  theory},\ }\href {https://doi.org/10.1016/j.cpc.2018.09.007} {\bibfield
  {journal} {\bibinfo  {journal} {Comp. Phys. Commun.}\ }\textbf {\bibinfo
  {volume} {235}},\ \bibinfo {pages} {388 } (\bibinfo {year}
  {2019})}\BibitemShut {NoStop}%
\bibitem [{\citenamefont {Hewson}\ and\ \citenamefont
  {Meyer}(2002)}]{Hewson02}%
  \BibitemOpen
  \bibfield  {author} {\bibinfo {author} {\bibfnamefont {A.~C.}\ \bibnamefont
  {Hewson}}\ and\ \bibinfo {author} {\bibfnamefont {D.}~\bibnamefont {Meyer}},\
  }\bibfield  {title} {\bibinfo {title} {Numerical renormalization group study
  of the {A}nderson-holstein impurity model},\ }\href
  {http://stacks.iop.org/0953-8984/14/i=3/a=312} {\bibfield  {journal}
  {\bibinfo  {journal} {J. Phys.: Condens. Matter}\ }\textbf {\bibinfo {volume}
  {14}},\ \bibinfo {pages} {427} (\bibinfo {year} {2002})}\BibitemShut
  {NoStop}%
\bibitem [{\citenamefont {Coleman}(2007)}]{coleman2006heavy}%
  \BibitemOpen
  \bibfield  {author} {\bibinfo {author} {\bibfnamefont {P.}~\bibnamefont
  {Coleman}},\ }\href@noop {} {\bibinfo {title} {Heavy fermions: electrons at
  the edge of magnetism}} (\bibinfo {year} {2007}),\ \bibinfo {note} {handbook
  of Magnetism and Advanced Magnetic Materials. Edited by H. Kronmuller and S.
  Parkin. Vol 1: Fundamentals and Theory. (John Wiley and Sons)}\BibitemShut
  {NoStop}%
\bibitem [{\citenamefont {Watzenb\"ock}\ \emph {et~al.}(2020)\citenamefont
  {Watzenb\"ock}, \citenamefont {Edelmann}, \citenamefont {Springer},
  \citenamefont {Sangiovanni},\ and\ \citenamefont {Toschi}}]{Watzenboeck2020}%
  \BibitemOpen
  \bibfield  {author} {\bibinfo {author} {\bibfnamefont {C.}~\bibnamefont
  {Watzenb\"ock}}, \bibinfo {author} {\bibfnamefont {M.}~\bibnamefont
  {Edelmann}}, \bibinfo {author} {\bibfnamefont {D.}~\bibnamefont {Springer}},
  \bibinfo {author} {\bibfnamefont {G.}~\bibnamefont {Sangiovanni}},\ and\
  \bibinfo {author} {\bibfnamefont {A.}~\bibnamefont {Toschi}},\ }\bibfield
  {title} {\bibinfo {title} {Characteristic timescales of the local moment
  dynamics in hund's metals},\ }\href
  {https://doi.org/10.1103/PhysRevLett.125.086402} {\bibfield  {journal}
  {\bibinfo  {journal} {Phys. Rev. Lett.}\ }\textbf {\bibinfo {volume} {125}},\
  \bibinfo {pages} {086402} (\bibinfo {year} {2020})}\BibitemShut {NoStop}%
\bibitem [{\citenamefont {Gaspard}\ and\ \citenamefont
  {Tomczak}(2021)}]{Tomczak2021}%
  \BibitemOpen
  \bibfield  {author} {\bibinfo {author} {\bibfnamefont {L.}~\bibnamefont
  {Gaspard}}\ and\ \bibinfo {author} {\bibfnamefont {J.~M.}\ \bibnamefont
  {Tomczak}},\ }\href@noop {} {\bibinfo {title} {Timescale of local moment
  screening across and above the {M}ott transition}} (\bibinfo {year} {2021}),\
  \Eprint {https://arxiv.org/abs/2112.02881} {arXiv:2112.02881
  [cond-mat.str-el]} \BibitemShut {NoStop}%
\bibitem [{\citenamefont {Watzenb\"ock}\ \emph {et~al.}(2021)\citenamefont
  {Watzenb\"ock}, \citenamefont {Fellinger}, \citenamefont {Held},\ and\
  \citenamefont {Toschi}}]{Watzenboeck2021}%
  \BibitemOpen
  \bibfield  {author} {\bibinfo {author} {\bibfnamefont {C.}~\bibnamefont
  {Watzenb\"ock}}, \bibinfo {author} {\bibfnamefont {M.}~\bibnamefont
  {Fellinger}}, \bibinfo {author} {\bibfnamefont {K.}~\bibnamefont {Held}},\
  and\ \bibinfo {author} {\bibfnamefont {A.}~\bibnamefont {Toschi}},\
  }\href@noop {} {\bibinfo {title} {Long-term memory magnetic correlations in
  the {H}ubbard model: A dynamical mean-field theory analysis}} (\bibinfo
  {year} {2021}),\ \Eprint {https://arxiv.org/abs/2112.02903} {arXiv:2112.02903
  [cond-mat.str-el]} \BibitemShut {NoStop}%
\bibitem [{\citenamefont {Kozik}\ \emph {et~al.}(2015)\citenamefont {Kozik},
  \citenamefont {Ferrero},\ and\ \citenamefont {Georges}}]{Kozik2015}%
  \BibitemOpen
  \bibfield  {author} {\bibinfo {author} {\bibfnamefont {E.}~\bibnamefont
  {Kozik}}, \bibinfo {author} {\bibfnamefont {M.}~\bibnamefont {Ferrero}},\
  and\ \bibinfo {author} {\bibfnamefont {A.}~\bibnamefont {Georges}},\
  }\bibfield  {title} {\bibinfo {title} {Nonexistence of the {L}uttinger-{W}ard
  functional and misleading convergence of skeleton diagrammatic series for
  {H}ubbard-like models},\ }\href
  {https://doi.org/10.1103/PhysRevLett.114.156402} {\bibfield  {journal}
  {\bibinfo  {journal} {Phys. Rev. Lett.}\ }\textbf {\bibinfo {volume} {114}},\
  \bibinfo {pages} {156402} (\bibinfo {year} {2015})}\BibitemShut {NoStop}%
\bibitem [{\citenamefont {Gunnarsson}\ \emph {et~al.}(2017)\citenamefont
  {Gunnarsson}, \citenamefont {Rohringer}, \citenamefont {Sch\"afer},
  \citenamefont {Sangiovanni},\ and\ \citenamefont {Toschi}}]{Gunnarsson2017}%
  \BibitemOpen
  \bibfield  {author} {\bibinfo {author} {\bibfnamefont {O.}~\bibnamefont
  {Gunnarsson}}, \bibinfo {author} {\bibfnamefont {G.}~\bibnamefont
  {Rohringer}}, \bibinfo {author} {\bibfnamefont {T.}~\bibnamefont
  {Sch\"afer}}, \bibinfo {author} {\bibfnamefont {G.}~\bibnamefont
  {Sangiovanni}},\ and\ \bibinfo {author} {\bibfnamefont {A.}~\bibnamefont
  {Toschi}},\ }\bibfield  {title} {\bibinfo {title} {Breakdown of traditional
  many-body theories for correlated electrons},\ }\href
  {https://doi.org/10.1103/PhysRevLett.119.056402} {\bibfield  {journal}
  {\bibinfo  {journal} {Phys. Rev. Lett.}\ }\textbf {\bibinfo {volume} {119}},\
  \bibinfo {pages} {056402} (\bibinfo {year} {2017})}\BibitemShut {NoStop}%
\bibitem [{\citenamefont {Reitner}\ \emph {et~al.}(2020)\citenamefont
  {Reitner}, \citenamefont {Chalupa}, \citenamefont {Del~Re}, \citenamefont
  {Springer}, \citenamefont {Ciuchi}, \citenamefont {Sangiovanni},\ and\
  \citenamefont {Toschi}}]{Reitner2020}%
  \BibitemOpen
  \bibfield  {author} {\bibinfo {author} {\bibfnamefont {M.}~\bibnamefont
  {Reitner}}, \bibinfo {author} {\bibfnamefont {P.}~\bibnamefont {Chalupa}},
  \bibinfo {author} {\bibfnamefont {L.}~\bibnamefont {Del~Re}}, \bibinfo
  {author} {\bibfnamefont {D.}~\bibnamefont {Springer}}, \bibinfo {author}
  {\bibfnamefont {S.}~\bibnamefont {Ciuchi}}, \bibinfo {author} {\bibfnamefont
  {G.}~\bibnamefont {Sangiovanni}},\ and\ \bibinfo {author} {\bibfnamefont
  {A.}~\bibnamefont {Toschi}},\ }\bibfield  {title} {\bibinfo {title}
  {Attractive effect of a strong electronic repulsion: The physics of vertex
  divergences},\ }\href {https://doi.org/10.1103/PhysRevLett.125.196403}
  {\bibfield  {journal} {\bibinfo  {journal} {Phys. Rev. Lett.}\ }\textbf
  {\bibinfo {volume} {125}},\ \bibinfo {pages} {196403} (\bibinfo {year}
  {2020})}\BibitemShut {NoStop}%
\bibitem [{\citenamefont {Sch\"afer}\ \emph {et~al.}(2013)\citenamefont
  {Sch\"afer}, \citenamefont {Rohringer}, \citenamefont {Gunnarsson},
  \citenamefont {Ciuchi}, \citenamefont {Sangiovanni},\ and\ \citenamefont
  {Toschi}}]{Schaefer2013}%
  \BibitemOpen
  \bibfield  {author} {\bibinfo {author} {\bibfnamefont {T.}~\bibnamefont
  {Sch\"afer}}, \bibinfo {author} {\bibfnamefont {G.}~\bibnamefont
  {Rohringer}}, \bibinfo {author} {\bibfnamefont {O.}~\bibnamefont
  {Gunnarsson}}, \bibinfo {author} {\bibfnamefont {S.}~\bibnamefont {Ciuchi}},
  \bibinfo {author} {\bibfnamefont {G.}~\bibnamefont {Sangiovanni}},\ and\
  \bibinfo {author} {\bibfnamefont {A.}~\bibnamefont {Toschi}},\ }\bibfield
  {title} {\bibinfo {title} {Divergent precursors of the {M}ott-{H}ubbard
  transition at the two-particle level},\ }\href
  {https://doi.org/10.1103/PhysRevLett.110.246405} {\bibfield  {journal}
  {\bibinfo  {journal} {Phys. Rev. Lett.}\ }\textbf {\bibinfo {volume} {110}},\
  \bibinfo {pages} {246405} (\bibinfo {year} {2013})}\BibitemShut {NoStop}%
\bibitem [{\citenamefont {Jani\v{s}}\ and\ \citenamefont
  {Pokorn\'y}(2014)}]{Janis2014}%
  \BibitemOpen
  \bibfield  {author} {\bibinfo {author} {\bibfnamefont {V.}~\bibnamefont
  {Jani\v{s}}}\ and\ \bibinfo {author} {\bibfnamefont {V.}~\bibnamefont
  {Pokorn\'y}},\ }\bibfield  {title} {\bibinfo {title} {Critical
  metal-insulator transition and divergence in a two-particle irreducible
  vertex in disordered and interacting electron systems},\ }\href
  {https://doi.org/10.1103/PhysRevB.90.045143} {\bibfield  {journal} {\bibinfo
  {journal} {Phys. Rev. B}\ }\textbf {\bibinfo {volume} {90}},\ \bibinfo
  {pages} {045143} (\bibinfo {year} {2014})}\BibitemShut {NoStop}%
\bibitem [{\citenamefont {Ribic}\ \emph {et~al.}(2016)\citenamefont {Ribic},
  \citenamefont {Rohringer},\ and\ \citenamefont {Held}}]{Ribic2016}%
  \BibitemOpen
  \bibfield  {author} {\bibinfo {author} {\bibfnamefont {T.}~\bibnamefont
  {Ribic}}, \bibinfo {author} {\bibfnamefont {G.}~\bibnamefont {Rohringer}},\
  and\ \bibinfo {author} {\bibfnamefont {K.}~\bibnamefont {Held}},\ }\bibfield
  {title} {\bibinfo {title} {{Nonlocal correlations and spectral properties of
  the Falicov-Kimball model}},\ }\href
  {https://doi.org/10.1103/PhysRevB.93.195105} {\bibfield  {journal} {\bibinfo
  {journal} {Phys. Rev. B}\ }\textbf {\bibinfo {volume} {93}},\ \bibinfo
  {pages} {195105} (\bibinfo {year} {2016})}\BibitemShut {NoStop}%
\bibitem [{\citenamefont {Sch\"afer}\ \emph {et~al.}(2016)\citenamefont
  {Sch\"afer}, \citenamefont {Ciuchi}, \citenamefont {Wallerberger},
  \citenamefont {Thunstr\"om}, \citenamefont {Gunnarsson}, \citenamefont
  {Sangiovanni}, \citenamefont {Rohringer},\ and\ \citenamefont
  {Toschi}}]{Schaefer2016c}%
  \BibitemOpen
  \bibfield  {author} {\bibinfo {author} {\bibfnamefont {T.}~\bibnamefont
  {Sch\"afer}}, \bibinfo {author} {\bibfnamefont {S.}~\bibnamefont {Ciuchi}},
  \bibinfo {author} {\bibfnamefont {M.}~\bibnamefont {Wallerberger}}, \bibinfo
  {author} {\bibfnamefont {P.}~\bibnamefont {Thunstr\"om}}, \bibinfo {author}
  {\bibfnamefont {O.}~\bibnamefont {Gunnarsson}}, \bibinfo {author}
  {\bibfnamefont {G.}~\bibnamefont {Sangiovanni}}, \bibinfo {author}
  {\bibfnamefont {G.}~\bibnamefont {Rohringer}},\ and\ \bibinfo {author}
  {\bibfnamefont {A.}~\bibnamefont {Toschi}},\ }\bibfield  {title} {\bibinfo
  {title} {Non-perturbative landscape of the {M}ott-{H}ubbard transition:
  Multiple divergence lines around the critical endpoint},\ }\href
  {https://doi.org/10.1103/PhysRevB.94.235108} {\bibfield  {journal} {\bibinfo
  {journal} {Phys. Rev. B}\ }\textbf {\bibinfo {volume} {94}},\ \bibinfo
  {pages} {235108} (\bibinfo {year} {2016})}\BibitemShut {NoStop}%
\bibitem [{\citenamefont {Vu\v{c}i\v{c}evi\'{c}}\ \emph
  {et~al.}(2018)\citenamefont {Vu\v{c}i\v{c}evi\'{c}}, \citenamefont
  {Wentzell}, \citenamefont {Ferrero},\ and\ \citenamefont
  {Parcollet}}]{Vucicevic2018}%
  \BibitemOpen
  \bibfield  {author} {\bibinfo {author} {\bibfnamefont {J.}~\bibnamefont
  {Vu\v{c}i\v{c}evi\'{c}}}, \bibinfo {author} {\bibfnamefont {N.}~\bibnamefont
  {Wentzell}}, \bibinfo {author} {\bibfnamefont {M.}~\bibnamefont {Ferrero}},\
  and\ \bibinfo {author} {\bibfnamefont {O.}~\bibnamefont {Parcollet}},\
  }\bibfield  {title} {\bibinfo {title} {Practical consequences of the
  {L}uttinger-{W}ard functional multivaluedness for cluster {D}{M}{F}{T}
  methods},\ }\href {https://doi.org/10.1103/PhysRevB.97.125141} {\bibfield
  {journal} {\bibinfo  {journal} {Phys. Rev. B}\ }\textbf {\bibinfo {volume}
  {97}},\ \bibinfo {pages} {125141} (\bibinfo {year} {2018})}\BibitemShut
  {NoStop}%
\bibitem [{\citenamefont {Thunstr\"om}\ \emph {et~al.}(2018)\citenamefont
  {Thunstr\"om}, \citenamefont {Gunnarsson}, \citenamefont {Ciuchi},\ and\
  \citenamefont {Rohringer}}]{Thunstroem2018}%
  \BibitemOpen
  \bibfield  {author} {\bibinfo {author} {\bibfnamefont {P.}~\bibnamefont
  {Thunstr\"om}}, \bibinfo {author} {\bibfnamefont {O.}~\bibnamefont
  {Gunnarsson}}, \bibinfo {author} {\bibfnamefont {S.}~\bibnamefont {Ciuchi}},\
  and\ \bibinfo {author} {\bibfnamefont {G.}~\bibnamefont {Rohringer}},\
  }\bibfield  {title} {\bibinfo {title} {Analytical investigation of
  singularities in two-particle irreducible vertex functions of the {H}ubbard
  atom},\ }\href {https://doi.org/10.1103/PhysRevB.98.235107} {\bibfield
  {journal} {\bibinfo  {journal} {Phys. Rev. B}\ }\textbf {\bibinfo {volume}
  {98}},\ \bibinfo {pages} {235107} (\bibinfo {year} {2018})}\BibitemShut
  {NoStop}%
\bibitem [{\citenamefont {Springer}\ \emph {et~al.}(2020)\citenamefont
  {Springer}, \citenamefont {Chalupa}, \citenamefont {Ciuchi}, \citenamefont
  {Sangiovanni},\ and\ \citenamefont {Toschi}}]{Springer2019}%
  \BibitemOpen
  \bibfield  {author} {\bibinfo {author} {\bibfnamefont {D.}~\bibnamefont
  {Springer}}, \bibinfo {author} {\bibfnamefont {P.}~\bibnamefont {Chalupa}},
  \bibinfo {author} {\bibfnamefont {S.}~\bibnamefont {Ciuchi}}, \bibinfo
  {author} {\bibfnamefont {G.}~\bibnamefont {Sangiovanni}},\ and\ \bibinfo
  {author} {\bibfnamefont {A.}~\bibnamefont {Toschi}},\ }\bibfield  {title}
  {\bibinfo {title} {Interplay between local response and vertex divergences in
  many-fermion systems with on-site attraction},\ }\href
  {https://doi.org/10.1103/PhysRevB.101.155148} {\bibfield  {journal} {\bibinfo
   {journal} {Phys. Rev. B}\ }\textbf {\bibinfo {volume} {101}},\ \bibinfo
  {pages} {155148} (\bibinfo {year} {2020})}\BibitemShut {NoStop}%
\bibitem [{\citenamefont {Melnick}\ and\ \citenamefont
  {Kotliar}(2020)}]{Kotliar2020}%
  \BibitemOpen
  \bibfield  {author} {\bibinfo {author} {\bibfnamefont {C.}~\bibnamefont
  {Melnick}}\ and\ \bibinfo {author} {\bibfnamefont {G.}~\bibnamefont
  {Kotliar}},\ }\bibfield  {title} {\bibinfo {title} {Fermi-liquid theory and
  divergences of the two-particle irreducible vertex in the periodic {A}nderson
  lattice},\ }\href {https://doi.org/10.1103/PhysRevB.101.165105} {\bibfield
  {journal} {\bibinfo  {journal} {Phys. Rev. B}\ }\textbf {\bibinfo {volume}
  {101}},\ \bibinfo {pages} {165105} (\bibinfo {year} {2020})}\BibitemShut
  {NoStop}%
\bibitem [{\citenamefont {Adler}(2022)}]{AdlerSBE}%
  \BibitemOpen
  \bibfield  {author} {\bibinfo {author} {\bibfnamefont {S.}~\bibnamefont
  {Adler}}} (\bibinfo {year} {2022}),\ \bibinfo {note}
  {(unpublished)}\BibitemShut {NoStop}%
\bibitem [{\citenamefont {Vilk}\ and\ \citenamefont
  {Tremblay}(1997)}]{Vilk1997}%
  \BibitemOpen
  \bibfield  {author} {\bibinfo {author} {\bibfnamefont {Y.~M.}\ \bibnamefont
  {Vilk}}\ and\ \bibinfo {author} {\bibfnamefont {A.-M.~S.}\ \bibnamefont
  {Tremblay}},\ }\bibfield  {title} {\bibinfo {title} {Non-perturbative
  many-body approach to the {H}ubbard model and single-particle pseudogap},\
  }\href {https://doi.org/10.1051/jp1:1997135} {\bibfield  {journal} {\bibinfo
  {journal} {J. Phys. I France}\ }\textbf {\bibinfo {volume} {7}},\ \bibinfo
  {pages} {1309} (\bibinfo {year} {1997})}\BibitemShut {NoStop}%
\bibitem [{\citenamefont {Rohringer}\ and\ \citenamefont
  {Toschi}(2016)}]{Rohringer2016}%
  \BibitemOpen
  \bibfield  {author} {\bibinfo {author} {\bibfnamefont {G.}~\bibnamefont
  {Rohringer}}\ and\ \bibinfo {author} {\bibfnamefont {A.}~\bibnamefont
  {Toschi}},\ }\bibfield  {title} {\bibinfo {title} {Impact of non-local
  correlations over different energy scales: A dynamical vertex approximation
  study},\ }\href {https://doi.org/10.1103/PhysRevB.94.125144} {\bibfield
  {journal} {\bibinfo  {journal} {Phys. Rev. B}\ }\textbf {\bibinfo {volume}
  {94}},\ \bibinfo {pages} {125144} (\bibinfo {year} {2016})}\BibitemShut
  {NoStop}%
\bibitem [{Note2()}]{Note2}%
  \BibitemOpen
  \bibinfo {note} {The QMC result was obtained using w2dynamics \cite
  {w2dynamics} with Worm sampling \cite {Gunacker15,Gunacker2016} and symmetric
  improved estimators \cite {Kaufmann2019}, designed to reduce the
  high-frequency noise, see further Appendix \ref {sec:APP-tech-QMC}. However,
  the noise cannot be suppressed completely, and thus the QMC result fluctuates
  around the PA and mfRG solution.}\BibitemShut {Stop}%
\bibitem [{\citenamefont {Krien}\ \emph {et~al.}(2017)\citenamefont {Krien},
  \citenamefont {van Loon}, \citenamefont {Hafermann}, \citenamefont {Otsuki},
  \citenamefont {Katsnelson},\ and\ \citenamefont {Lichtenstein}}]{Krien2017}%
  \BibitemOpen
  \bibfield  {author} {\bibinfo {author} {\bibfnamefont {F.}~\bibnamefont
  {Krien}}, \bibinfo {author} {\bibfnamefont {E.~G. C.~P.}\ \bibnamefont {van
  Loon}}, \bibinfo {author} {\bibfnamefont {H.}~\bibnamefont {Hafermann}},
  \bibinfo {author} {\bibfnamefont {J.}~\bibnamefont {Otsuki}}, \bibinfo
  {author} {\bibfnamefont {M.~I.}\ \bibnamefont {Katsnelson}},\ and\ \bibinfo
  {author} {\bibfnamefont {A.~I.}\ \bibnamefont {Lichtenstein}},\ }\bibfield
  {title} {\bibinfo {title} {Conservation in two-particle self-consistent
  extensions of dynamical mean-field theory},\ }\href
  {https://doi.org/10.1103/PhysRevB.96.075155} {\bibfield  {journal} {\bibinfo
  {journal} {Phys. Rev. B}\ }\textbf {\bibinfo {volume} {96}},\ \bibinfo
  {pages} {075155} (\bibinfo {year} {2017})}\BibitemShut {NoStop}%
\bibitem [{\citenamefont {Krien}(2018)}]{KrienThesis}%
  \BibitemOpen
  \bibfield  {author} {\bibinfo {author} {\bibfnamefont {F.}~\bibnamefont
  {Krien}},\ }\emph {\bibinfo {title} {Conserving dynamical mean-field
  approaches to strongly correlated systems}},\ \href
  {https://ediss.sub.uni-hamburg.de/handle/ediss/7726} {Ph.D. thesis},\
  \bibinfo  {school} {Universit\"at Hamburg} (\bibinfo {year}
  {2018})\BibitemShut {NoStop}%
\bibitem [{\citenamefont {Hafermann}\ \emph {et~al.}(2014)\citenamefont
  {Hafermann}, \citenamefont {van Loon}, \citenamefont {Katsnelson},
  \citenamefont {Lichtenstein},\ and\ \citenamefont
  {Parcollet}}]{Hafermann2014a}%
  \BibitemOpen
  \bibfield  {author} {\bibinfo {author} {\bibfnamefont {H.}~\bibnamefont
  {Hafermann}}, \bibinfo {author} {\bibfnamefont {E.~G. C.~P.}\ \bibnamefont
  {van Loon}}, \bibinfo {author} {\bibfnamefont {M.~I.}\ \bibnamefont
  {Katsnelson}}, \bibinfo {author} {\bibfnamefont {A.~I.}\ \bibnamefont
  {Lichtenstein}},\ and\ \bibinfo {author} {\bibfnamefont {O.}~\bibnamefont
  {Parcollet}},\ }\bibfield  {title} {\bibinfo {title} {Collective charge
  excitations of strongly correlated electrons, vertex corrections, and gauge
  invariance},\ }\href {https://doi.org/10.1103/PhysRevB.90.235105} {\bibfield
  {journal} {\bibinfo  {journal} {Phys. Rev. B}\ }\textbf {\bibinfo {volume}
  {90}},\ \bibinfo {pages} {235105} (\bibinfo {year} {2014})}\BibitemShut
  {NoStop}%
\bibitem [{\citenamefont {Baym}(1962)}]{Baym1962}%
  \BibitemOpen
  \bibfield  {author} {\bibinfo {author} {\bibfnamefont {G.}~\bibnamefont
  {Baym}},\ }\bibfield  {title} {\bibinfo {title} {Self-consistent
  approximations in many-body systems},\ }\href
  {https://doi.org/10.1103/PhysRev.127.1391} {\bibfield  {journal} {\bibinfo
  {journal} {Phys. Rev.}\ }\textbf {\bibinfo {volume} {127}},\ \bibinfo {pages}
  {1391} (\bibinfo {year} {1962})}\BibitemShut {NoStop}%
\bibitem [{\citenamefont {Kopietz}\ \emph
  {et~al.}(2010{\natexlab{b}})\citenamefont {Kopietz}, \citenamefont
  {Bartosch}, \citenamefont {Costa}, \citenamefont {Isidori},\ and\
  \citenamefont {Ferraz}}]{Kopietz2010a}%
  \BibitemOpen
  \bibfield  {author} {\bibinfo {author} {\bibfnamefont {P.}~\bibnamefont
  {Kopietz}}, \bibinfo {author} {\bibfnamefont {L.}~\bibnamefont {Bartosch}},
  \bibinfo {author} {\bibfnamefont {L.}~\bibnamefont {Costa}}, \bibinfo
  {author} {\bibfnamefont {A.}~\bibnamefont {Isidori}},\ and\ \bibinfo {author}
  {\bibfnamefont {A.}~\bibnamefont {Ferraz}},\ }\bibfield  {title} {\bibinfo
  {title} {Ward identities for the {A}nderson impurity model: derivation via
  functional methods and the exact renormalization group},\ }\href
  {https://doi.org/10.1088/1751-8113/43/38/385004} {\bibfield  {journal}
  {\bibinfo  {journal} {J. Phys. A}\ }\textbf {\bibinfo {volume} {43}},\
  \bibinfo {pages} {385004} (\bibinfo {year} {2010}{\natexlab{b}})}\BibitemShut
  {NoStop}%
\bibitem [{\citenamefont {Toschi}\ \emph {et~al.}(2007)\citenamefont {Toschi},
  \citenamefont {Katanin},\ and\ \citenamefont {Held}}]{Toschi2007}%
  \BibitemOpen
  \bibfield  {author} {\bibinfo {author} {\bibfnamefont {A.}~\bibnamefont
  {Toschi}}, \bibinfo {author} {\bibfnamefont {A.~A.}\ \bibnamefont
  {Katanin}},\ and\ \bibinfo {author} {\bibfnamefont {K.}~\bibnamefont
  {Held}},\ }\bibfield  {title} {\bibinfo {title} {Dynamical vertex
  approximation; a step beyond dynamical mean-field theory},\ }\href
  {https://doi.org/10.1103/PhysRevB.75.045118} {\bibfield  {journal} {\bibinfo
  {journal} {Phys Rev. B}\ }\textbf {\bibinfo {volume} {75}},\ \bibinfo {pages}
  {045118} (\bibinfo {year} {2007})}\BibitemShut {NoStop}%
\bibitem [{\citenamefont {Ayral}\ and\ \citenamefont
  {Parcollet}(2016)}]{Ayral2016}%
  \BibitemOpen
  \bibfield  {author} {\bibinfo {author} {\bibfnamefont {T.}~\bibnamefont
  {Ayral}}\ and\ \bibinfo {author} {\bibfnamefont {O.}~\bibnamefont
  {Parcollet}},\ }\bibfield  {title} {\bibinfo {title} {Mott physics and
  collective modes: An atomic approximation of the four-particle irreducible
  functional},\ }\href {https://doi.org/10.1103/PhysRevB.94.075159} {\bibfield
  {journal} {\bibinfo  {journal} {Phys. Rev. B}\ }\textbf {\bibinfo {volume}
  {94}},\ \bibinfo {pages} {075159} (\bibinfo {year} {2016})}\BibitemShut
  {NoStop}%
\bibitem [{\citenamefont {Hille}(2020)}]{Hille2020Thesis}%
  \BibitemOpen
  \bibfield  {author} {\bibinfo {author} {\bibfnamefont {C.~U.}\ \bibnamefont
  {Hille}},\ }\emph {\bibinfo {title} {The role of the self-energy in the
  functional renormalization group description of interacting Fermi systems}},\
  \href {https://doi.org/http://dx.doi.org/10.15496/publikation-46212} {Ph.D.
  thesis},\ \bibinfo  {school} {Eberhard Karls Universit\"at T\"ubingen}
  (\bibinfo {year} {2020})\BibitemShut {NoStop}%
\bibitem [{\citenamefont {Gunacker}\ \emph {et~al.}(2015)\citenamefont
  {Gunacker}, \citenamefont {Wallerberger}, \citenamefont {Gull}, \citenamefont
  {Hausoel}, \citenamefont {Sangiovanni},\ and\ \citenamefont
  {Held}}]{Gunacker15}%
  \BibitemOpen
  \bibfield  {author} {\bibinfo {author} {\bibfnamefont {P.}~\bibnamefont
  {Gunacker}}, \bibinfo {author} {\bibfnamefont {M.}~\bibnamefont
  {Wallerberger}}, \bibinfo {author} {\bibfnamefont {E.}~\bibnamefont {Gull}},
  \bibinfo {author} {\bibfnamefont {A.}~\bibnamefont {Hausoel}}, \bibinfo
  {author} {\bibfnamefont {G.}~\bibnamefont {Sangiovanni}},\ and\ \bibinfo
  {author} {\bibfnamefont {K.}~\bibnamefont {Held}},\ }\bibfield  {title}
  {\bibinfo {title} {Continuous-time quantum {M}onte {C}arlo using worm
  sampling},\ }\href {https://doi.org/10.1103/PhysRevB.92.155102} {\bibfield
  {journal} {\bibinfo  {journal} {Phys. Rev. B}\ }\textbf {\bibinfo {volume}
  {92}},\ \bibinfo {pages} {155102} (\bibinfo {year} {2015})}\BibitemShut
  {NoStop}%
\bibitem [{\citenamefont {Gunacker}\ \emph {et~al.}(2016)\citenamefont
  {Gunacker}, \citenamefont {Wallerberger}, \citenamefont {Ribic},
  \citenamefont {Hausoel}, \citenamefont {Sangiovanni},\ and\ \citenamefont
  {Held}}]{Gunacker2016}%
  \BibitemOpen
  \bibfield  {author} {\bibinfo {author} {\bibfnamefont {P.}~\bibnamefont
  {Gunacker}}, \bibinfo {author} {\bibfnamefont {M.}~\bibnamefont
  {Wallerberger}}, \bibinfo {author} {\bibfnamefont {T.}~\bibnamefont {Ribic}},
  \bibinfo {author} {\bibfnamefont {A.}~\bibnamefont {Hausoel}}, \bibinfo
  {author} {\bibfnamefont {G.}~\bibnamefont {Sangiovanni}},\ and\ \bibinfo
  {author} {\bibfnamefont {K.}~\bibnamefont {Held}},\ }\bibfield  {title}
  {\bibinfo {title} {Worm-improved estimators in continuous-time quantum
  {M}onte {C}arlo},\ }\href {https://doi.org/10.1103/PhysRevB.94.125153}
  {\bibfield  {journal} {\bibinfo  {journal} {Phys. Rev. B}\ }\textbf {\bibinfo
  {volume} {94}},\ \bibinfo {pages} {125153} (\bibinfo {year}
  {2016})}\BibitemShut {NoStop}%
\bibitem [{\citenamefont {Kaufmann}\ \emph {et~al.}(2019)\citenamefont
  {Kaufmann}, \citenamefont {Gunacker}, \citenamefont {Kowalski}, \citenamefont
  {Sangiovanni},\ and\ \citenamefont {Held}}]{Kaufmann2019}%
  \BibitemOpen
  \bibfield  {author} {\bibinfo {author} {\bibfnamefont {J.}~\bibnamefont
  {Kaufmann}}, \bibinfo {author} {\bibfnamefont {P.}~\bibnamefont {Gunacker}},
  \bibinfo {author} {\bibfnamefont {A.}~\bibnamefont {Kowalski}}, \bibinfo
  {author} {\bibfnamefont {G.}~\bibnamefont {Sangiovanni}},\ and\ \bibinfo
  {author} {\bibfnamefont {K.}~\bibnamefont {Held}},\ }\bibfield  {title}
  {\bibinfo {title} {Symmetric improved estimators for continuous-time quantum
  {M}onte {C}arlo},\ }\href {https://doi.org/10.1103/PhysRevB.100.075119}
  {\bibfield  {journal} {\bibinfo  {journal} {Phys. Rev. B}\ }\textbf {\bibinfo
  {volume} {100}},\ \bibinfo {pages} {075119} (\bibinfo {year}
  {2019})}\BibitemShut {NoStop}%
\bibitem [{\citenamefont {Kugler}(2018)}]{Kugler2018c}%
  \BibitemOpen
  \bibfield  {author} {\bibinfo {author} {\bibfnamefont {F.~B.}\ \bibnamefont
  {Kugler}},\ }\bibfield  {title} {\bibinfo {title} {Counting {F}eynman
  diagrams via many-body relations},\ }\href
  {https://doi.org/10.1103/PhysRevE.98.023303} {\bibfield  {journal} {\bibinfo
  {journal} {Phys. Rev. E}\ }\textbf {\bibinfo {volume} {98}},\ \bibinfo
  {pages} {023303} (\bibinfo {year} {2018})}\BibitemShut {NoStop}%
\end{thebibliography}%

\end{document}